\documentclass[twocolumn]{aa} 
\usepackage[varg]{txfonts}
\usepackage{natbib}
\usepackage{url}
\PassOptionsToPackage{hyphens}{url}
\usepackage{graphicx}
\usepackage{subfigure}
\usepackage{amsmath}
\usepackage{txfonts}
\usepackage{xcolor}
\usepackage{listings} 
\usepackage{times}
\usepackage{multirow}
\usepackage{float}
\usepackage{epstopdf}
\usepackage{longtable}
\usepackage{soul}
\usepackage[colorlinks=true,linkcolor=red,citecolor=blue, breaklinks=true]{hyperref}
\restylefloat{figure}  
\usepackage{natbib,twoopt}
\usepackage[breaklinks=true]{hyperref} 
\bibpunct{(}{)}{;}{a}{}{,} 
\makeatletter
\newcommandtwoopt{\citeads}[3][][]{\href{http://adsabs.harvard.edu/abs/#3}%
{\def\hyper@linkstart##1##2{}%
\let\hyper@linkend\@empty\citealp[#1][#2]{#3}}}
\newcommandtwoopt{\citepads}[3][][]{\href{http://adsabs.harvard.edu/abs/#3}%
{\def\hyper@linkstart##1##2{}%
\let\hyper@linkend\@empty\citep[#1][#2]{#3}}}
\newcommandtwoopt{\citetads}[3][][]{\href{http://adsabs.harvard.edu/abs/#3}%
{\def\hyper@linkstart##1##2{}%
\let\hyper@linkend\@empty\citet[#1][#2]{#3}}}
\newcommandtwoopt{\citeyearads}[3][][]%
{\href{http://adsabs.harvard.edu/abs/#3}
{\def\hyper@linkstart##1##2{}%
\let\hyper@linkend\@empty\citeyear[#1][#2]{#3}}}
\makeatother
\newcommand{\hi}{\ion{H}{i}}
\newcommand{\hii}{\ion{H}{ii}}

\newcommand{\herschel}{{\it Herschel}}

\def\NHUNIT{\ifmmode {\rm \,cm^{-2}} \else $\rm \,cm^{-2}$ \fi} 
\newcommand{\mum}{$\mu\rm m$}
\def\nhh{\ifmmode N_{\rm H_{2}}\else $N_{\rm H_{2}}$\fi} 
\def\nhhc{\ifmmode N_{\rm H_{2}}^0\else $N_{\rm H_{2}}^0$\fi} 
\def\nhhbg{\ifmmode N_{\rm H_{2}}^{\rm bg}\else $N_{\rm H_{2}}^{\rm bg}$\fi} 
\def\nh{\ifmmode N_{\rm H}\else $N_{\rm H}$\fi}
\def\ml{\ifmmode M_{\rm line}\else $M_{\rm line}$\fi}  
\def\sunpc{\ifmmode \rm M_\odot/\rm pc\else $\rm M_\odot/\rm pc$\fi}  
\def\rout{\ifmmode R_{\rm out}\else $R_{\rm out}$\fi}  
\def\av{\ifmmode A_{\rm V}\else $A_{\rm V}$\fi}   
\def\fwhmdec{\ifmmode FHWM_{\rm dec}\else $FWHM_{\rm dec}$\fi}   
\def\rflat{\ifmmode R_{\rm flat}\else $R_{\rm flat}$\fi}   
\def\kms{\ifmmode {km\,s$^{-1}$}\else km\,s$^{-1}$\fi}  

\def\arcm{\ifmmode {^{\scriptstyle\prime}}
          \else $^{\scriptstyle\prime}$\fi}
\newdimen\sa  \newdimen\sb
\def\parcs{\sa=.07em \sb=.03em
     \ifmmode \hbox{\rlap{.}}^{\scriptstyle\prime\kern -\sb\prime}\hbox{\kern -\sa}
     \else \rlap{.}$^{\scriptstyle\prime\kern -\sb\prime}$\kern -\sa\fi}
\def\parcm{\sa=.08em \sb=.03em
     \ifmmode \hbox{\rlap{.}\kern\sa}^{\scriptstyle\prime}\hbox{\kern-\sb}
     \else \rlap{.}\kern\sa$^{\scriptstyle\prime}$\kern-\sb\fi}
\def\parcd{\sa=.08em \sb=.03em
     \ifmmode \hbox{\rlap{.}\kern\sa}^{\scriptstyle\circ}\hbox{\kern-\sb}
     \else \rlap{.}\kern\sa$^{\scriptstyle\circ}$\kern-\sb\fi}
\def\rev{}
\def\revbis{}
\begin{document} 

\title{Velocity structure of the 50\,pc-long NGC~6334 filamentary cloud}
\subtitle{Hints of multiple compressions and their impact on the cloud properties?}

 \titlerunning{Velocity structure of the NGC~6334 filamentary cloud}
 
   \author{
     Doris Arzoumanian\inst{1,2,3}
             \and
      Delphine Russeil\inst{1}
      \and
      Annie Zavagno\inst{1}
              \and
      Michael Chun-Yuan Chen\inst{4}
          \and
           Philippe Andr\'e\inst{5}
                \and
               Shu-ichiro~Inutsuka\inst{6}
                   \and
                    Yoshiaki Misugi\inst{6}
                 \and
                 \'Alvaro~S\'anchez-Monge\inst{7}
      \and
       Peter Schilke\inst{7}
       \and
       Alexander Men'shchikov\inst{5}
          \and
       Mikito Kohno\inst{8}
            }

   \institute{$^1$Aix Marseille Univ, CNRS, CNES, LAM, Marseille, France\\ 
                $^2$Instituto de Astrof\'isica e Ci{\^e}ncias do Espa\c{c}o, Universidade do Porto, CAUP, Rua das Estrelas, 4150-762 Porto, Portugal\\
                $^3$Division of Science, National Astronomical Observatory of Japan, 2-21-1 Osawa, Mitaka, Tokyo 181-8588, Japan\\
  \email{doris.arzoumanian@nao.ac.jp} \\
                 $^4$Department for Physics, Engineering Physics and Astrophysics, Queen's University, Kingston, ON, K7L 3N6, Canada\\
                 $^5$Laboratoire d'Astrophysique (AIM), CEA/DRF, CNRS, Universit\'e Paris-Saclay, Universit\'e Paris Diderot, Sorbonne Paris Cit\'e, 91191 Gif-sur-Yvette, France\\
                 $^6$Department of Physics, Graduate School of Science, Nagoya University, Furo-cho, Chikusa-ku, Nagoya, Aichi 464-8602, Japan\\
                 $^7$I.\ Physikalisches Institut, Universit\"at zu K\"oln, Z\"ulpicher Str.\ 77, 50937 K\"oln, Germany\\
                 $^8$Astronomy Section, Nagoya City Science Museum, 2-17-1 Sakae, Naka-ku, Nagoya, Aichi 460-0008, Japan
                  }
     \date{}



\abstract{
{\it Context.} The interstellar medium is observed to be organised in filamentary structures, as well as neutral (\hi) and ionized (\hii)   bubbles.  
The expanding nature of these bubbles makes them shape their surrounding medium and possibly play a role in the formation and evolution of interstellar filaments. The impact of the expansion of these bubbles on the interstellar medium is not well understood.  \\ 
{\it Aims.} We aim to describe the kinematics of a filamentary molecular cloud forming high-mass stars and hosting multiple \hii\ regions, to study the possible environmental impact  on the properties of molecular filaments.   \\ 
{\it Methods.} We present APEX $^{13}$CO and C$^{18}$O($2-1$) mapping observations of the $10\times50$\,pc NGC~6334 molecular cloud complex. We investigate the gas velocity structure along and across the 50\,pc-long cloud and towards velocity-coherent-filaments (VCFs).\\ 
{\it Results.} The NGC~6334 complex is observed to have a coherent velocity structure 
 {\rev  smoothly varying by} $\sim5$\,\kms\ over its 50\,pc elongation parallel to the Galactic plane. We  identify a sample of {\rev 75 VCFs  in  the C$^{18}$O($2-1$) position-position-velocity cube and present the properties of 47 VCFs with a length $\gtrsim1\,$pc (5 beams). }
We {\rev measure} a wealth of velocity gradients along the VCFs. 
The amplitudes of these velocity gradients and the velocity dispersion measured along the crests  increase with the  column density of the VCFs.  
We derive the column density and velocity power spectra of the VCFs. These power spectra are well represented with power laws showing similar slopes {\rev for both quantities (with a mean of about $-2$),   albeit some differ by up to a factor of two}.
The position velocity diagrams perpendicular to three VCFs (selected for being in different physical environments) show  {\rev the V-shaped velocity pattern, corresponding to a bent structure in velocity space with the filament at the tip of the V surrounded by an extended structure connected to it with a velocity gradient. }
{\revbis This velocity structure is qualitatively similar to that resulting from numerical simulations} of filament formation from large-scale compression from propagating shock fronts.  
 In addition, the radial profiles perpendicular to these VCFs 
 hint to  small-scale internal impacts from neighbouring \hii\ bubbles on two of them, while the third is mostly unaffected.\\
{\it Conclusions.}  
{\rev The observed opposite curvature in velocity space (V- and $\Lambda$-shaped)}
towards the  VCFs points to 
various origins of  large-scale external compressions from propagating \hi\ bubbles.
 This {\rev suggests the plausible } importance of multiple \hi\  compressions, separated in space and time,  in the formation and evolution of molecular clouds and their star formation history. These latter atomic compressions due to past and distant star formation events are complemented by the impact of \hii\ bubbles from present time and local star formation activity. 
}

\keywords{stars: formation -- ISM: clouds -- ISM: structure  -- submillimeter: ISM }

\maketitle

\section{Introduction}\label{intro} 

Observations of the interstellar medium (ISM), especially its molecular component, reveal the impressive organization of the dust and gas into complex networks of filaments  {\rev \citep[e.g.,][]{Schneider1979,Molinari2010,Menshchikov2010,Umemoto2017,Mattern2018,Schisano2020}.} In molecular clouds, gravitationally unstable filaments are now identified as the main birthplaces of individual solar-type stars  \citep[e.g.,][]{Andre2010,Andre2014,Tafalla2015}, while the hubs formed {\rev at} their intersections are associated with stellar clusters and high-mass stars \citep[][]{Myers2009,Schneider2012,Peretto2014,Kumar2020,Kumar2021}. 
Filamentary structures and hubs are also observed in the ISM of external galaxies \citep[e.g., the Magellanic clouds,][]{Fukui2019}. This extragalactic filamentary ISM, now revealed thanks to  high-angular resolution observations with  ALMA {\rev(Atacama Large Millimeter Array),}  seem to be associated with the star formation process as in our Galaxy, suggesting that filaments are also important for star formation in galaxies in general.
An important question, which is still highly debated, is to understand the formation and evolution of these filaments 
and to describe the physical processes leading to their fragmentation into  star forming cores. 

The matter cycle in the ISM is regulated by  heating and cooling processes, and  the compression and expansion of this interstellar material. 
{\rev  There are two broad types of compressions and expansions that affects both distant ($\sim100$\,pc scale) and local  (a few parsecs) environments:}
 1) Large-scale ($\sim100$\,pc) external, mostly neutral, compressions from expansion of  \hi\ shells or supershells \citep[e.g.,][]{Dawson2011,Bracco2020} generated mainly by supernovae, and 2) small-scale ($\sim1-5$\,pc) local and internal compressions, mostly ionized,  due to present-time stellar feedback from mainly \hii\ regions, stellar winds, and outflows \citep[e.g.,][]{Russeil2016}. 

Recent theoretical models 
{\rev clearly demonstrate the role of expanding} 
  \hi\ shells for the formation of molecular clouds \citep[e.g.,][]{Hennebelle2008,Heitsch2009,Inoue2009}. In particular, these models stress the importance of  
multiple compressions 
for the formation  of magnetised filamentary molecular clouds
 \citep[e.g.,][]{Inutsuka2015,Iwasaki2019}.
The typical timescale of such compressions is estimated in the Galactic disk to be on average $\sim1$~Myr \citep{McKee1977,Inutsuka2015}. Thus, the formation of molecular clouds may last from a few to $\sim10$\,Myr or more \citep[see, e.g.,][]{Kobayashi2017}. 
These  successive  compressions %
{\rev may} continuously alter the density, velocity, and magnetic field structures of the clouds having also a strong impact on  the formation of new generation of filaments and consequently that  of stars. 
While the first generation of stars form and impact their (local) surroundings (through outflow, jets, winds, and ionising radiation), new cold matter is continuously assembled replenishing the sites of star formation, i.e., filaments and hubs. This matter replenishment may be channeled from within the cloud itself through molecular filaments towards  dense ridge-like main-filaments \citep[][]{Schneider2010,Palmeirim2013}  or towards hubs {\rev \citep[][]{Myers2009,Peretto2013,Peretto2014,Trevino-Morales2019}.} Matter can  also be brought into the system (the cloud)  by a new event of external collision  \citep[e.g.,][]{Fukui2018}. 
\citet[][]{Arzoumanian2018} identified, {\rev in position-velocity (PV) diagrams, extended structures with mean line-of-sight (LOS) velocities offseted with respect to and connected to the velocity of a filament. They suggested a}  multi-interaction scenario where sheet-like extended structures interact, in space and time, with a star forming filament and are responsible for its compression or disruption. %
\citet[][]{Arzoumanian2018} also identified a {\rev bent velocity structure in the PV space. They showed that such a  V- or $\Lambda$-shaped velocity structure can  result from 
 the filament formation process by shock compression as  proposed by the theoretical model of \citet{Inoue2018}.  In this latter model } a filament is formed 
due to convergence of a flow of matter generated by the bending of the ambient magnetic field structure induced by an interstellar shock compression \citep[see also][]{Inoue2013,Vaidya2013}. This velocity structure has also been observed towards another filament \citep[the Musca filament in][]{Bonne2020}.
More recently, {\rev in a theoretical study,}  \citet[][]{Abe2021}  
 {\rev proposed a classification of filament formation mechanisms
 }
resulting from the variation of the relative importance between the shock velocity, the turbulence, and the magnetic field strength \citep[{\rev see also the theoretical study by}][]{Che-Yu-Chen2020}. 

\begin{figure*}[!h]
   \centering
     \resizebox{21.cm}{!}{
     \hspace{-2.cm}
\includegraphics[angle=0]{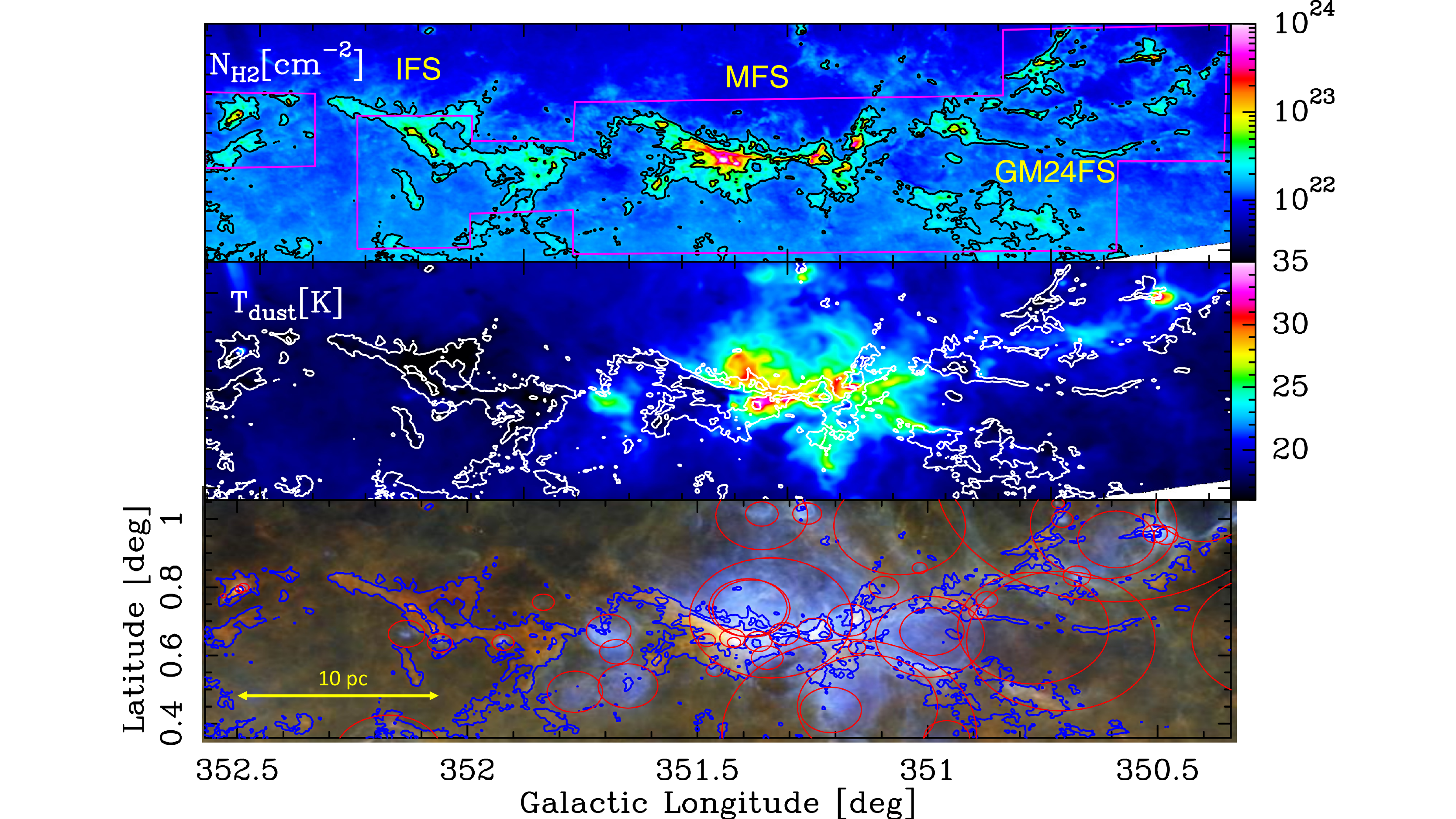} 
}
\vspace{-.4cm}
  \caption{ 
  {\it Top}:   Column density map of the NGC~6334 cloud  derived from \herschel\ HOBYS data \citep{Motte2010} as explained in Sect.\,\ref{Herschel}. The map is at the resolution of 18\parcs2 and the contours are at column densities of 1.8, and 5\,$\times10^{22}\,\NHUNIT$. 
 The coverage of the APEX observations is shown in magenta.  
 The  three identified subregions are indicated  (in yellow) from east to west: the inter-filament system (IFS) for $352\parcd3>l>351\parcd8$, the main-filament system (MFS) for $351\parcd8>l>351^\circ$, and the GM 24 filament system (GM24FS) for $351^\circ>l>350\parcd4$. 
        {\it Middle}:   Dust temperature map at 36\parcs3 (cf., Sect.\,\ref{Herschel}). The  contours are the same as in the top panel.  
{\it Bottom}:  Colour composite image of NGC~6334 derived from \herschel\ data, where the
 70\,\mum, 160\,\mum, and 250\,\mum\ intensities are represented in blue, green, and red, respectively.
  This image has been obtained from \citet{Causi2016}. The contours trace the column density as in the top panel. {\rev The red circles show the  positions and sizes of the \hii\ regions from \citet{AndersonLD2014} and \citet{Langston2000}, and are the same as those plotted on Fig.\,1 of  \citet{Russeil2016}.}
}          
  \label{ColdensTempMaps}
    \end{figure*}

To make progress in our understanding of the impact of these two types of compressions; internal-ionised and external-neutral, on  star-forming molecular clouds, we here analyse the velocity structure of the NGC~6334  high-mass star-forming complex as a whole and at smaller scales towards dense filaments both, in the vicinity and away from local stellar feedback and \hii\ regions.
NGC~6334 is a well studied molecular cloud complex \citep[see][for an extensive review]{Persi2008}
 at a relatively nearby distance of $1.3\pm0.3$\,kpc \citep{Chibueze2014}.
NGC~6334 looks like a "Cat's Paw" with the grouping of a large number of  \hii\ regions   \citep[][{\rev and see Fig.\,\ref{ColdensTempMaps}}]{Persi2010,Russeil2016}, surrounding an elongated filamentary cloud %
very bright at (sub)millimeter wavelengths \citep{Kraemer1999,Matthews2008,Russeil2013,Zernickel2013,Tige2017}  
and  actively forming  high-mass stars 
 \citep{Sandell2000,McCutcheon2000,Munoz2007,Qiu2011}.
 This filamentary cloud is dominated by  a 10\,pc long main-filament %
with a line mass  ranging from $\ml\sim500\,\sunpc$ to $\sim2000\,\sunpc$ over  its  crest  \citep{Andre2016} 
and it is fragmented into a series of  cores with a mean mass $\sim10\,$M$_\odot$ 
  \citep[e.g.,][]{Shimajiri2019}. 
The filament inner width is observed to be on the order of 0.1\,pc  \citep{Andre2016},  compatible with the findings derived from statistical analysis of dust continuum $Herschel$ observations of nearby and less massive filaments  \citep[][]{Arzoumanian2011,Arzoumanian2019,Koch2015}. This similarity suggests  that the 
gravitational fragmentation of $\sim0.1$\,pc-wide  filaments may also be the main mode of intermediate-mass star formation  \citep{Shimajiri2019,Andre2019}.\\

In this paper, we study the velocity structure of the $\sim50$\,pc long NGC~6334 high-mass star-forming region as traced by $^{12}$CO($2-1$), $^{13}$CO($2-1$), and C$^{18}$O($2-1$) molecular line emission. 
This paper is organized as follows: 
In Sect.\,\ref{obs}, we describe the observational data used in the analysis. 
In Sect.\,\ref{ana1}, we present the velocity structure observed towards the 50\,pc-long NGC~6334 cloud.
{\rev In Sects.\,\ref{ana2},\,\ref{ana3}, and \ref{ana_selected}} we analyze in more details the velocity structure along and across a sample of velocity-coherent-filaments  identified towards the studied field. 
In Sect.\,\ref{disc}, we discuss the possible physical origin of the observed velocity structures of the cloud and filaments, and suggest a scenario for the dynamical formation   of this high-mass star-forming  filamentary cloud. We give a summary of the analysis and results in Sect.\,\ref{Summary}. 

\section{Observations}\label{obs}
\subsection{$^{13}$CO($2-1$) and C$^{18}$O($2-1$) APEX Data}

The APEX-SHeFI \citep{Vassilev2008} OTF mapping observations of the $^{13}$CO($2-1$) and C$^{18}$O($2-1$) lines were conducted in the frame of the programme 091.F-9512(A). The data reduction and the first analysis of these data have been performed by \cite{Zernickel2015}.
We summarise here the data characteristics. The spectral range is 216.8$-$220.8\,GHz with a spectral resolution of 88.5\,kHz (corresponding to a
velocity resolution of 0.12\,km/s at 220\,GHz). The spectra were smoothed to a velocity resolution of 0.3\,km/s to give a final average root mean square (rms) noise per 0.3\,km/s-channel of 0.44\,K for $^{13}$CO and 0.55\,K for C$^{18}$O. 
The final angular resolution of both  $^{13}$CO and C$^{18}$O cubes is 30\parcs2 {\revbis (0.19\,pc at the 1.3\,kpc distance of the cloud).} Both cubes were reprojected on the same grid  with a pixel size of 14\parcs3. 
The longitude range of the map is from $350\parcd4$  to $352\parcd6$ covering the well known NGC~6334 main star-forming region but also the GM~24 region on the west and the inter-filament region toward NGC~6357 on the east (see Fig.\,\ref{ColdensTempMaps}).   
\cite{Zernickel2015} underlines that due to the offset position contamination by foreground and background clouds, spectral  absorption features at $-141$, $-88$, $-33$, $-17$ and $+24$\,km/s are seen over the whole map \citep[see Fig.\,4.1 of][]{Zernickel2015}. 
In the following, we will focus the analysis to the [$-20,+10$]\,km/s velocity range that covers the {\rev kinematics} of the gas  associated with NGC~6334.

\subsection{$^{12}$CO($2-1$) NANTEN2 Data}

We included in the analysis the  $^{12}$CO($2-1$)  velocity cubes obtained with the  4\,m millimeter/sub-millimeter radio telescope of Nagoya University (NANTEN2) and described in \citet{Fukui2018}. The final beam size of the data cube we used in this work is 90\arcsec\ {\revbis (0.57\,pc at the 1.3\,kpc distance of the cloud)},  
with a velocity resolution of 0.08\,\kms, and a typical rms noise level of  1.1\,K\,ch$^{-1}$.

\subsection{\hi\  Data}

We also complemented the molecular line data with the atomic \hi\ cube observed as part of the Southern Galactic Plane Survey\footnote{\url{https://www.atnf.csiro.au/research/HI/sgps/fits_files.html}} and described in \citet{McClure-Griffiths2005} and \citet{Haverkorn2006}. This data cube has angular and spectral resolutions of 117\arcsec\ {\revbis (0.73\,pc at the  distance of the cloud)}  and 0.82\,\kms, respectively.

      \begin{figure*}[!h]
   \centering
     \resizebox{19.cm}{!}{
\includegraphics[angle=0]{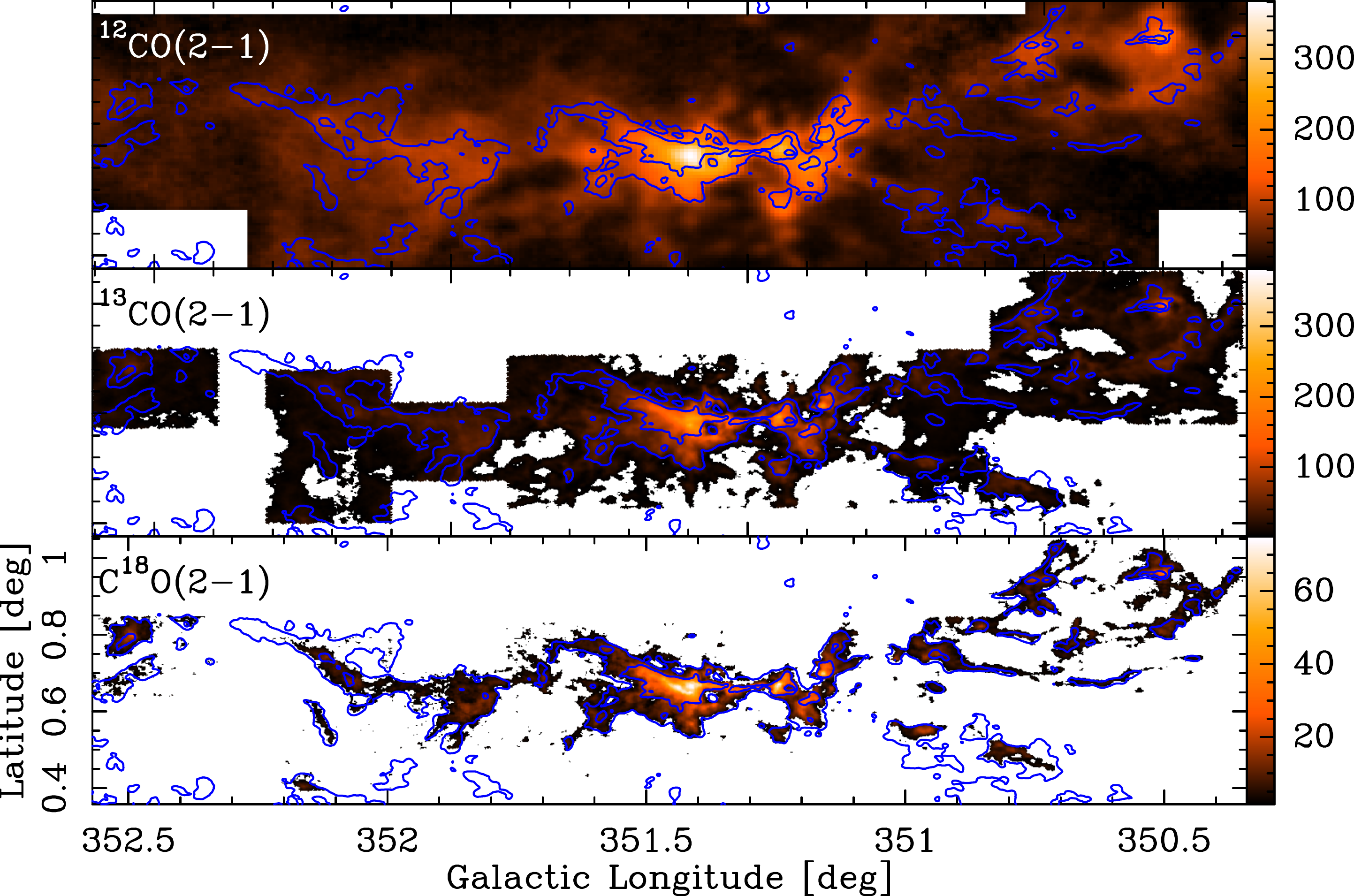}
}
\vspace{-0.3cm}
  \caption{ Integrated intensity maps, in units of K\,kms$^{-1}$, of the $^{12}$CO($2-1$), $^{13}$CO($2-1$), and C$^{18}$O($2-1$) emission, from top to bottom. The $^{13}$CO($2-1$) and C$^{18}$O($2-1$) maps have a spatial resolution of 30\parcs2 (corresponding to 0.19\,pc at the 1.3\,kpc distance of the cloud) and the emission is integrated over the LSR velocity range $-12$ to 4\,kms$^{-1}$. The $^{12}$CO($2-1$) map has a spatial resolution of 90\arcsec\ and the emission is integrated over the LSR velocity range $-25$ to 12\,kms$^{-1}$. 
  The blue contours trace column densities of 1.8 and 5.0\,$\times10^{22}\NHUNIT$ derived from \herschel\ data (at 36\parcs3) and are the same on all three panels.  
}          
  \label{mom0maps}
    \end{figure*}
  
    \subsection{$Herschel$ column density and dust temperature maps}\label{Herschel}

In the analysis presented in this paper, we also make use of the column density (\nhh) and dust temperature maps   ($T_{\rm dust}$)
derived from \herschel\  imaging data taken as part of the HOBYS  key programme \citep[][]{Motte2010}.  

The \nhh\  and $T_{\rm dust}$ maps  were produced following the same procedure as described in some of the papers analysing \herschel\ data such as \citet{Konyves2015} and \citet{Schisano2020}. 
These \nhh\ maps  were calculated  adopting a mean molecular weight per hydrogen molecule $\mu_{\rm H_{2}}=2.8$ \citep[e.g.,][]{Kauffmann2008} and  
 have an estimated accuracy of better than $\sim 50\%$ \citep[see][]{Konyves2015,Roy2013,Roy2014}.
We derived both standard \nhh\ maps at the $36\parcs3$ (half-power beam width -- HPBW) resolution of \herschel/SPIRE  500\,$\mu$m data 
and ``high-resolution''  \nhh\ maps at  the $18\parcs2$ resolution of \herschel/SPIRE  250\,$\mu$m data.
{\revbis The $18\parcs2$ resolution  corresponds to 0.11\,pc at the 1.3\,kpc distance of the cloud.}
The multi-scale decomposition method used to derive \herschel\ 
column density maps at $18\parcs2$ resolution is described in detail in Appendix~A of \citet{Palmeirim2013}. 
We here used the module \emph{hires} of the \emph{getsf} extraction code
 \citep{Men'shchikov2021} to produce these \nhh\  and $T_{\rm dust}$ maps. 
We also smoothed the $18\parcs2$ \nhh\ map to the resolution of the APEX CO maps at $30\parcs2$. 
In this paper all the \nhh\  and $T_{\rm dust}$ values correspond to those derived from \herschel\ data.
Figure\,\ref{ColdensTempMaps} shows the 
gas column density and the dust temperature maps towards the NGC~6334 complex. The footprint of the APEX maps is also shown.  

In the remaining of the paper we refer to the entire region studied in this work as  the NGC~6334 complex. We further identify three subregions as indicated in 
Fig.\,\ref{ColdensTempMaps}-top: the inter-filament system (IFS) for $352\parcd3>l>351\parcd8$, the main filament system (MFS) for $351\parcd8>l>351^\circ$, and the GM 24 filament system (GM24FS) for $351^\circ>l>350\parcd4$. 
The IFS refers to the filamentary region that connects the NGC~6334 main star-forming region to the NGC~6357 region at $l\sim353^\circ$ \citep{Russeil2010}. The GM24FS corresponds to the filamentary structures observed in the west towards the GM 24 nebula \citep{Fukui2018A}. 

\begin{figure}[!h]
   \centering
     \resizebox{9.cm}{!}{
\includegraphics[angle=0]{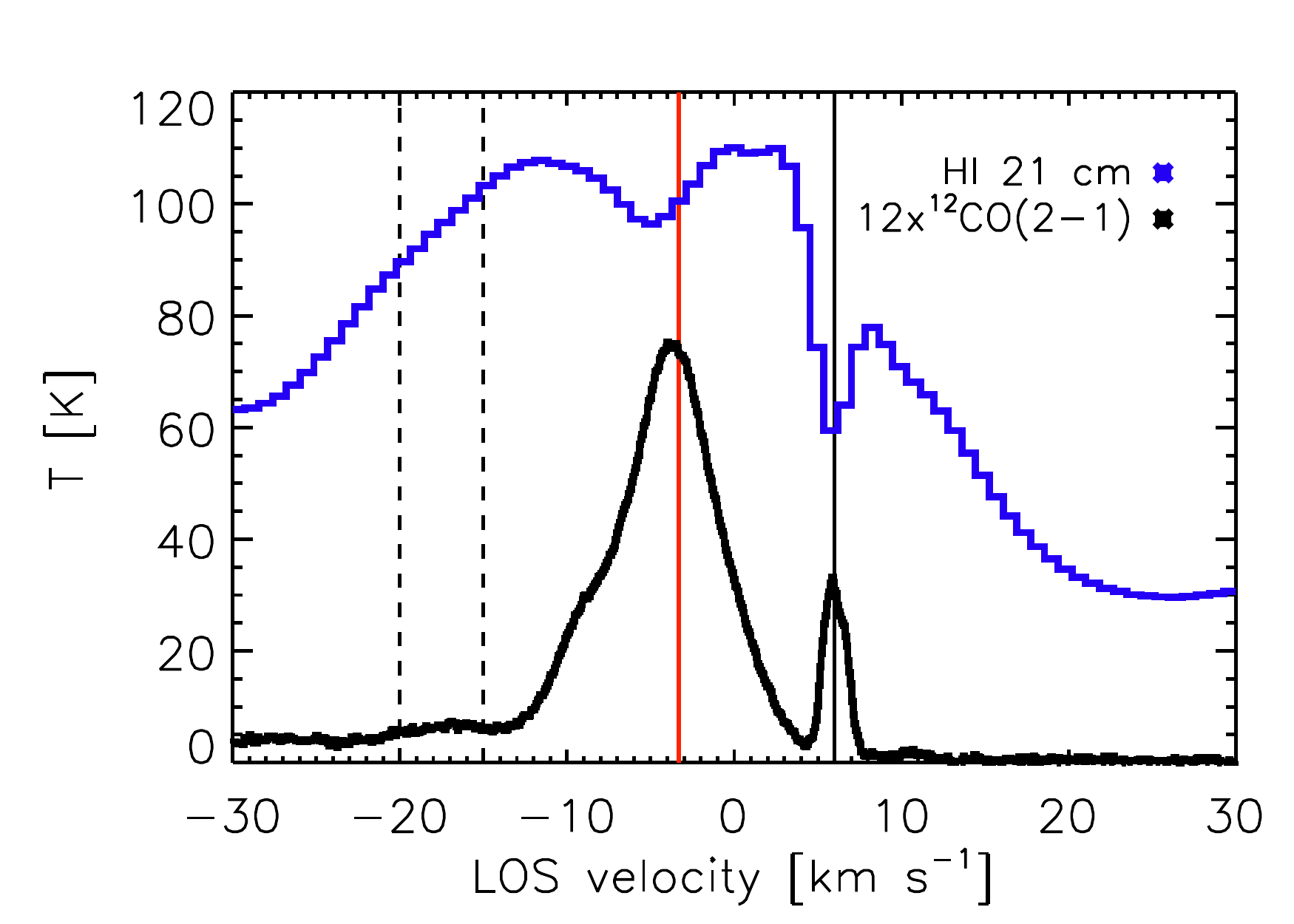}}
   \resizebox{9.cm}{!}{
\includegraphics[angle=0]{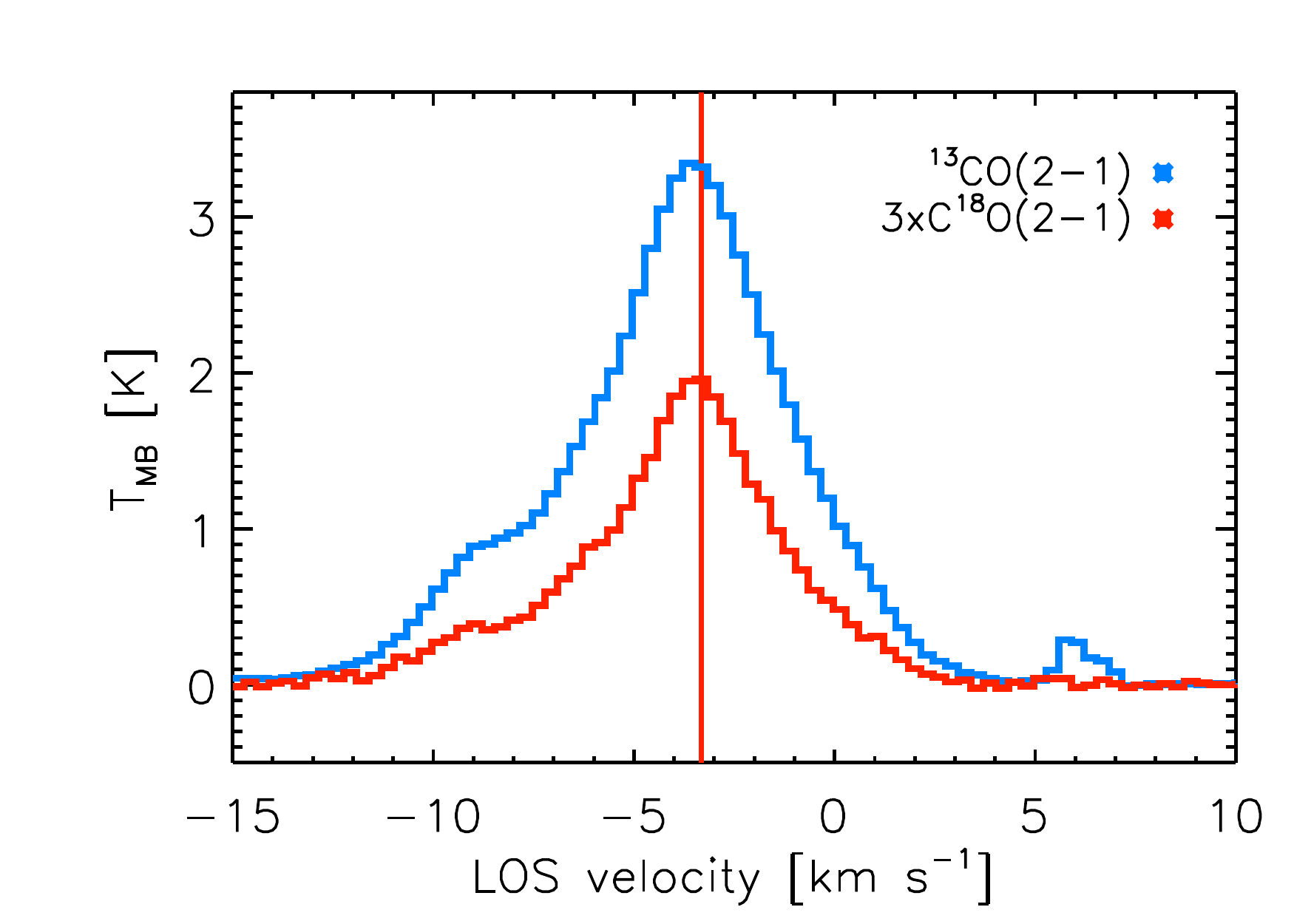}
}
\vspace{-0.3cm}
  \caption{ Positionally averaged spectra for the observed molecular and atomic emission. 
   The spectra are averaged  across the same observed region defined by the footprint of the APEX maps (cf., Fig.\,\ref{mom0maps}).  The vertical lines are as follow: The red line indicates the peak position of the  $^{13}$CO($2-1$) and C$^{18}$O($2-1$)  emission at $-3.3$\,\kms. The solid black line gives the position of the extended emission at $6$\,\kms, observed in emission and absorption, in $^{12}$CO($2-1$) and HI, respectively. The dashed black lines indicate the extent of the cloud emission identified in the $^{12}$CO($2-1$) {\bf V1} velocity range (cf., Sect.\,\ref{PV}). 
}          
  \label{IntSpecta}
    \end{figure}

 \section{Velocity structure of the 50\,pc long NGC~6334 cloud}\label{ana1}

  \subsection{Velocity integrated intensity maps}\label{mom0}    
  
  Figure\,\ref{mom0maps} presents the velocity-integrated intensity maps for the three lines observed towards the NGC~6334 molecular cloud complex. For the $^{13}$CO($2-1$) and C$^{18}$O($2-1$) lines the emission is integrated for the local standard of rest (LSR) velocity range between $-12$ and 4\,kms$^{-1}$, which encompasses the bulk of the emission associated with the complex as can be seen on the spectra averaged across the field {\rev  (Fig.\,\ref{IntSpecta}-bottom).}
  Gaussian fits to the average spectra of the $^{13}$CO($2-1$) and C$^{18}$O($2-1$) lines yield a mean LSR velocity of the cloud of about $-3.6$\,\kms\ and a  FWHM velocity width of about 5\,\kms\ (similar for the two lines). This indicates the  coherence in velocity of  the $50\,$pc long NGC~6334 complex analysed here.

  The C$^{18}$O($2-1$) integrated emission traces the elongated filamentary structures of the cloud
  with column densities $\nhh>1.8\,\times10^{22}\NHUNIT$ as  derived from \herschel\ data. The $^{13}$CO and $^{12}$CO integrated intensity maps show more extended structures towards the cloud.

We derived  the $^{13}$CO$(2-1)$/C$^{18}$O$(2-1)$ ratio maps from the  peak intensity observed over the LSR velocity range $-12$ to 4\,kms$^{-1}$,  and estimated the mean optical depth of the lines assuming a mean value of the abundance ratio  [$^{13}$CO]/[C$^{18}$O]=5.5 \citep[][]{Wilson1994} in the local ISM {\rev (see details in Appendix\,\ref{App1}).}
As shown in Fig.\,\ref{RatioOpacity}, the C$^{18}$O$(2-1)$ emission is optically thin all over the observed field. The $^{13}$CO$(2-1)$ emission is mostly optically thin but shows optical depth values up to $\sim4$ in the densest regions. 
The $^{12}$CO$(2-1)$ emission is optically thick. 

  This filamentary molecular cloud is associated with atomic medium as traced with the atomic \hi\ 21\,cm line. The mean  \hi\ spectrum (Fig.\,\ref{IntSpecta}-top) shows a dip at velocities about $-3.6$\,\kms\ associated with the molecular gas of the NGC~6334 complex.   The observed extended emission of the  \hi\ spectrum  around the velocities of the molecular cloud suggests that {\rev the} latter is embedded in and probably formed from a large scale atomic medium spanning a wider range in velocities.

 \subsection{Position velocity diagrams}\label{PV}    
 
\begin{figure*}[!h]
   \centering
     \resizebox{16.cm}{!}{
\includegraphics[angle=0]{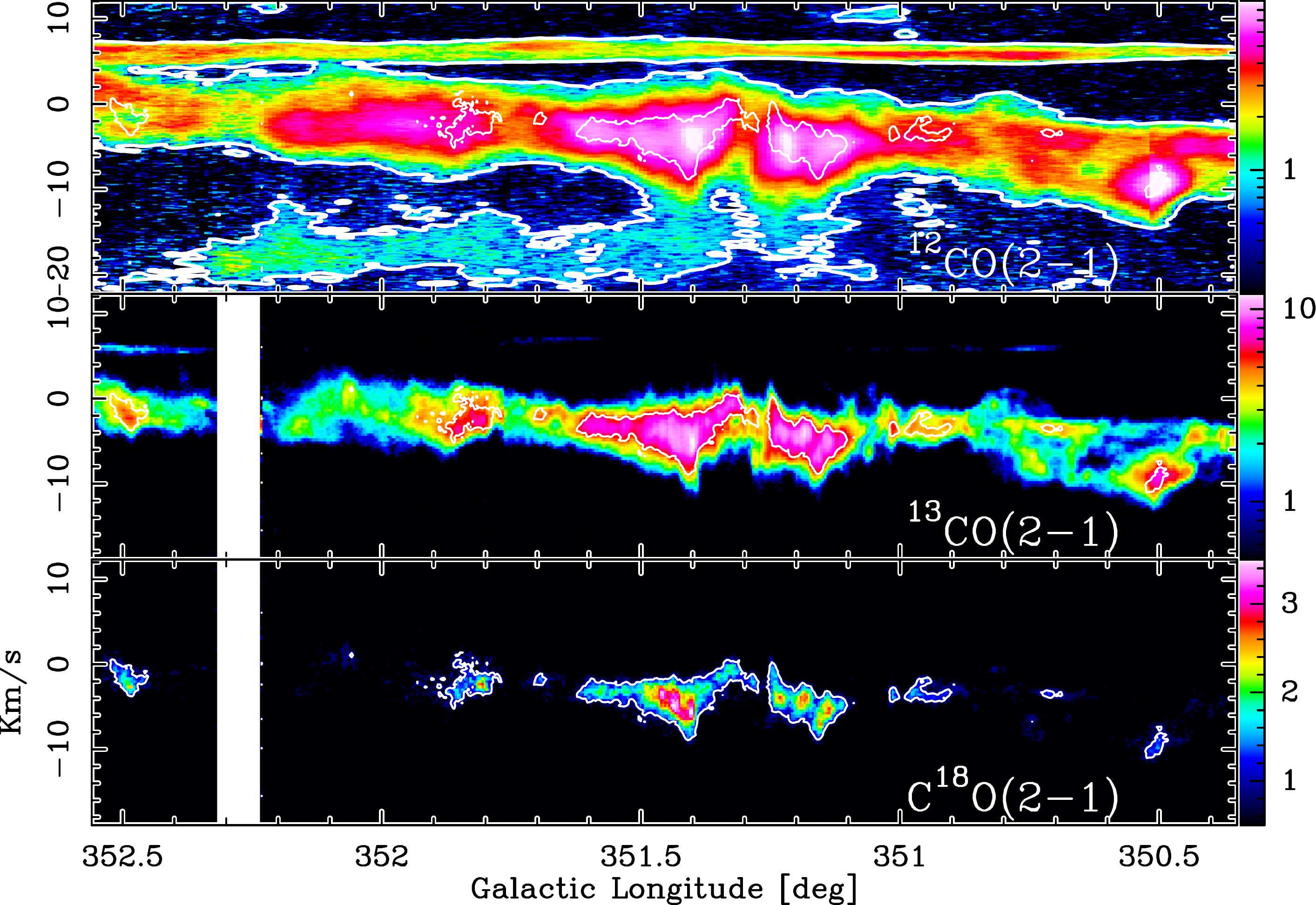}
}
  \caption{ Position$-$velocity (PV) diagrams, in units of K(T$_{\rm MB}$), for the $^{12}$CO($2-1$), $^{13}$CO($2-1$), and C$^{18}$O($2-1$) emission, from top to bottom. The $^{13}$CO($2-1$) and C$^{18}$O($2-1$) maps have a spatial resolution of 30\parcs2. The $^{12}$CO($2-1$) map has a spatial resolution of 90\arcsec.   
   These PV diagrams are averaged on latitude across the cloud for $0\parcd4\le b\le1^\circ$. 
The white contours indicate the C$^{18}$O($2-1$) intensity at 1\,K and are the same for the three panels. 
 {\rev  On the top panel, the  thick white contour corresponds to the $^{12}$CO($2-1$) intensity at 0.6\,K.}
}          
  \label{PVmaps}
    \end{figure*}

 Figure\,\ref{PVmaps} shows position-velocity (PV) diagrams along the Galactic longitude 
and averaged in latitude. 
  These PV diagrams trace a {\rev coherent velocity structure smoothly varying 
  from $\sim0$\,\kms\ to $\sim-5,-10$\,\kms\ from east to west over the $\sim50\,$pc ($\sim2\parcd2$) long filamentary cloud,} which is impressively coherent in velocity with an overall velocity range of only $\sim5\,$\kms\ (FWHM of the averaged $^{13}$CO and C$^{18}$O spectra shown in Fig.\,\ref{IntSpecta}).
 {\rev Such large scale ($\gtrsim10\,$pc) velocity coherent elongated (collections of) filamentary structures has also been identified recently in a number of  observational studies  and  referred to as giant molecular filaments \citep[e.g.,][]{Ragan2014,Kohno2021} or Galactic bones  \citep[e.g.,][]{Goodman2014,Zucker2015}.} 
  {\rev \citet{Zernickel2015} have investigated the possible link between the observed $\sim0.1\,$\kms\,pc$^{-1}$ velocity gradient along the NGC 6334 complex and the velocity structure expected from the Galactic rotation and concluded that this velocity gradient is larger than what would be expected from Galactic rotation models at the location of the NGC 6334 complex. Hence, this velocity gradient may not  result from the Galactic rotation and may have a different origin.}

\begin{figure*}[!h]
   \centering
     \resizebox{19.5cm}{!}{
\hspace{-.2cm}
\includegraphics[angle=0]{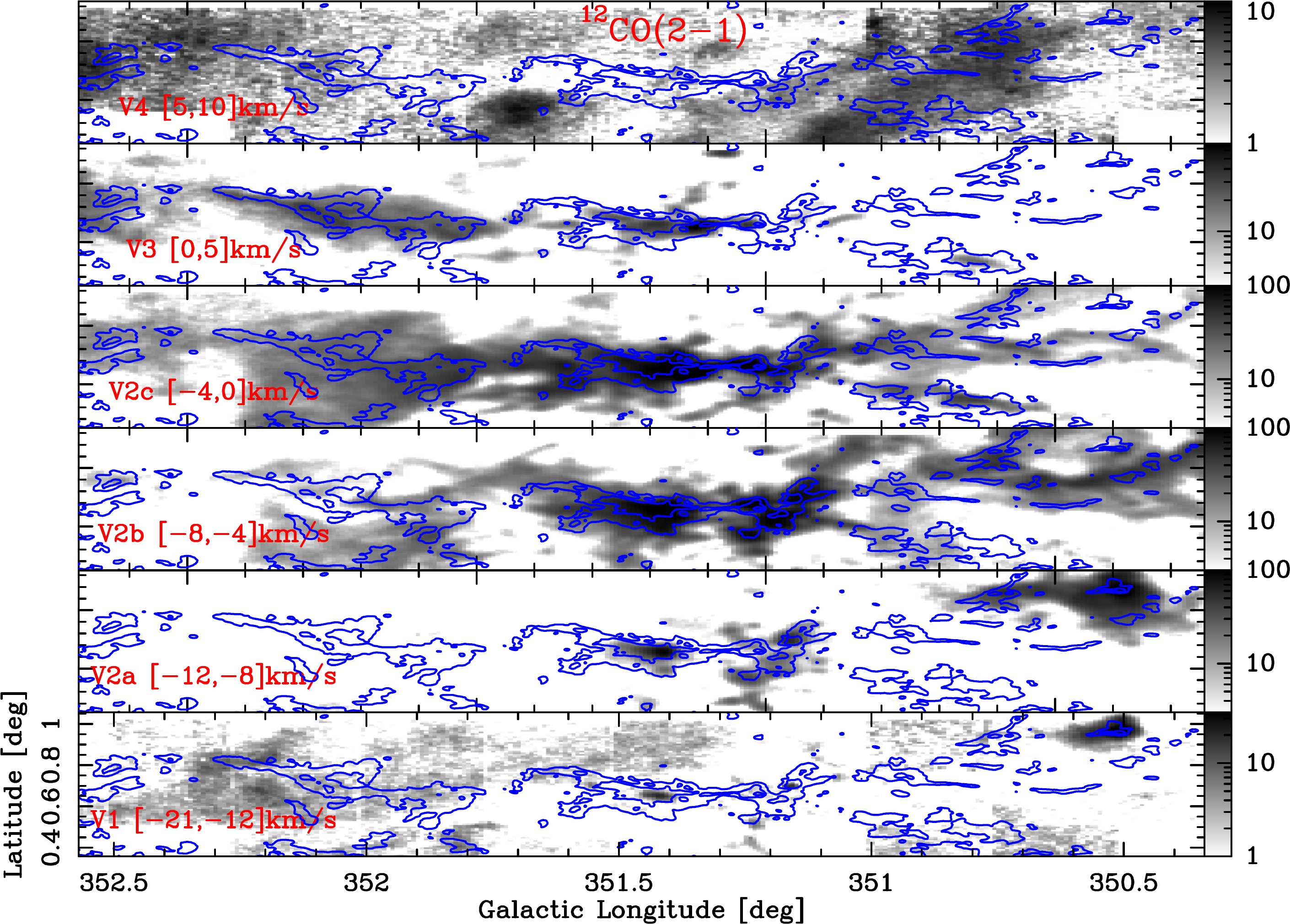}
\hspace{.15cm}
\includegraphics[angle=0]{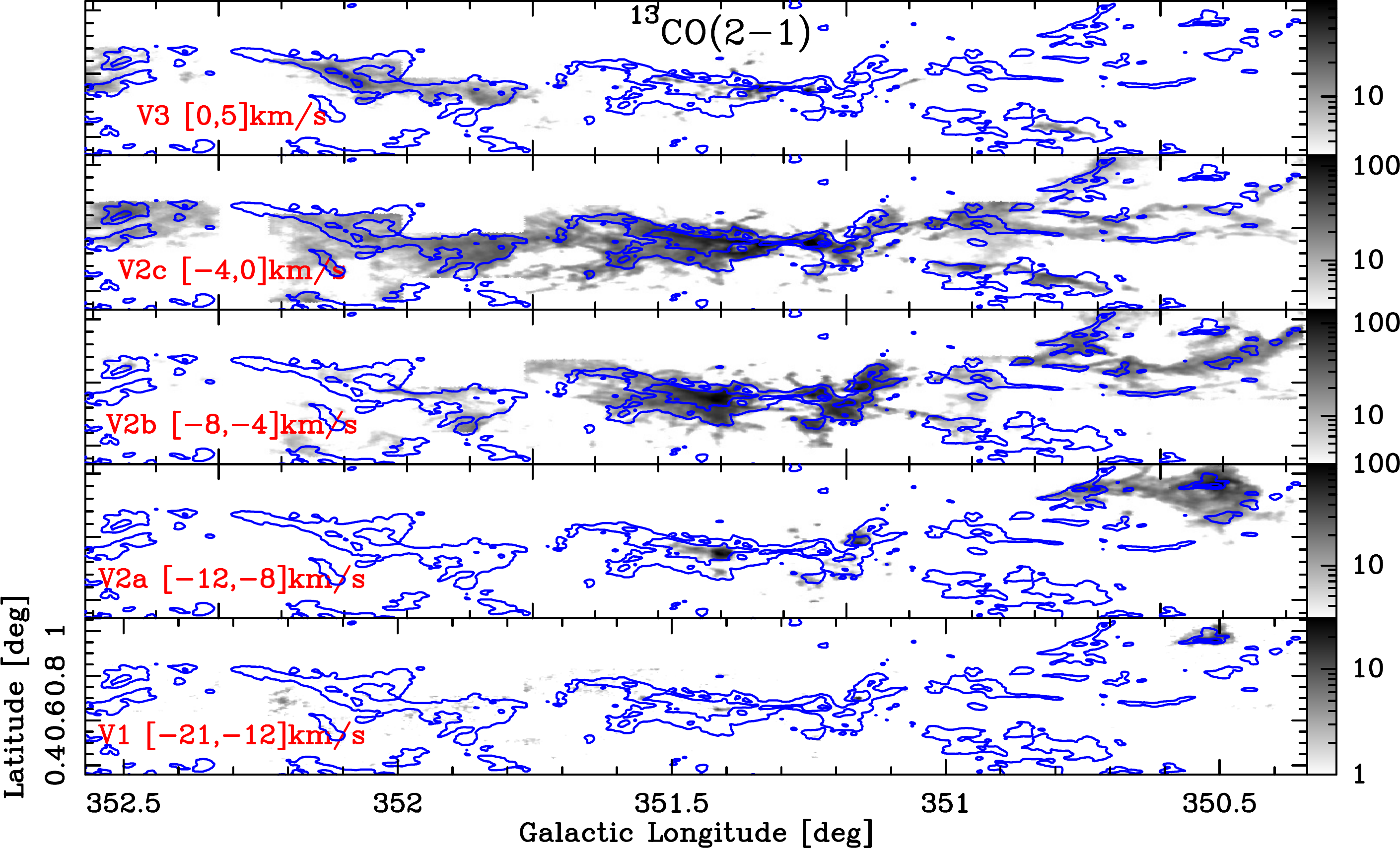}}
\vspace{-.5cm}
  \caption{ Velocity channel maps in units of K\,kms$^{-1}$ for the $^{12}$CO($2-1$) and $^{13}$CO($2-1$) emission on the left and right, respectively. 
 The velocity ranges of the channel maps are indicated on the bottom left of each panel. 
Note that the range of the colour scale (indicated by the bar for each panel) is different for each plot. This scale has been chosen to represent the dynamical range of the emission for each velocity channel map. 
}          
  \label{Channelmaps}
    \end{figure*}

Averaging the emission in latitude increases the signal to noise ratio  ($S/N$), which makes it possible to detect the faint $^{12}$CO$(2-1)$ emission at LRS velocities between $-20$ and $-12$\,\kms. 
{\revbis This extended component} spanning velocities from about  $-20$ to $-12$\,\kms\ where it connects to the velocity of the denser gas of the NGC~6334 cloud has been suggested by \citet{Fukui2018} to trace an interaction event (started a few Myr ago) between the cloud seen now at $\sim-20$\,\kms\  and the NGC~6334 complex at the mean velocity of  $\sim-3$\,\kms\ {\rev (velocity channel maps are shown in Fig.\,\ref{Channelmaps} and described in Sect.\,\ref{VchanMaps}). }
{\rev Bridge features connecting two clouds in velocity space have been used in observations to trace physical interactions  between clouds  with distinct bulk velocities \citep[e.g.,][]{Torii2011,Fukui2018,Enokiya2021} and have been reproduced in numerical simulations of cloud-cloud collision   \citep[][]{Takahira2014,Haworth2015a,Haworth2015b,Priestley2021}. }
{\rev We note, however, that such velocity bridges are not unambiguous signatures of cloud-cloud collision. Additional evidences of physical connection, e.g., similar distance estimate and complementarity in spatial distribution of the velocity components of the two clouds, are also needed to further confirm the interaction \citep[cf., e.g.,][]{Fukui2018CCC}. 
}

{\rev The $^{12}$CO$(2-1)$ PV map also shows a narrow component  at 6\,\kms, which is also detected in $^{13}$CO$(2-1)$ at Galactic Longitudes around $l\sim 350\parcd7, 351\parcd7$, and $352\parcd5$.} 
{\rev This  narrow component in velocity space has been reported by \citet{Radhakrishnan1972} and \citet{Russeil2016}, from the analysis of  \hi\ spectra, to 
be associated with a foreground extended cloud  layer close to the Sun and not associated with the NGC~6334 complex. }
 
 \subsection{Velocity channel maps}\label{VchanMaps}

Figure\,\ref{Channelmaps} presents velocity channel maps derived from the $^{12}$CO($2-1$)  and $^{13}$CO($2-1$) cubes. We do not show here the channel maps derived from the C$^{18}$O($2-1$) cubes that are very similar to the $^{13}$CO channel maps. 

The {\bf V1} channel map, is derived from the integration of the intensity over the velocity range $[-21,-12]\,$\kms\  {\rev tracing the cloud with the blueshifted velocity suggested to be interacting with the NGC~6334 complex} \citep[see Sect.\,\ref{PV}, and][]{Fukui2018}. The {\bf V2} channel maps (V2a, V2b, and V2c) trace the main emission of NGC~6334 in the  velocity range $[-12,0]\,$\kms, where the 50\,pc  filamentary structures are detected.  
The {\bf V3} channel map traces the main extension of the cloud  towards the east and the bulk emission of the inter-region filaments connecting the NGC~6334 and NGC 6357 clouds \citep{Russeil2010}. We here call this extension IFS (see Fig.\,\ref{ColdensTempMaps}-top). The NGC~6357 complex is located on the east around the $353^\circ$ Galactic longitude and is observed to be connected in velocity to the NGC~6334 complex \citep{Fukui2018}.  The {\bf V4} channel map derived from the integration of the emission in the  velocity range $[5,10]\,$\kms, traces extended structures at a mean velocity of  $\sim6$\,\kms\ (see Fig.\,\ref{Channelmaps}-top), {\rev suggested to be a foreground gas structure close to the Sun \citep{Radhakrishnan1972,Russeil2016}.}

\section{Identification of velocity-coherent-filaments (VCFs)}\label{ana2}

In the following, 
we  describe {\rev the multi-velocity-component Gaussian fitting of the observed lines. We then}  present the
method used to trace elongated crests in three-dimension (3D) position-position-velocity (PPV) space and identify velocity-coherent-filaments (VCF). 
 {\rev For this analysis, we use the  C$^{18}$O($2-1$) cube at the spatial resolution of 30\parcs2 (projected onto a grid with 14\parcs3 pixel size) and 0.3\,\kms\ velocity resolution. We choose to analyse the C$^{18}$O($2-1$) emission, which is optically thin over the cloud  (cf., Appendix\,\ref{App1}).}

    \begin{figure*}[!h]
   \centering
     \resizebox{19.cm}{!}{
     \includegraphics[angle=0]{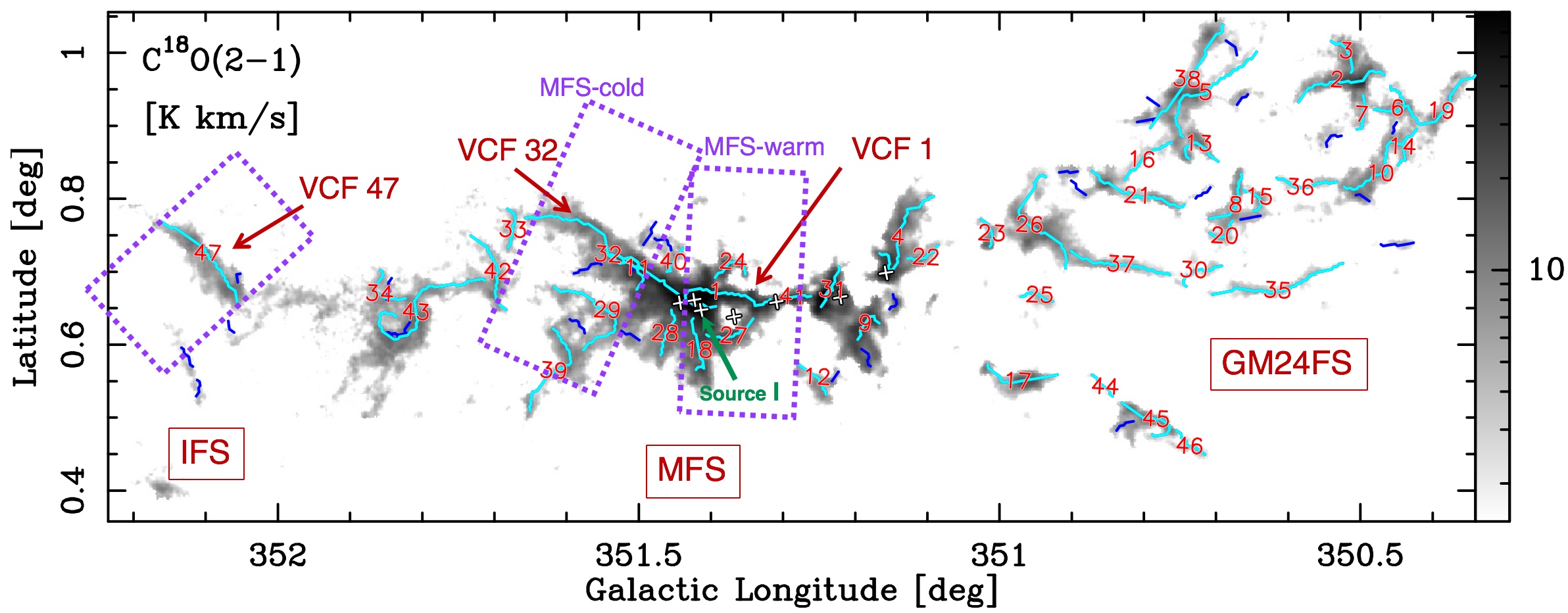}}
\vspace{-.2cm}
  \caption{
Integrated intensity map, in units of K\,kms$^{-1}$, of the C$^{18}$O($2-1$) emission (Same as in Fig.\,\ref{mom0maps}). The coloured curves show the crests of the velocity coherent filaments (VCFs) identified using CRISPy. The crests with total length ($L$) larger than 10 pixels are traced in cyan and numbered. 
{\rev The VCF~1, 32, and 47} are identified with red arrows. The dotted purple boxes indicate the regions from which the averaged radial profiles and the PV maps are derived (see Sects.\,\ref{ana4a} and \ref{ana4b}). 
{\rev 
The white crosses indicate the seven infrared high-mass star-forming regions, which are usually referred to as sources  I(N), E, I, II, III, IV, and V, from east to west \citep{Persi2008,Arzoumanian2021}. {\revbis The source I is indicated with the green arrow.}}
}          
  \label{MapSkel}
    \end{figure*}

 \begin{figure*}[!h]
   \centering
   \resizebox{19.cm}{!}{  \hspace{-.8cm}
     \includegraphics[angle=0]{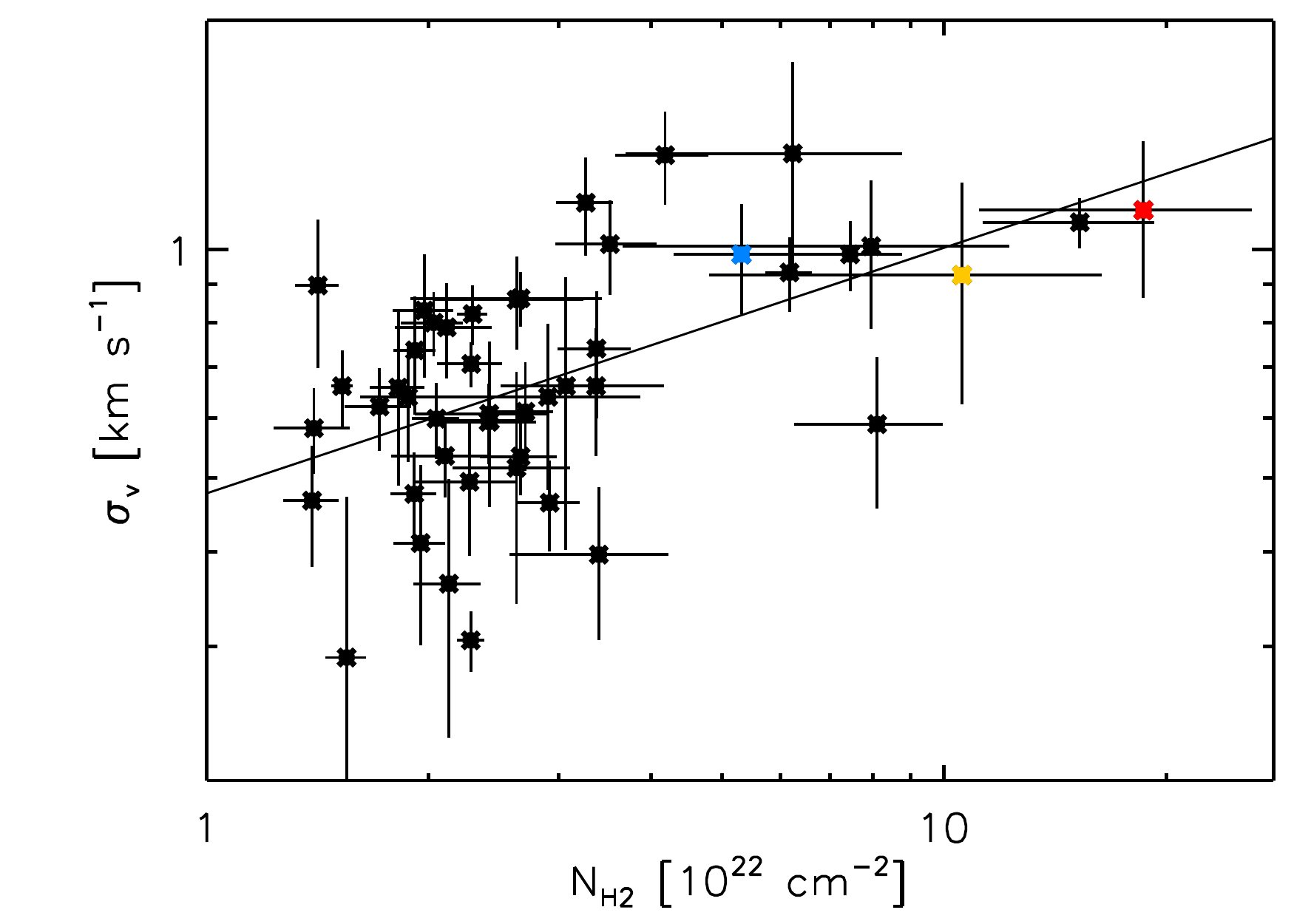}
     \hspace{-.5cm}
     \includegraphics[angle=0]{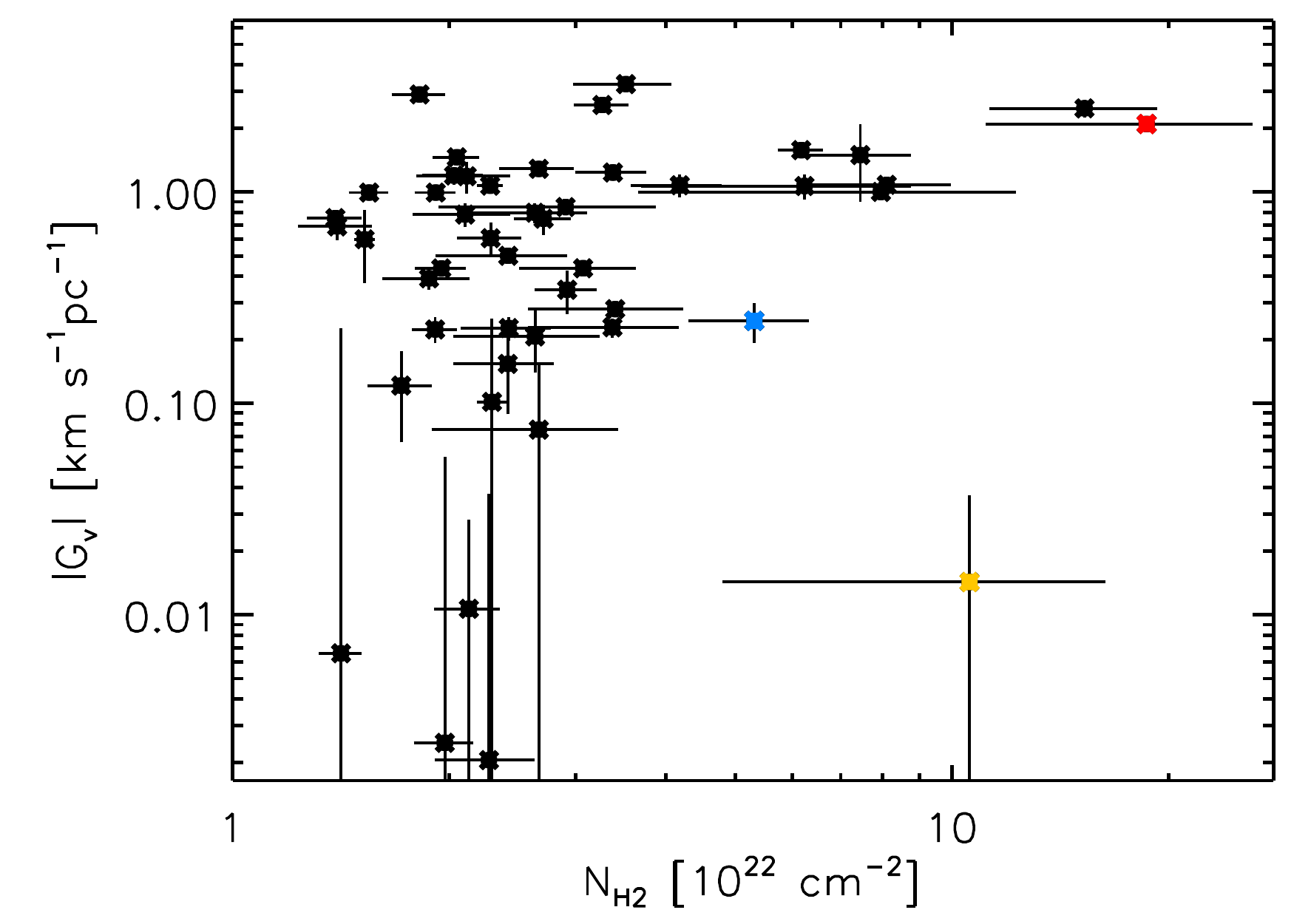}
               \hspace{-.8cm}
     \includegraphics[angle=0]{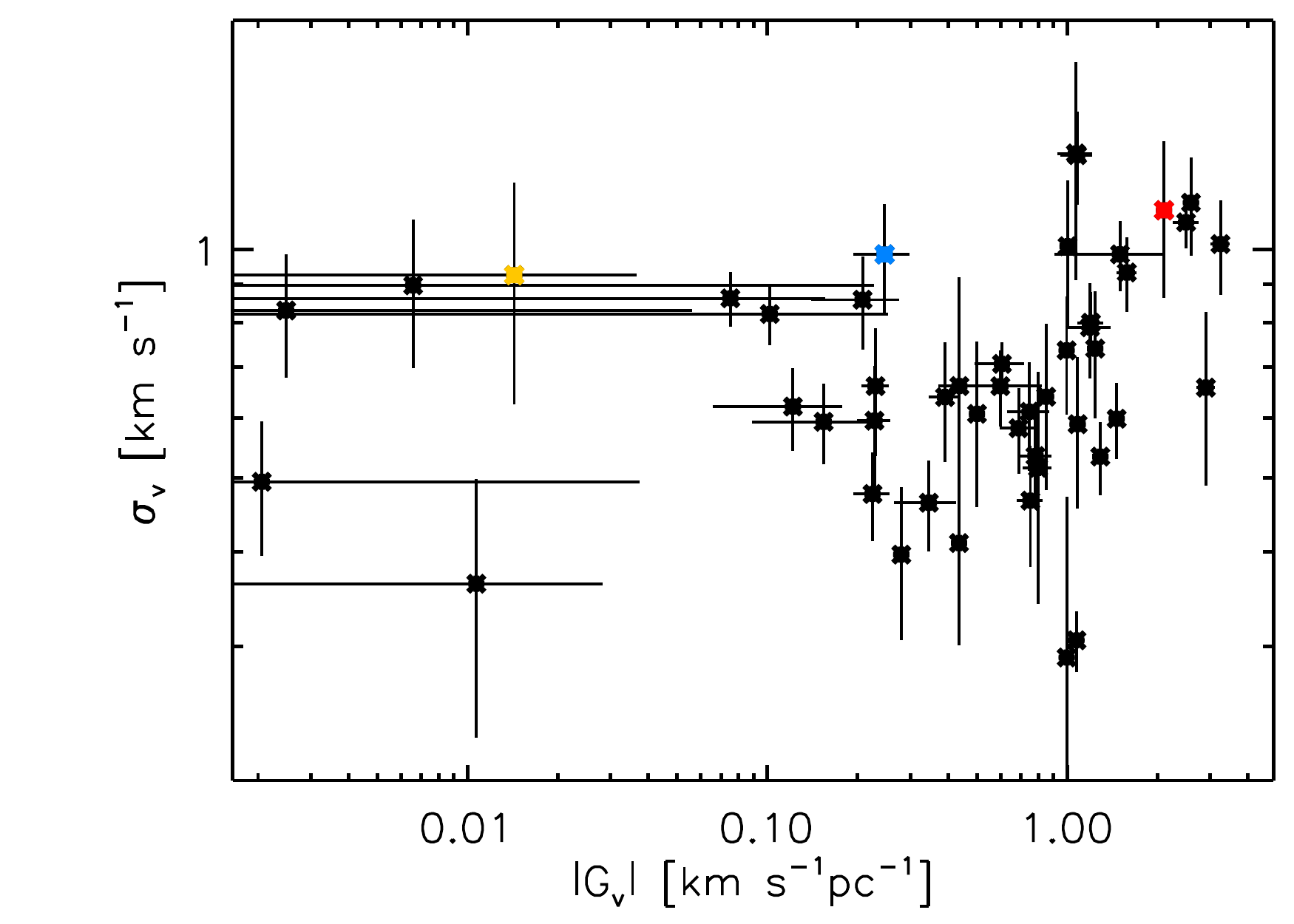}
}
\vspace{-.4cm}
  \caption{
{\rev  Scatter plots of measured parameters  for the 47 VCFs. The values are averaged along the crests of the VCFs.  
The data points in red,  yellow, and blue indicate the  VCFs 1, 32, and 47, respectively.}
  {\it Left:}   Velocity dispersion versus column density. 
   The horizontal and vertical bars indicate the $1\sigma$ dispersions of $N_{\rm H_2}$ and  $\sigma_{v}$ about their average value along the crests.   
  The diagonal line shows the best power-law fit $\sigma_{v}\propto\,N_{\rm H_2}^{\,0.32}$.  
   {\it Middle:} Velocity gradient {\revbis G$_{v}$} (given in absolute value) versus column density. The horizontal  bars indicate the $1\sigma$ dispersion of $N_{\rm H_2}$ (same as the left panel). The vertical  bars give the $1\sigma$ uncertainty of the linear fits to the velocity fluctuations along the crests.
   {\rev {\it Right:}  Velocity dispersion versus {\revbis G$_{v}$ (given in absolute value).}
   } 
}          
  \label{sigmaV_NH2}
    \end{figure*}

\subsection{Multi-component spectral fitting}\label{ana2a}

We modelled the observed cube by least-square fitting of the C$^{18}$O($2-1$) spectrum at each pixel position of the 2D position-position (PP) space. We assume that the observed spectrum is the sum of several velocity components with line shapes close to  Gaussian functions. We thus use a  multi-Gaussian function with $n$ velocity components. The model spectra, at each pixel of the ($x,y$) PP space, is thus the sum of $n$ Gaussians
\begin{equation}\label{Eq:EqGauss}
T_{\rm MB}^{\rm model}(v) = \sum_{i}^{n} T^{\rm peak}_i\exp\left[-\frac{\,(v-v_i)^2}{2\,\sigma_{v_i}^2}\right]\,,
\end{equation}
where  $T^{\rm peak}_i$,  $v_i$, and $\sigma_{v_i}$ are the line peak temperature [K], the centroid velocity [\kms], and the velocity dispersion [\kms] of the $i^{\rm th}$ velocity component. 
{\rev The first step is to identify the  total number of velocity components, $n$, of each spectrum (see also Appendix\,\ref{App2}). This is done  using the first and second derivatives of the spectrum with emission above a signal-to-noise ratio $S/N=T_{\rm MB}/\sigma_{T_{\rm MB}}>4$, where $\sigma_{T_{\rm MB}}$ is the rms of the spectrum calculated from emission free velocities. The identified $n$-components are fitted using Eq.\,\ref{Eq:EqGauss}.  
Then, 1) the fitted velocity components with peak temperature  $T^{\rm peak}_i<4\sigma_{T_{\rm MB}}$ and 2) the components $v_i$ with a minimal separation with its neighbouring components $v_{i+1}$ smaller than 5 times the velocity resolution of 0.3\,\kms, i.e., $|v_{i}-v_{i+1}|<1.5$\,\kms, are discarded. 
The algorithm continues iterating until the remaining fitted components satisfy the $S/N$ and the velocity separation conditions (see Fig.\,\ref{Spectra_Gfit} for an example).
}
{\rev We performed several tests comparing the observed and modeled spectra to find the  parameters, which reproduce the best the observations, and set a $S/N=4$ for the peak intensity and a minimum velocity separation of 5 times the velocity resolution of 0.3\,\kms. }
In Appendix\,\ref{App2}, we show the velocity integrated intensity maps of the observed and modelled cubes, as well as the ratio map between the observed and model cubes $R(x,y,v)=T_{\rm MB}(x,y,v)/T_{\rm MB}^{\rm model}(x,y,v)$ . The ratio  map has a median, mean, and standard deviation of 1.02, 1.10, and 0.46, respectively, suggesting that the modelled  cube represents well the observations. A total of 13283 spectra were fitted. As can be seen on Fig.\,\ref{MapVelComp}, more than $88\%$ of the spectra are fitted with a single velocity component. Only $\sim10\%$ and $\sim2\%$ of the spectra required two or three velocity components to describe the observations, respectively. 

\subsection{VCF identification}\label{ana2b}

We traced the crests of elongated filamentary structures running CRISPy  on the modelled PPV cube. {\rev CRISPy \citep{ChenMike2020} is based on the Subspace Constrained Mean Shift Algorithm   \citep[developed by][]{Chen2014Generalized,Chen2015}.} 

We run CRISPy setting a minimum intensity detection threshold of 2\,K ($S/N\sim4$) and a smoothing length of 2 pixels ($\sim$ beam). 
We subsequently gridded the raw CRISPy result onto our modelled cube and prune away extra branches in each ridge from its longest path using {\rev the algorithm and} the methods by \citet{ChenMike2020}.
{\revbis  This 3D graph-theory-based pruning method is based on the 2D software developed by \citet{Koch2015} and the current implementation of this 3D version simply prunes all structures that are not part of the longest path without the need for user inputs.}
{\rev Further details on the pruning methods are detailed in those papers.} 

{\rev Then, the pruned 3D skeleton is referenced to sort the modelled velocity components
into velocity-coherent-filaments (VCF) using the method  described in \citet{ChenMike2020},} 
with a maximum velocity separation 
of 0.9\,\kms\ ($\sim$3 times the velocity resolution). This analysis yields a total of 75 VCFs  identified towards the whole studied region at the angular  and spectral resolutions of 30\parcs2  and 0.3\,\kms, respectively. 
{\rev 
In the following, we present the analysis of  47 VCFs out of the 75 VCFs, that have a length larger than $\sim5\times$\,beam  (0.9\,pc and 10 pixels). We selected this length to ensure the sampled VCFs have a minimum aspect  ratio of 1:5 at our angular resolution thus distinctively longer than what is considered a core. 
This selection also provides a minimum 
number of 10 pixels to measure statistical properties (e.g., mean, dispersion) along the crests of the VCFs.}

\begin{figure*}[!h]
   \centering
     \resizebox{19cm}{!}{  \hspace{-.8cm}
 \includegraphics[angle=0]{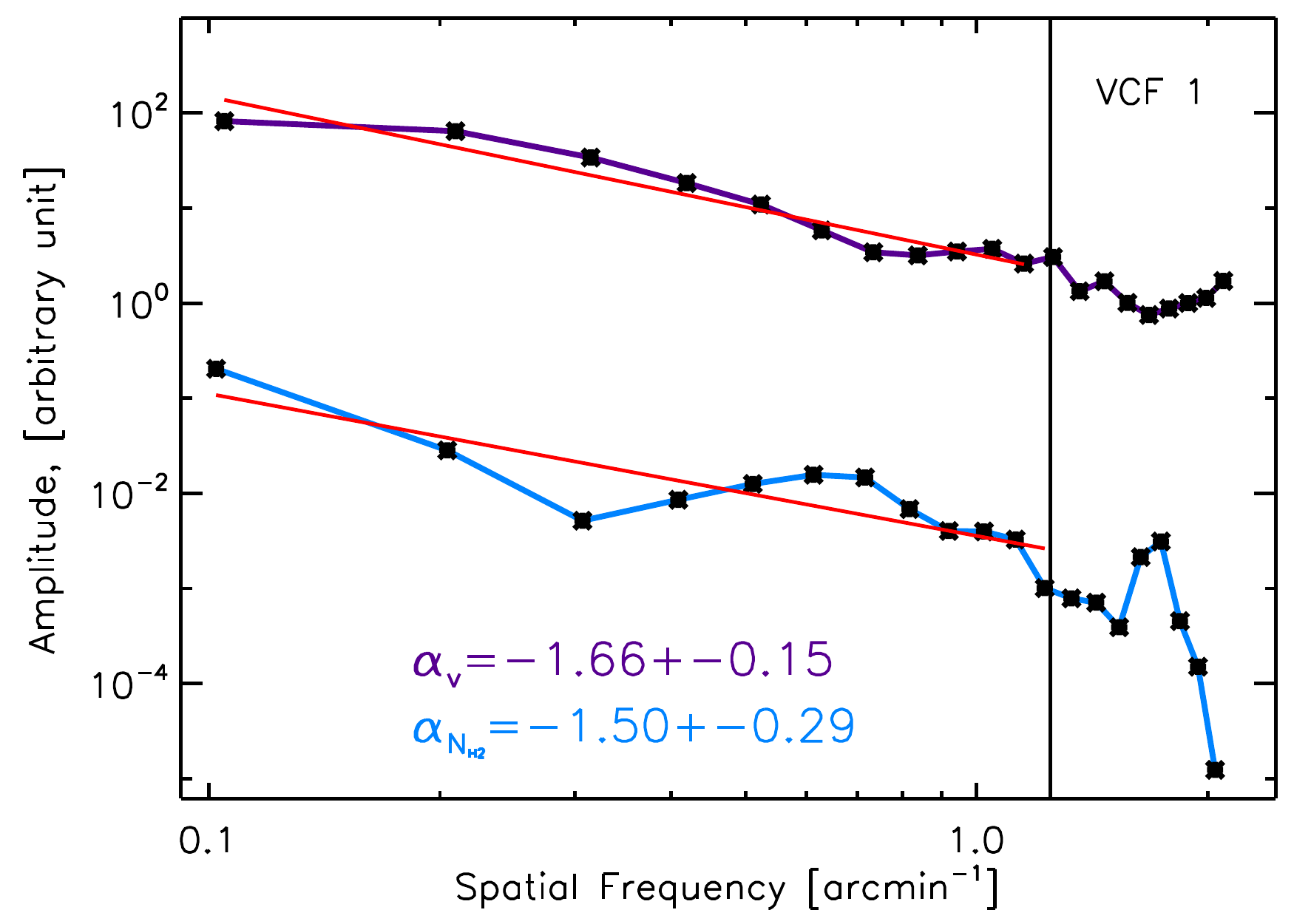}
 \includegraphics[angle=0]{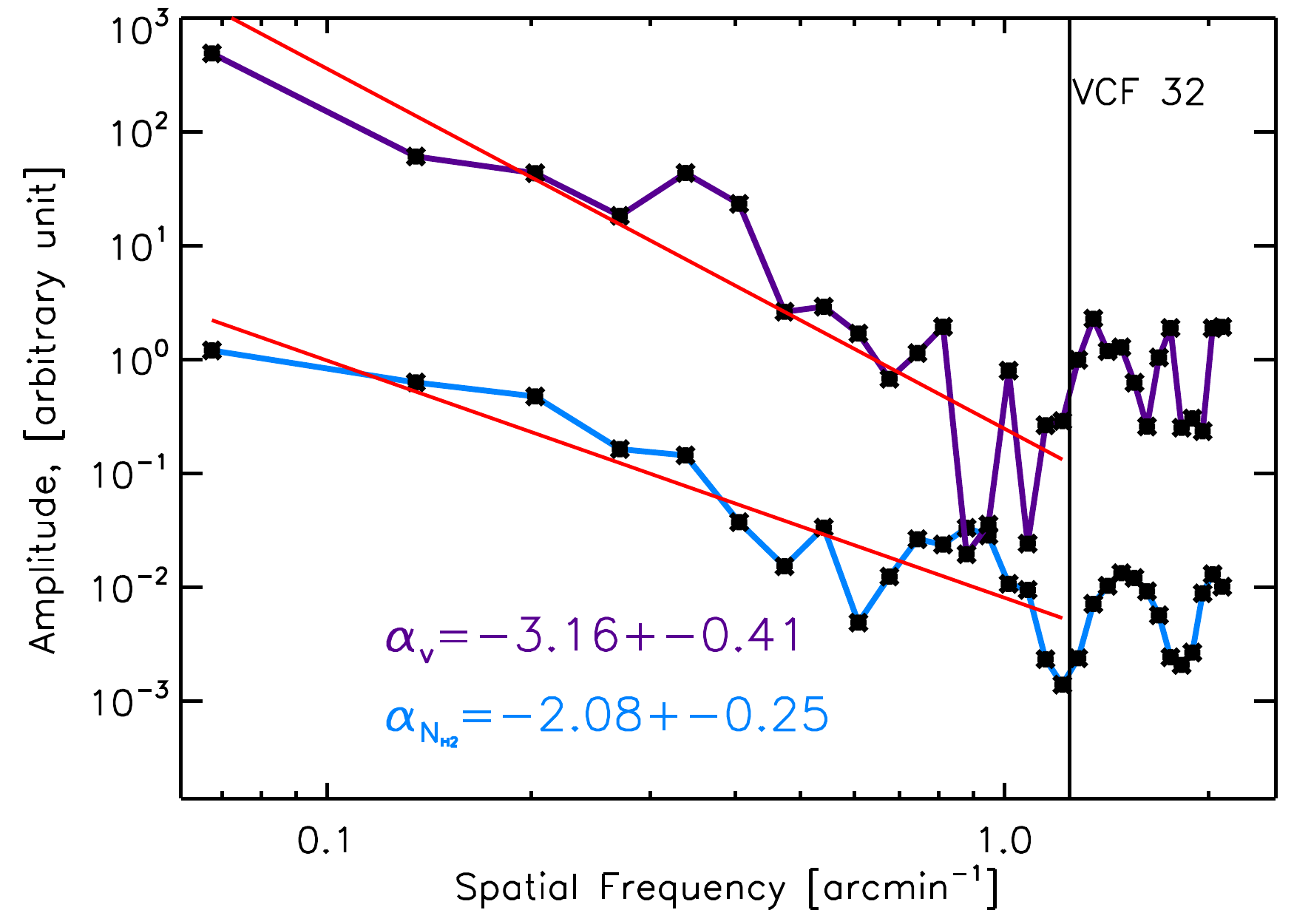}   
         \includegraphics[angle=0]{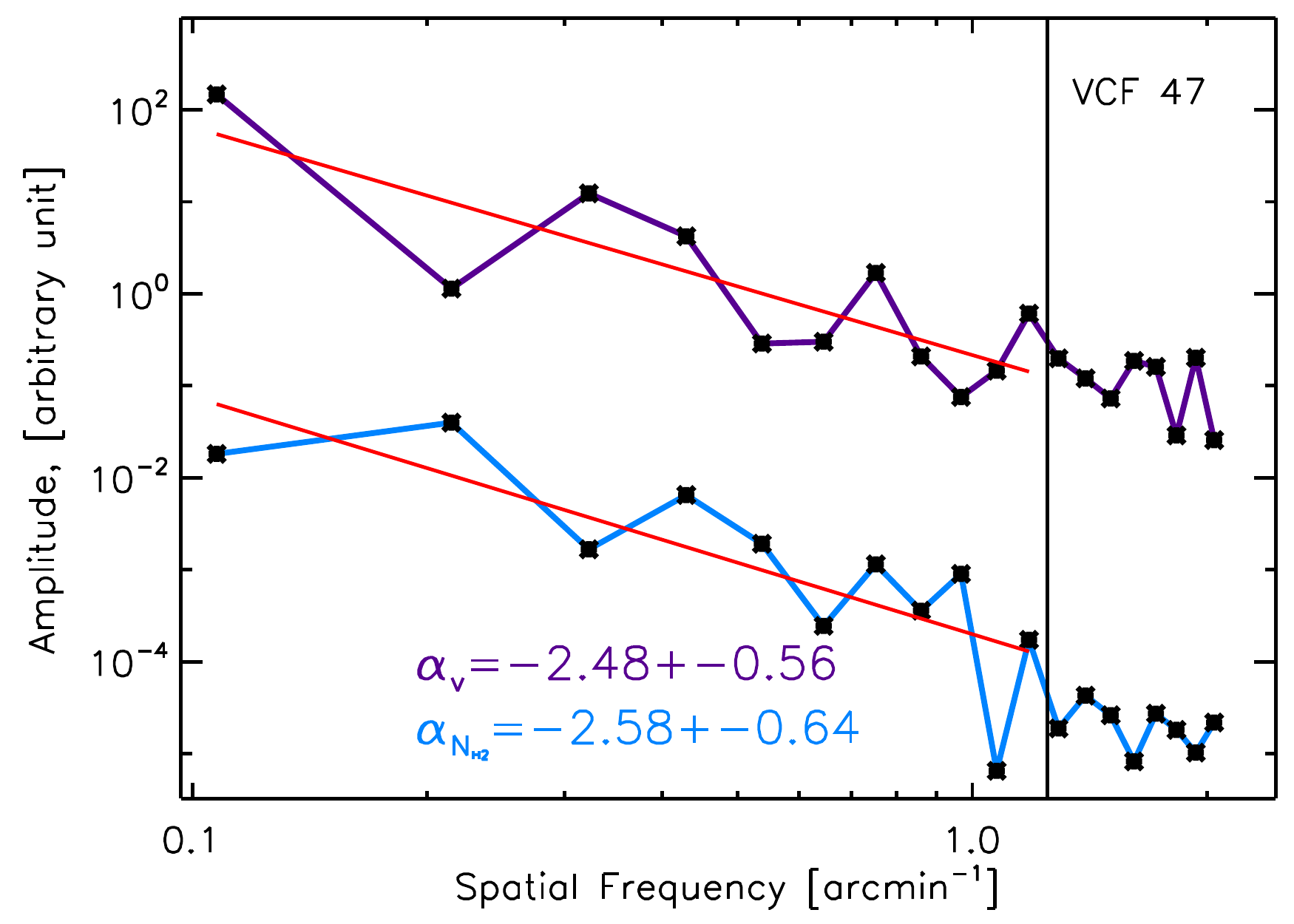}}
\vspace{-.5cm}
  \caption{
  Power spectra of the column density (blue) and the velocity (purple) for the {\rev VCF crests 1 (left), 32 (middle),  and 47 (right).} 
  The amplitudes of the power spectra are scaled so they can be plotted on the same panel.
  The red lines  show the best linear fit to the power spectra and the values of the slopes are indicated on the corresponding panels. 
  The vertical black solid lines indicate the $30\parcs$2  angular resolution of the data equivalent to $\sim1.2$\,arcmin$^{-1}$, which is the highest frequency data point used to fit the power spectra.
}          
  \label{PS1-47}
    \end{figure*}
    
\section{Statistical properties of the identified 47 VCFs}\label{ana3}

Figure\,\ref{MapSkel}  shows the derived  crests of the 75 VCFs over-plotted on the C$^{18}$O($2-1$) velocity integrated intensity map.  
Table\,\ref{tab:Tparam}  summarizes the main properties of the 47 VCFs. 

\subsection{Mean properties along the VCF crests}\label{ana3a}

The mean length of the VCFs is $\sim2$\,pc with some of them as long as $\sim5$\,pc. 
They span a column density range of an order of magnitude about a mean value of $\sim4\times10^{22}\,\NHUNIT$ (as derived from \herschel\ data). 
The {\rev LOS averaged} dust temperature map as derived from \herschel\ data indicates a mean value of $\sim20$\,K, which is on average larger than the dust temperature observed towards low-mass star forming regions of the Gould Belt \citep[e.g.,][]{Arzoumanian2019}, indicating the possible influence of  young (massive) stars in heating the dense gas (at least the external layers of the filaments). This heating may be originating from the surroundings  of (e.g., neighbouring  \hii\ regions) or from inside  (young massive-star or star cluster forming along)  the filaments {\rev (see discussion in Sect.\,\ref{ana4a} below). }
{\rev These filaments are mostly  thermally supercritical with line masses ($M_{\rm line}$) larger than the critical equilibrium value for isothermal cylinders   $M_{\rm line,crit}= 2\,c_{\rm s}^2/{\rm G}\sim28\,\sunpc$  for a gas temperature of $20\,$K \citep[e.g.][]{Ostriker1964,Inutsuka1997}.}
{\rev The  mass per unit length of the VCFs is calculated as $M_{\rm line}=\mu_{\rm H_2}m_{\rm H}  \nhh^{\rm bs} \times W_{\rm VCF}$, where  $W_{\rm VCF}=0.13$\,pc is the VCF width\footnote{{\rev This $0.13$\,pc width value corresponds to an effective filament width of $1.3\times0.1$\,pc measured by \citet{Arzoumanian2019} to enclose the mass of a $0.1$\,pc wide filament with non-Gaussian wings beyond the $0.1$\,pc inner width.}}, 
$\mu_{\rm H_2}=2.8$  is the mean molecular weight per hydrogen molecule,  $m_{\rm H}$ is the mass of a hydrogen atom, and  $\nhh^{\rm bs}=\nhh-1\times10^{22}\,\NHUNIT$ is the background subtracted  column density (as derived from \herschel\ data) averaged along the VCF. The background value $\sim1\times10^{22}\,\NHUNIT$ is  the mean \nhh\ towards the regions of the molecular cloud  not traced by 
 C$^{18}$O \citep[see also the discussion about the similar background value estimated in this region by][]{Arzoumanian2021}. }

{\rev Figure\,\ref{sigmaV_NH2}-left shows the mean  velocity dispersions of the VCFs as a function of their mean column density. The mean velocity dispersion corresponds to the mean value of the fitted velocity dispersions (i.e., line width) of the corresponding velocity component at each position along the VCF.}
The mean velocity dispersions of the VCFs show an increasing trend as a function of their mean column density with a best fit linear relation of $\sigma_{v}\propto\,N_{\rm H_2}^{\,0.32}$ (Fig.\,\ref{sigmaV_NH2}-left).  This relation is  similar to what was derived for the sample of thermally supercritical filaments presented in \citet[][]{Arzoumanian2013} and was interpreted as resulting from {\rev the evolution of the filaments by accretion of surrounding matter.}  
We  {\rev measure} a wealth of velocity gradients along the VCF crests. We quantify the velocity gradients with linear 1D fits of the observed velocity along the crests (see figure\,\ref{plotsMeanParam}-middle). 
{\rev We find a large scatter in the measured velocity gradients for filaments with $N_{\rm H_2}\lesssim5\times10^{22}\,\NHUNIT$  and a tentative increasing trend for filaments with $N_{\rm H_2}\gtrsim5\times10^{22}\,\NHUNIT$ (Fig.\,\ref{sigmaV_NH2}-middle). }
{\rev The filaments with larger velocity gradients tend to also show larger velocity dispersions. We note, however, that the observed velocities are line-of-sight projected velocities and  small velocity gradients may be due to 1) the orientation of the filament close  to the plane-of-the-sky or 2) the variation of the sign of the gradient along the crest as it is the case for the VCF 32 
(see Fig.\,\ref{sigmaV_NH2}-middle and Sect.\,\ref{ana_selected}). }

\subsection{Column density and velocity power spectra of the VCFs}\label{ana3b}
  
\begin{figure}[!h]
   \resizebox{8cm}{!}{
     \includegraphics[angle=0]{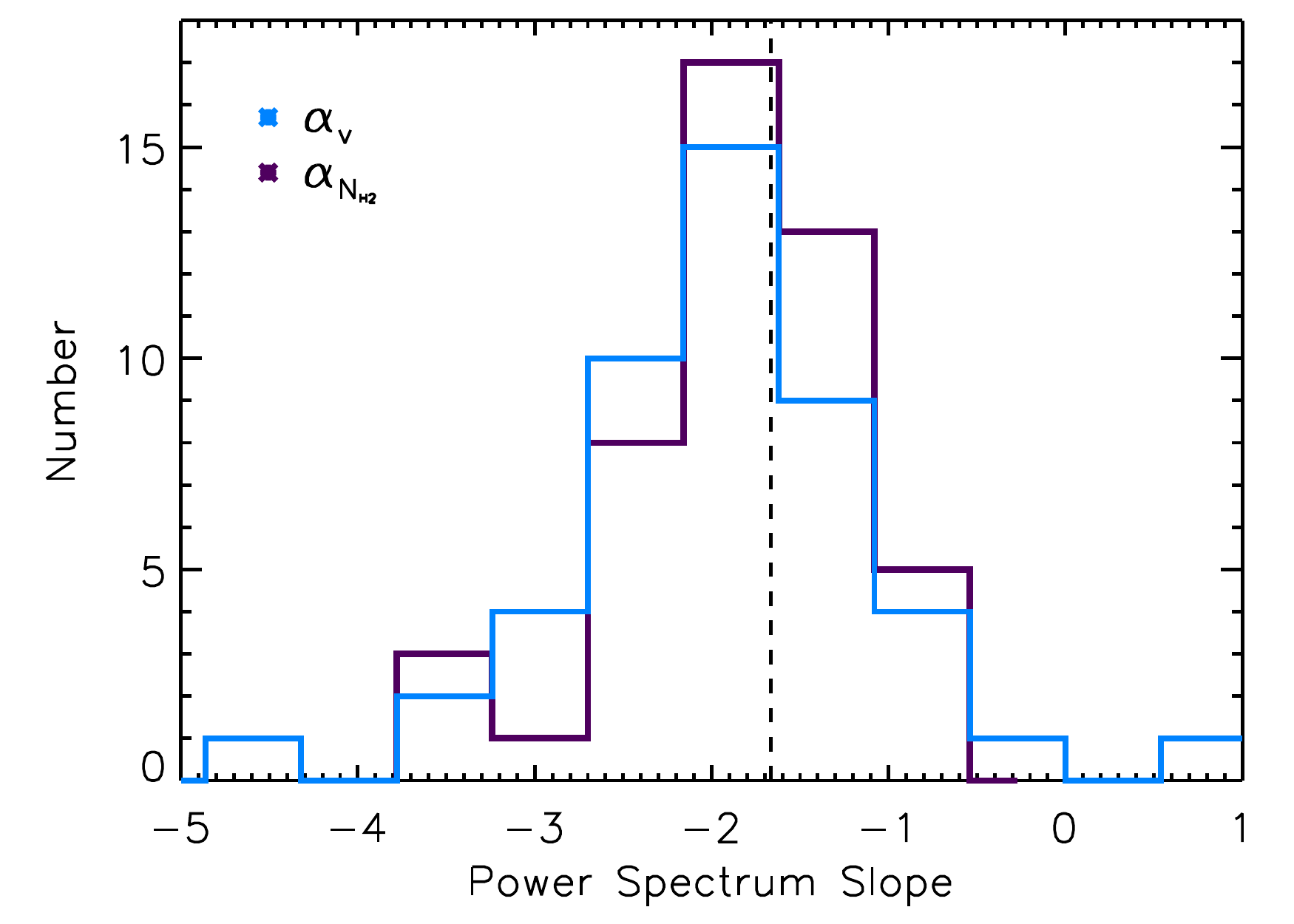}}
  \resizebox{8cm}{!}{
     \includegraphics[angle=0]{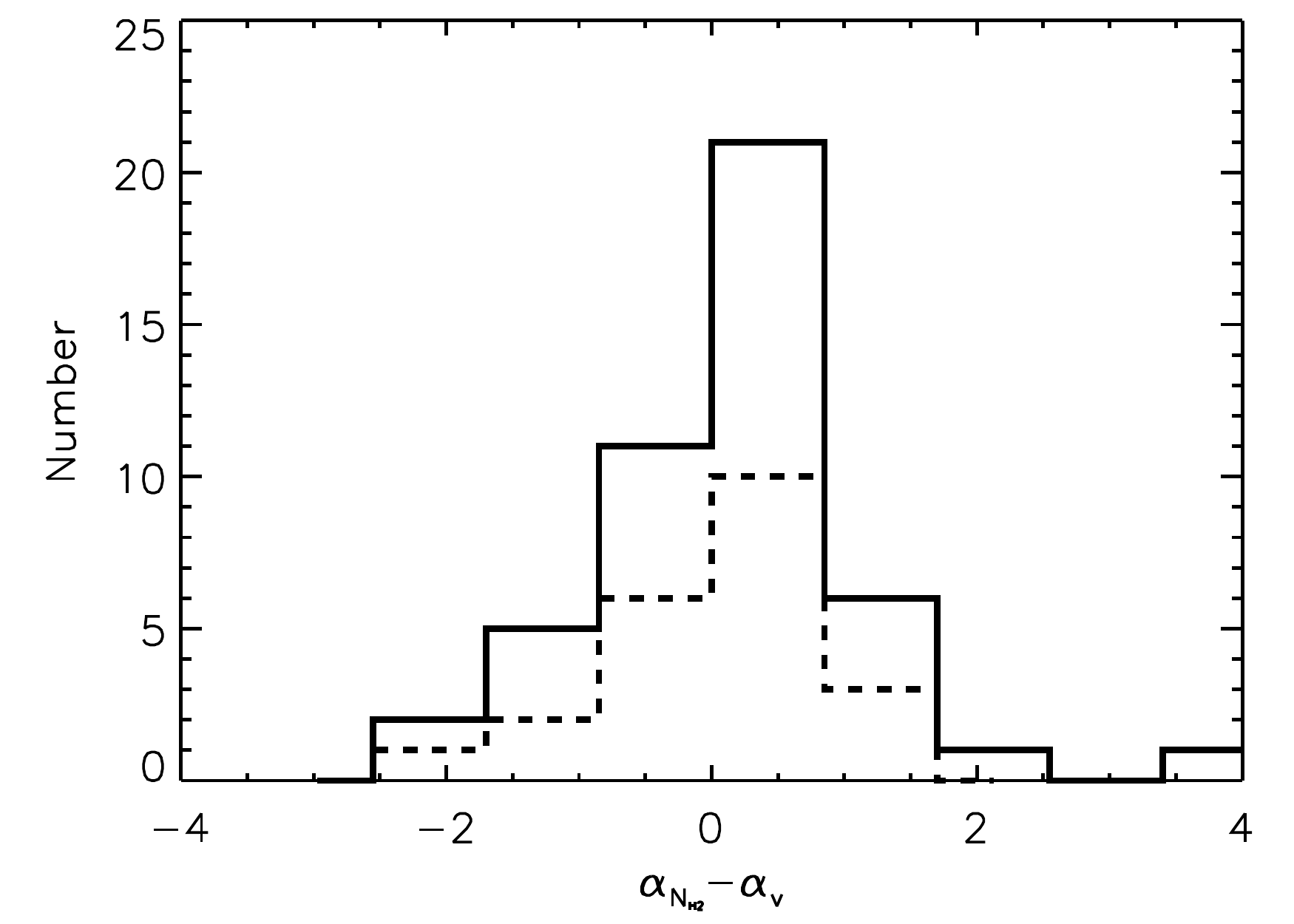}
}
\vspace{-.2cm}
  \caption{ {\rev  {\it Top:} Distribution of the power spectrum slopes of the column density ($\alpha_{N_{\rm H_2}}$) and velocity ($\alpha_v$) in purple and blue, respectively, for the 47 VCFs.} {\revbis The dashed  vertical line indicates the $-5/3$ value. }   
{\rev  {\it Bottom:}  The difference between the power spectrum slopes of the column density and velocity ($\alpha_{N_{\rm H_2}}-\alpha_v$). The solid and dashed histograms are for the 47 VCFs and the 22 VCFs with $L>10$\,pixels and $L>20$\,pixels,  mean and standard   deviation of, $0.15\pm0.99$ and  $0.07\pm0.81$, respectively. }
}           
  \label{PSslopHisto}
    \end{figure}

In order to quantify the observed column density and velocity fluctuations, 
 we here present the analysis of the one-dimensional (1D)  power spectrum along the  crests of the identified VCFs. 
The power spectrum $P(k)$  of $Y(l)$ (one of the VCF properties) 
 is proportional to the square of its Fourier transform and can be expressed in 1D as
\begin{equation}
P(k)=\frac{1}{L}|\tilde Y(k)|^2,
\end{equation}
where $\tilde Y(k)=\int Y(l)\exp^{-2i\pi k l}dl$ is the Fourier transform of $Y(l)$,  $l$ is the spatial position along the crest, $k$ is the angular frequency, and $L=\int dl$ is the total VCF length. 
Figure\,\ref{PS1-47}  shows the power spectra of  $N_{\rm H_2}(l)$ and $v(l)$ of {\rev three VCF~1, 32 and 47 (see also Sect.\,\ref{ana_selected} for the analysis of these three selected VCFs)}. 
No characteristic scales can be seen on the power spectra of both  $N_{\rm H_2}$ and $v$, which are well represented by power laws down to the angular resolution of the data ($30\parcs$2 or $s\sim1.2$\,arcmin$^{-1}$).
The observed power spectra are fitted with a power law function {\rev in log-log space} where $k=1.2$\,arcmin$^{-1}$  (the smallest resolved scale) is the highest frequency data point used for the fit. 
{\rev Figure\,\ref{PSslopHisto}-top shows the distribution of the $\alpha_{N_{\rm H_2}}$ and  $\alpha_v$ slopes of the $N_{\rm H_2}(l)$ and $v(l)$  power spectra, with mean values of $\sim-1.9\pm0.7$ and $\sim-2.0\pm0.9$,  respectively (see Table\,\ref{tab:Tparam}). 
Figure\,\ref{PSslopHisto}-bottom shows that the statistical difference between the $\alpha_{N_{\rm H_2}}$ and  $\alpha_v$  slope values for the same filaments are close to zero with a mean value of $0.15\pm0.99$. This difference is $0.07\pm0.81$ and closer to zero when only the VCFs with length $L>20$\,pixels are considered. A sub-selection of VCFs with  $L>20$\,pixels (10 beams)  ensures a statistically {\revbis larger} number of resolution elements to describe the fluctuations along the VCF crests. 
}

The best fit slope value of the $N_{\rm H_2}$  fluctuation along the VCF~1 is also similar to the value found by \citet{Arzoumanian2021} for a longer section of the filament, namely the combined crests of the VCF 41 and 1, and at higher angular resolution (down to $14\arcsec$). {\rev We discuss possible implications of this analysis in Sect.\,\ref{disc} below.}

\section{Properties along and across the crests of selected VCFs}\label{ana_selected}

\begin{figure*}[!h]
   \centering
     \resizebox{17cm}{!}{ 
                    \includegraphics[angle=0]{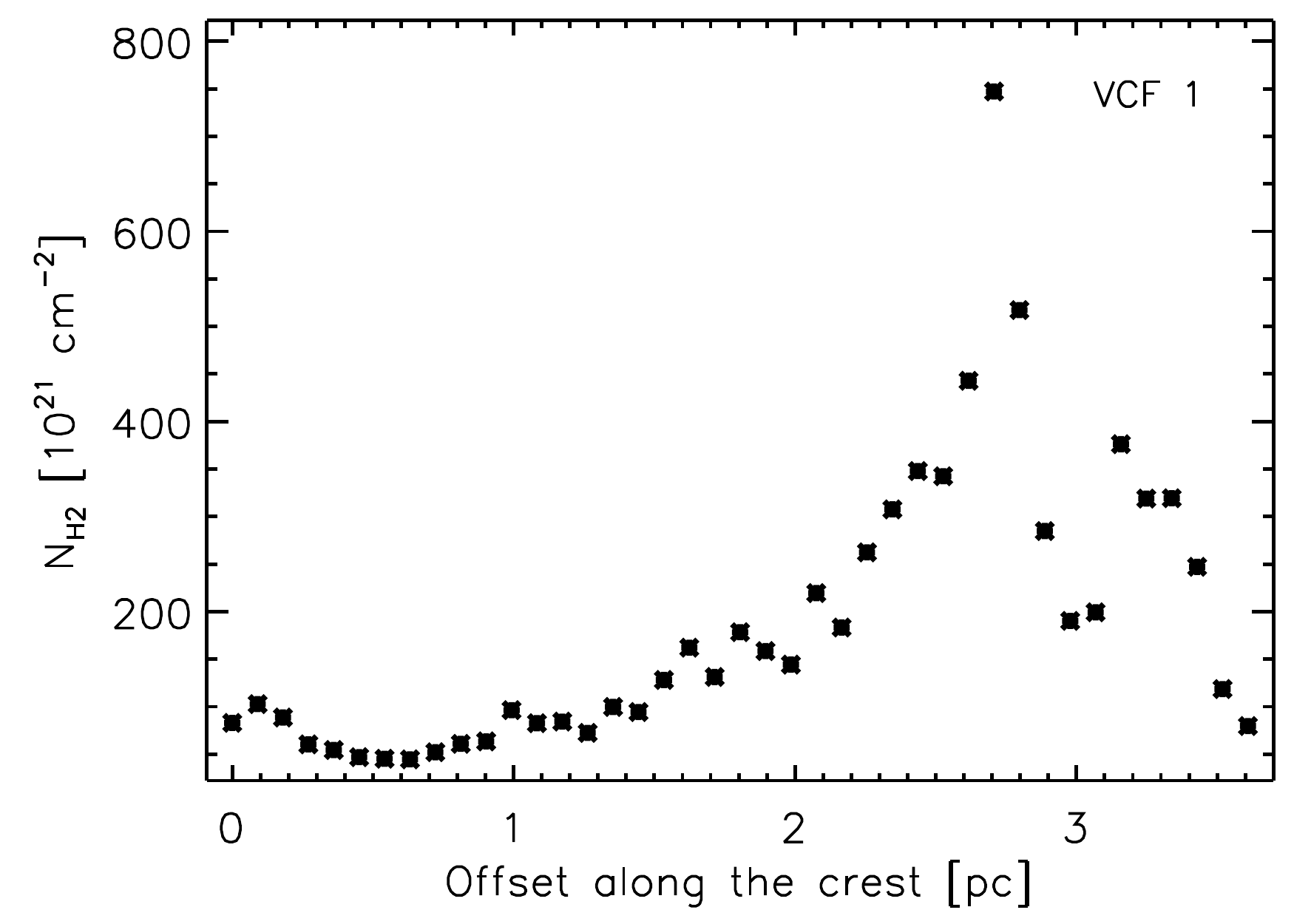}
                            \includegraphics[angle=0]{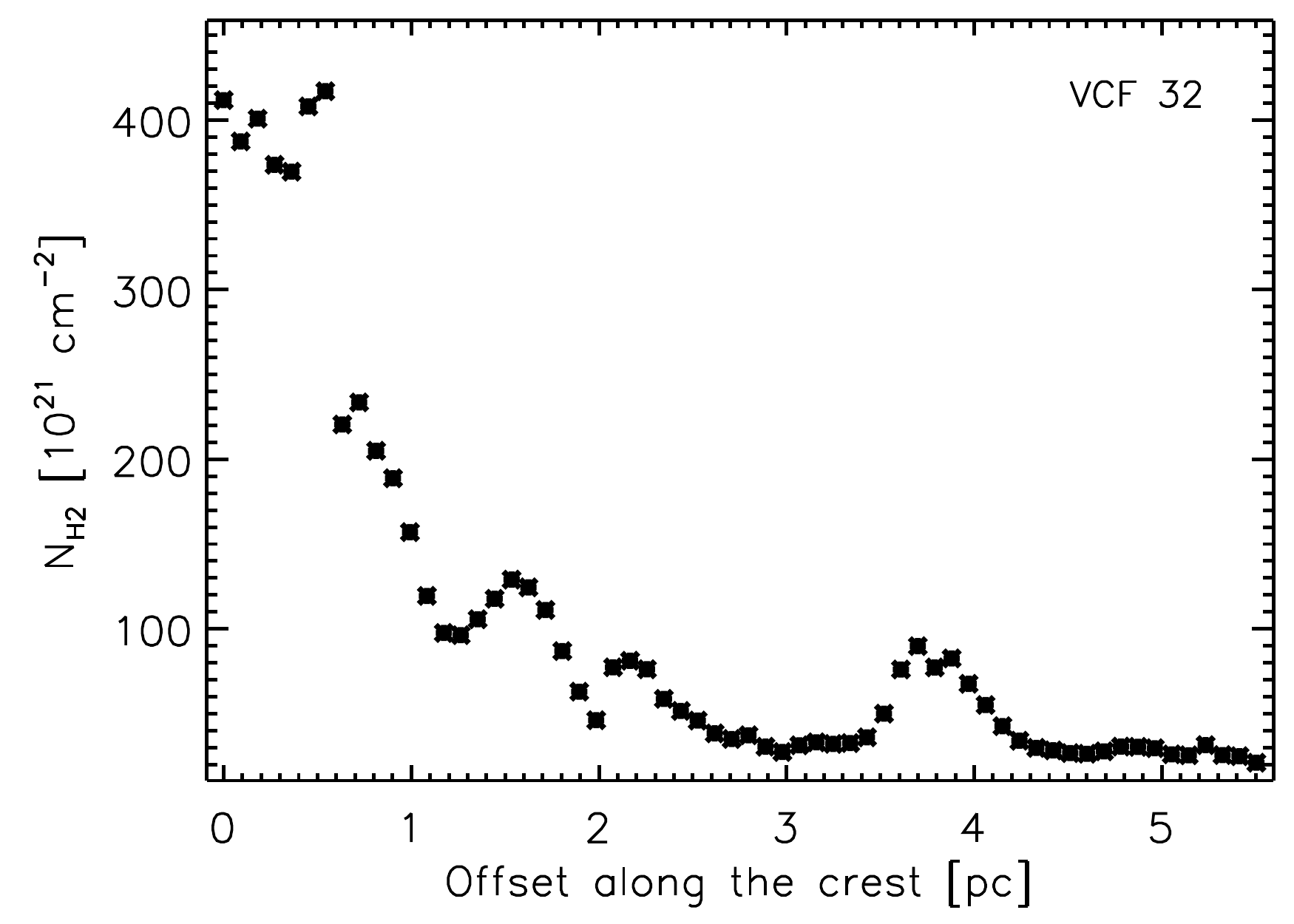}
     \includegraphics[angle=0]{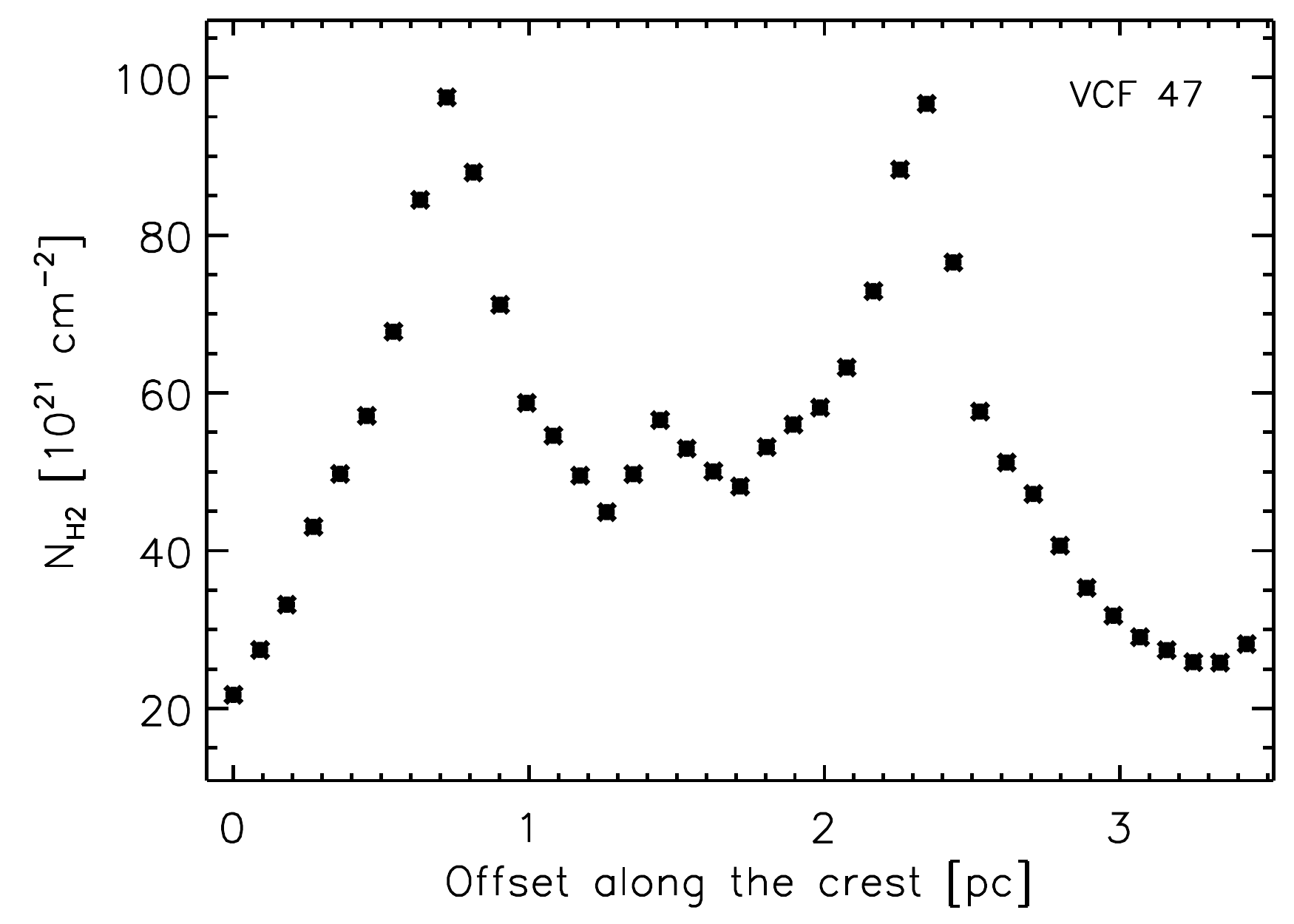}
               }
                  \resizebox{17cm}{!}{   
                                                \includegraphics[angle=0]{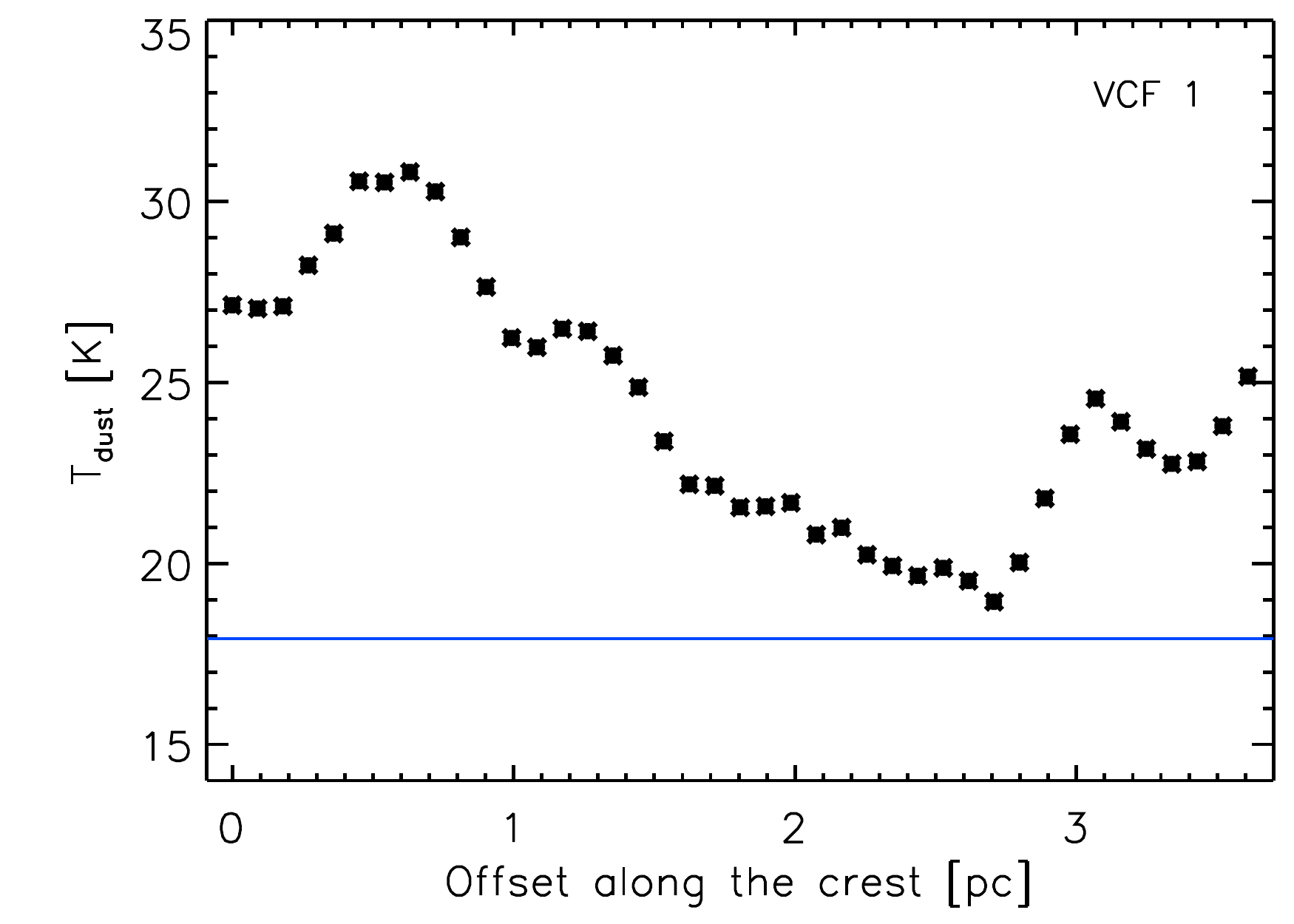}
                                                      \includegraphics[angle=0]{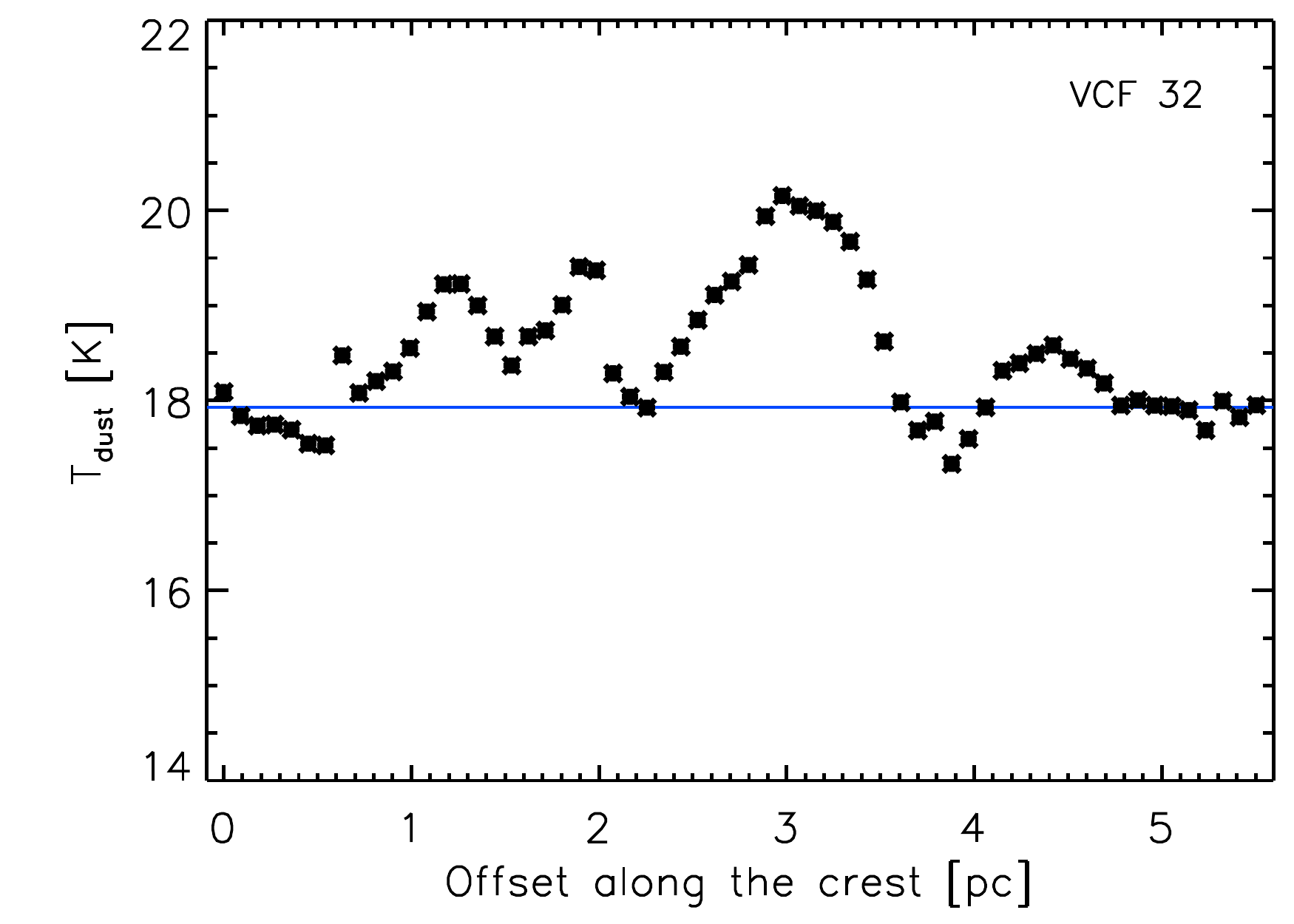}
          \includegraphics[angle=0]{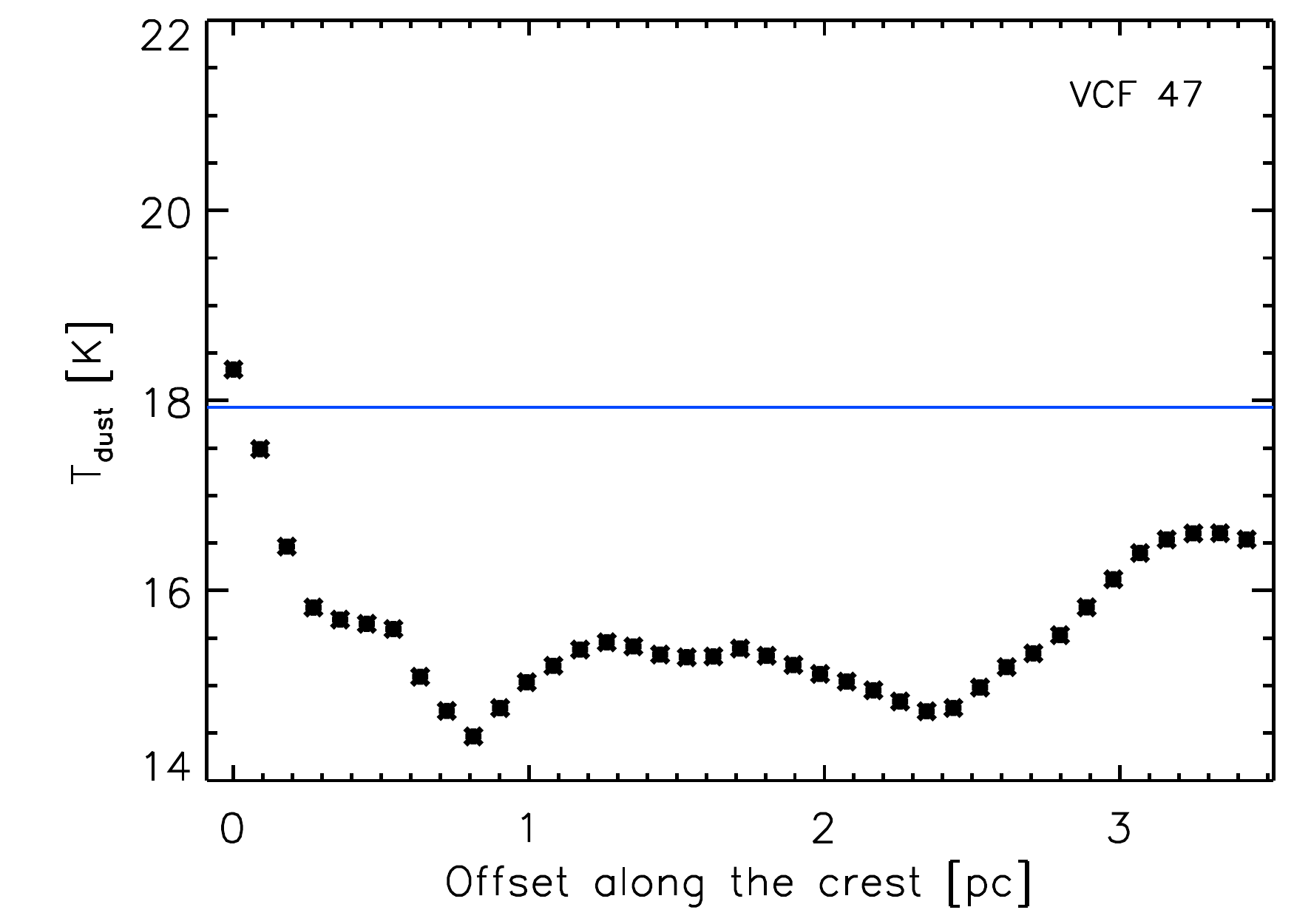}
                              }
                         \resizebox{17cm}{!}{            
                                          \includegraphics[angle=0]{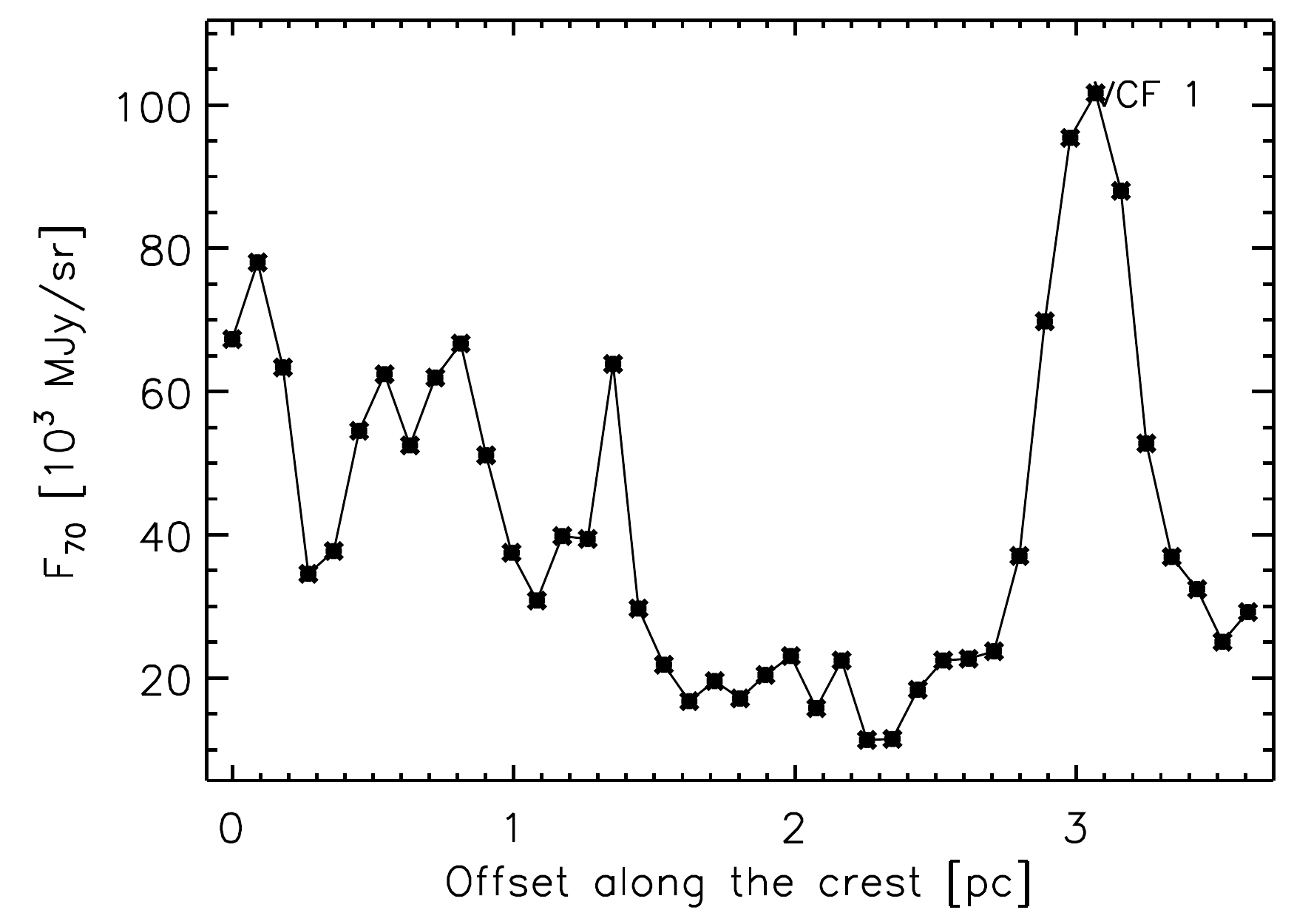}
              \includegraphics[angle=0]{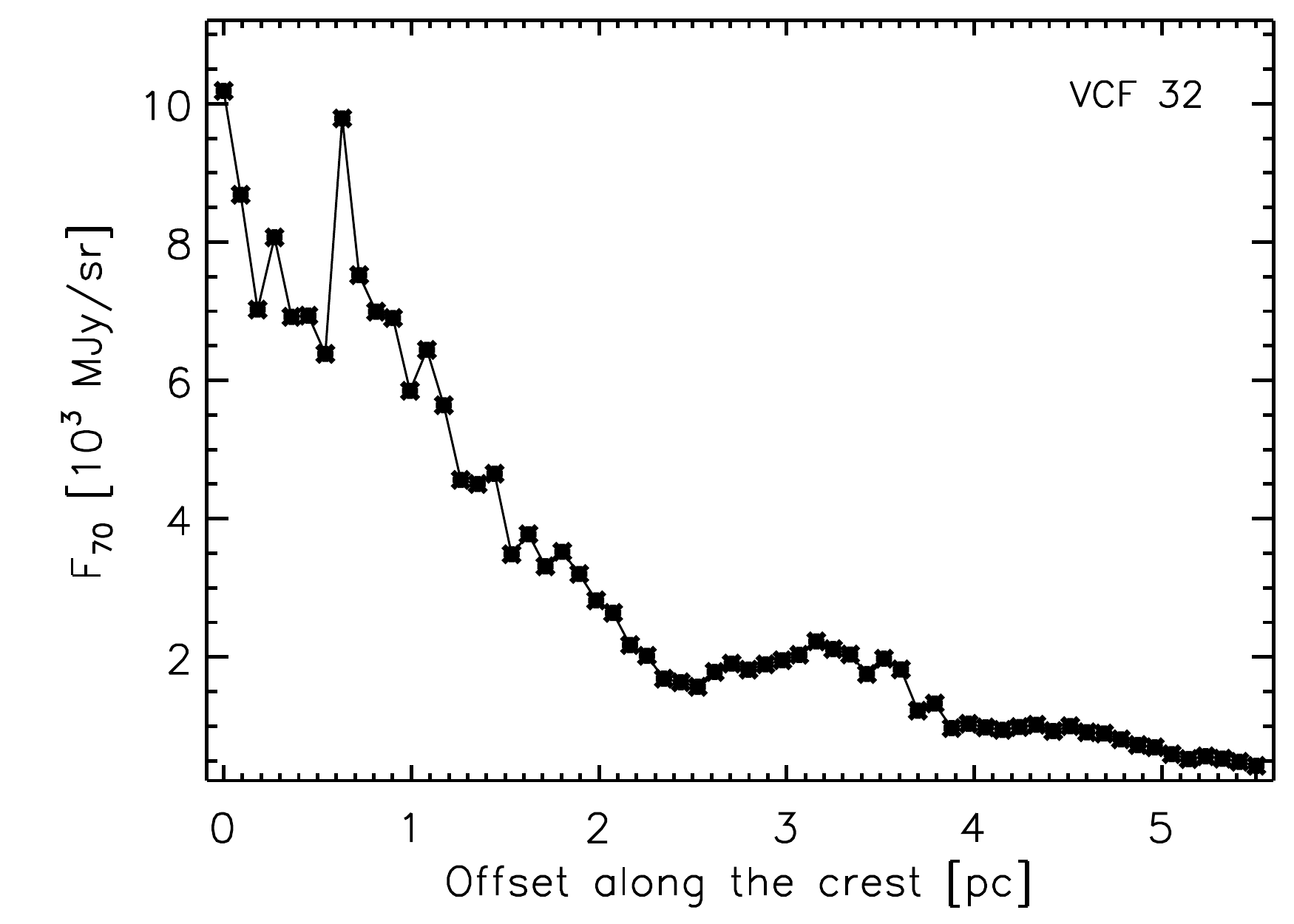}
           \includegraphics[angle=0]{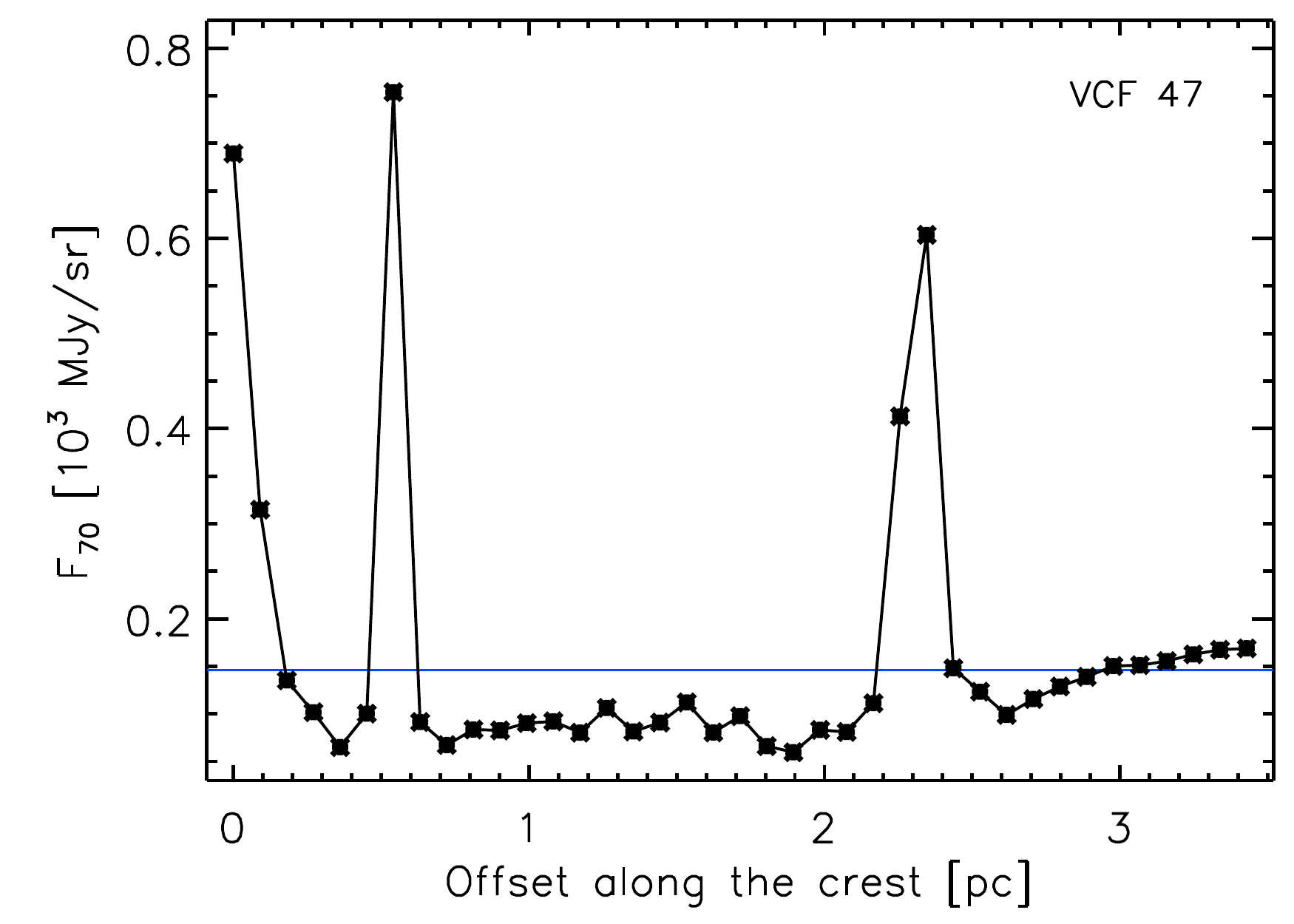}
                 }
                      \resizebox{17cm}{!}{            
                                                    \includegraphics[angle=0]{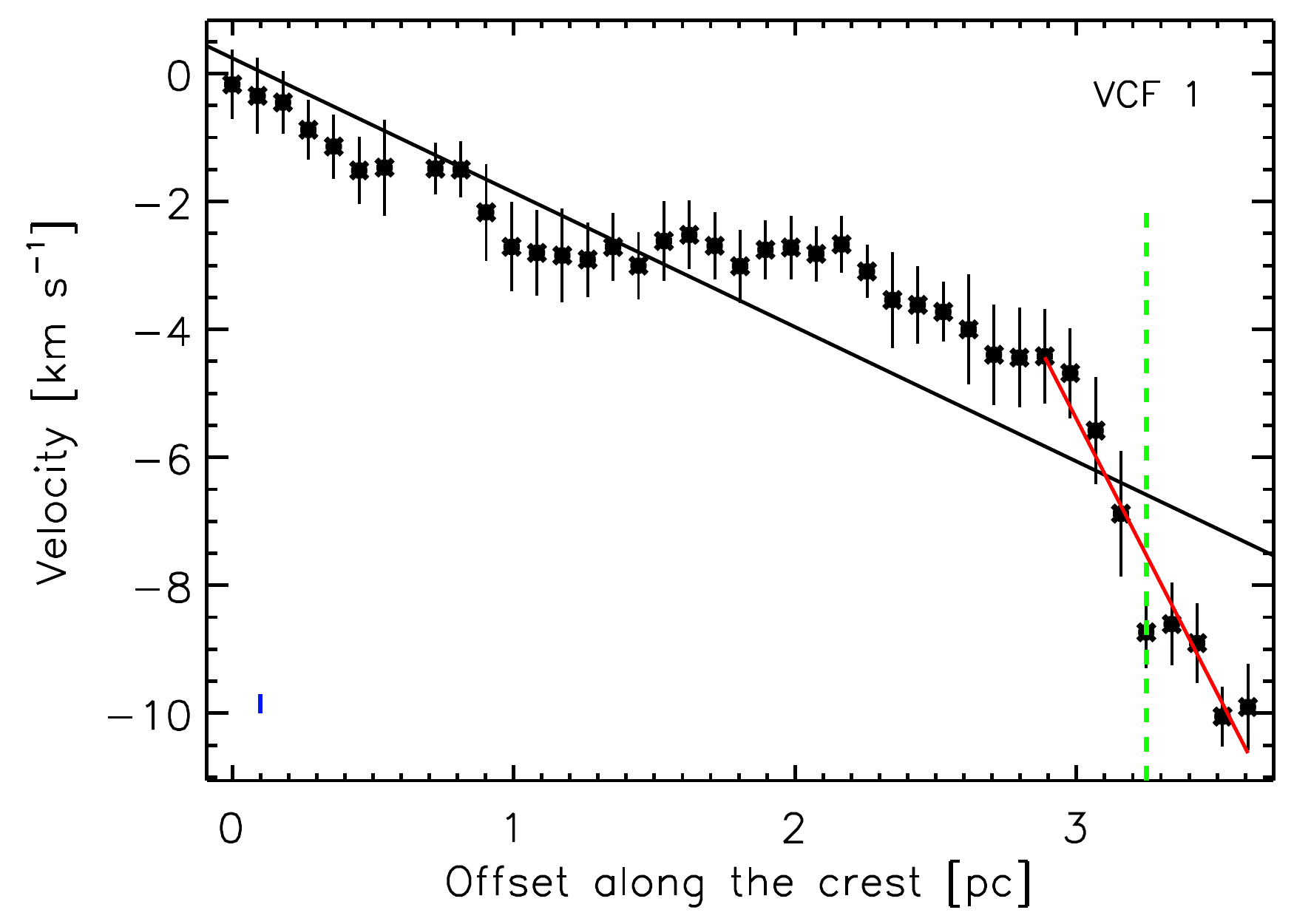}   
                       \includegraphics[angle=0]{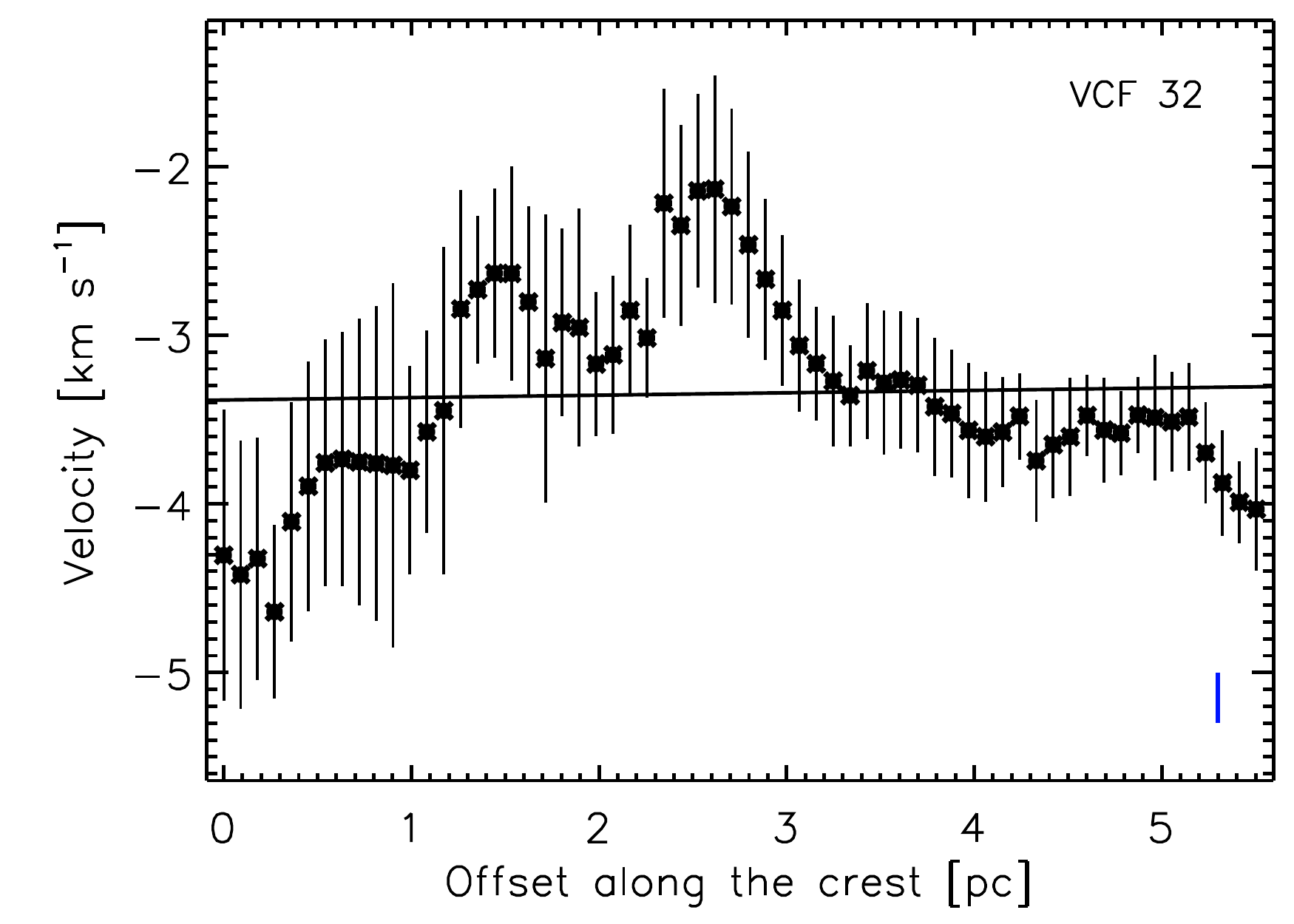}  
               \includegraphics[angle=0]{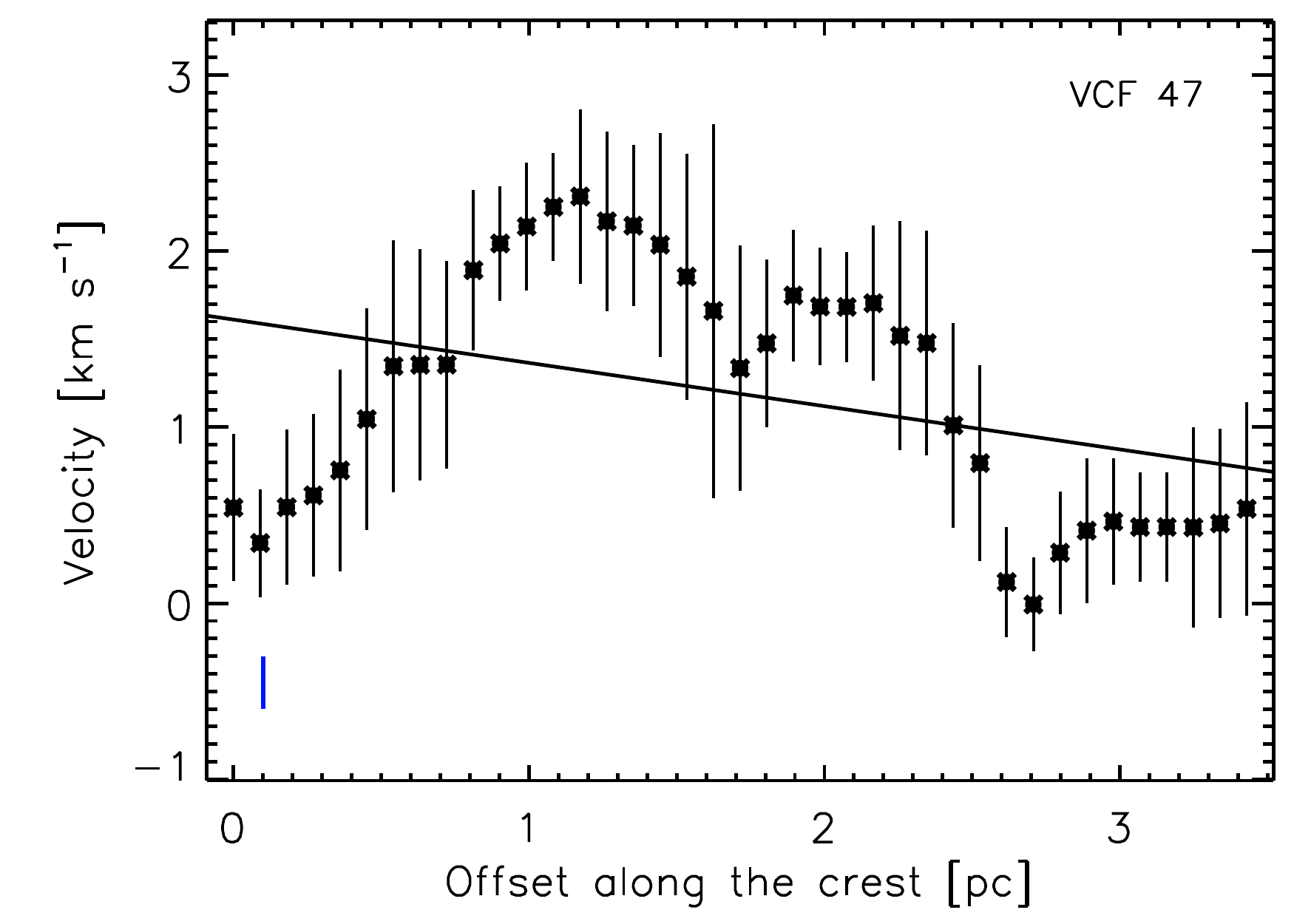}     
                              }
                             \resizebox{17cm}{!}{  
                                                                   \includegraphics[angle=0]{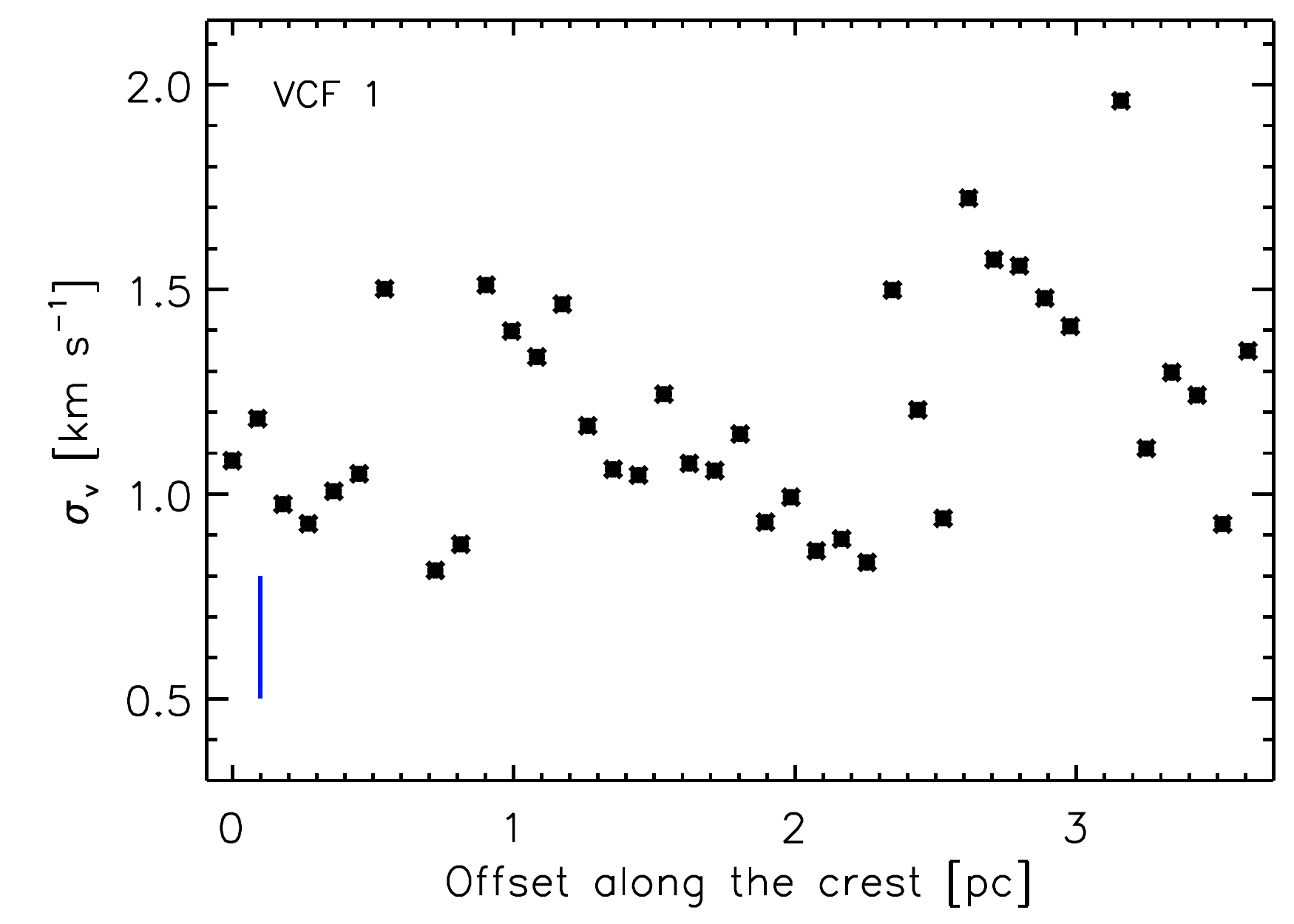}
                                      \includegraphics[angle=0]{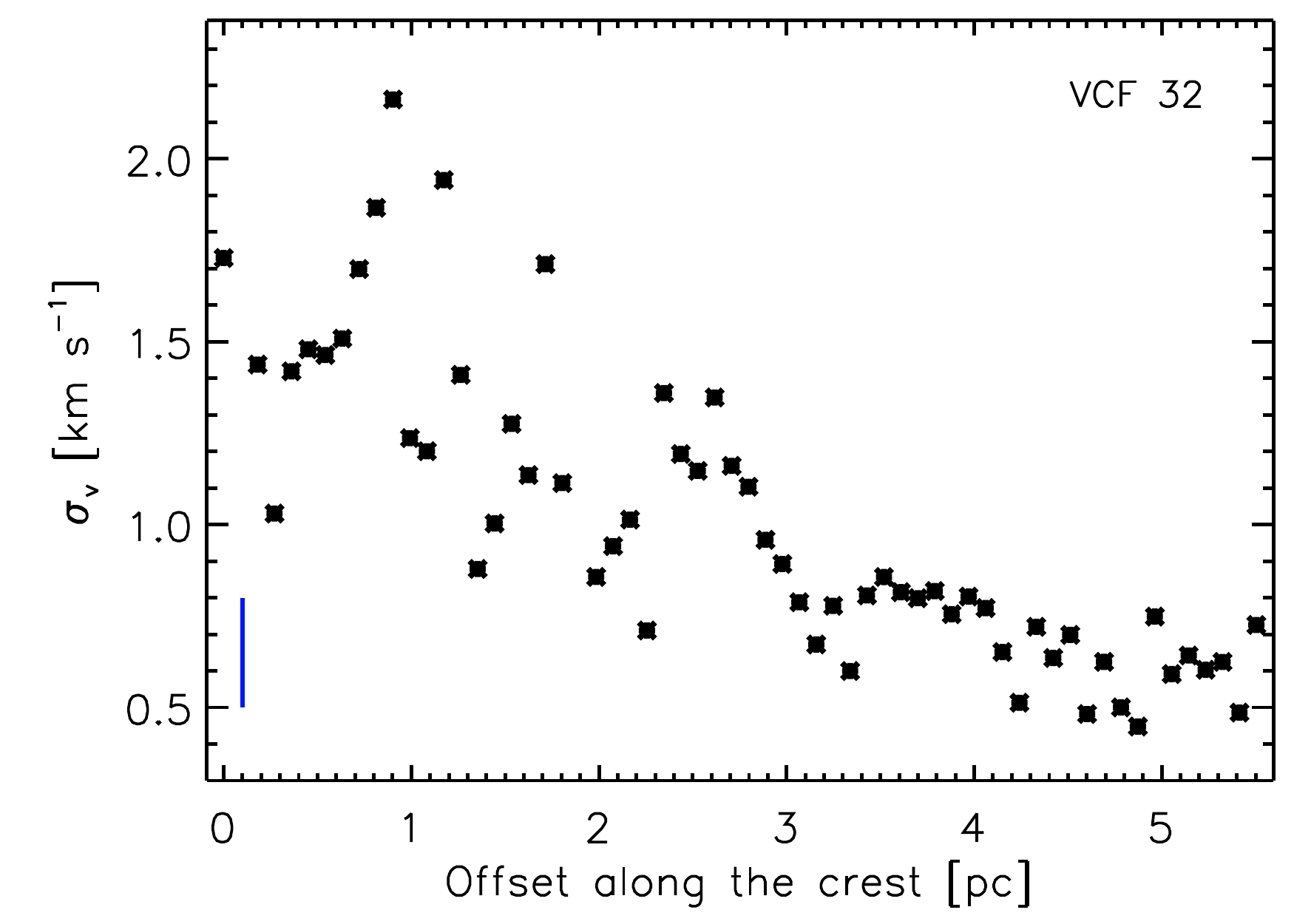}
                                      \includegraphics[angle=0]{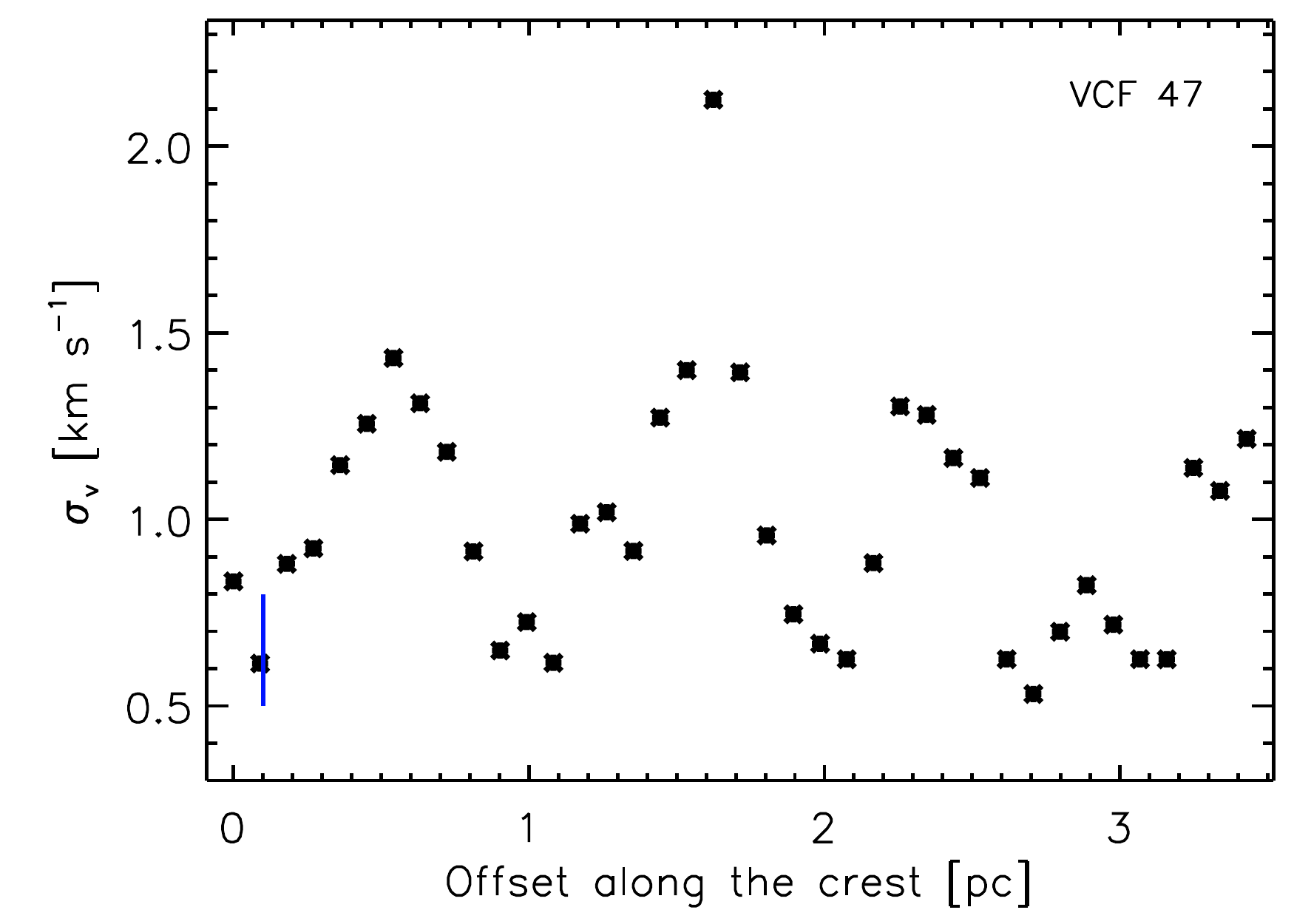}
                                      }
  \caption{
  {\small
  Properties along the {\rev VCF crests 1, 32, and 47, from left to right}. 
  The zero offsets correspond to the most western-southern end of the crests (as seen on Fig.\,\ref{MapSkel}).  
 {\rev Rows from top to bottom are: (1) Column density as derived from \herschel\ data.  (2) LOS averaged dust temperature as derived from \herschel\ data. The horizontal blue lines indicate the mean background temperature around the VFCs of $\sim17.9\pm0.2$\,K.
  (3) PACS 70\,\mum\ flux. The horizontal blue line indicates the mean background emission of $\sim146\pm23$\,MJy/sr.
(4) Velocity. The black vertical lines on each data points show the velocity dispersion (same as the values shown in row 5). The black lines show a linear fit to these data points.
The velocity resolution of 0.3\,\kms\ is indicated by the vertical blue lines. 
For the VCF~1, a linear fit to the velocity $>2.8$\,\kms\ {\revbis over a length of 0.7\,pc} is indicated in red and corresponds to the crest crossing the 
young star cluster identified as source I \citep[see, e.g.,][]{Persi2008}.
 {\revbis The position of source I is indicated with the vertical dashed green line (see also the green arrow  on Fig.\,\ref{MapSkel}). }
(5) Velocity dispersion. The velocity resolution of 0.3\kms\ is indicated by the vertical blue lines. }
  }          }
  \label{plotsMeanParam}
    \end{figure*}

Figure\,\ref{plotsMeanParam} shows the variation of the properties along the crests of {\rev three VCFs 1, 32, and 47,} which are part of the subregions IFS and MFS, respectively (see Fig.\,\ref{MapSkel}). {\rev We select  these VCFs because they are located in different environments, } yet part of the same molecular cloud and connected in velocity (see Fig.\,\ref{Channelmaps}).  We compare {\rev below} their observed properties to study the environmental impact on the formation and evolution of filament systems.

   \begin{figure*}[!h]
   \centering
     \resizebox{19.cm}{!}{  \hspace{-.8cm}
            \includegraphics[angle=0]{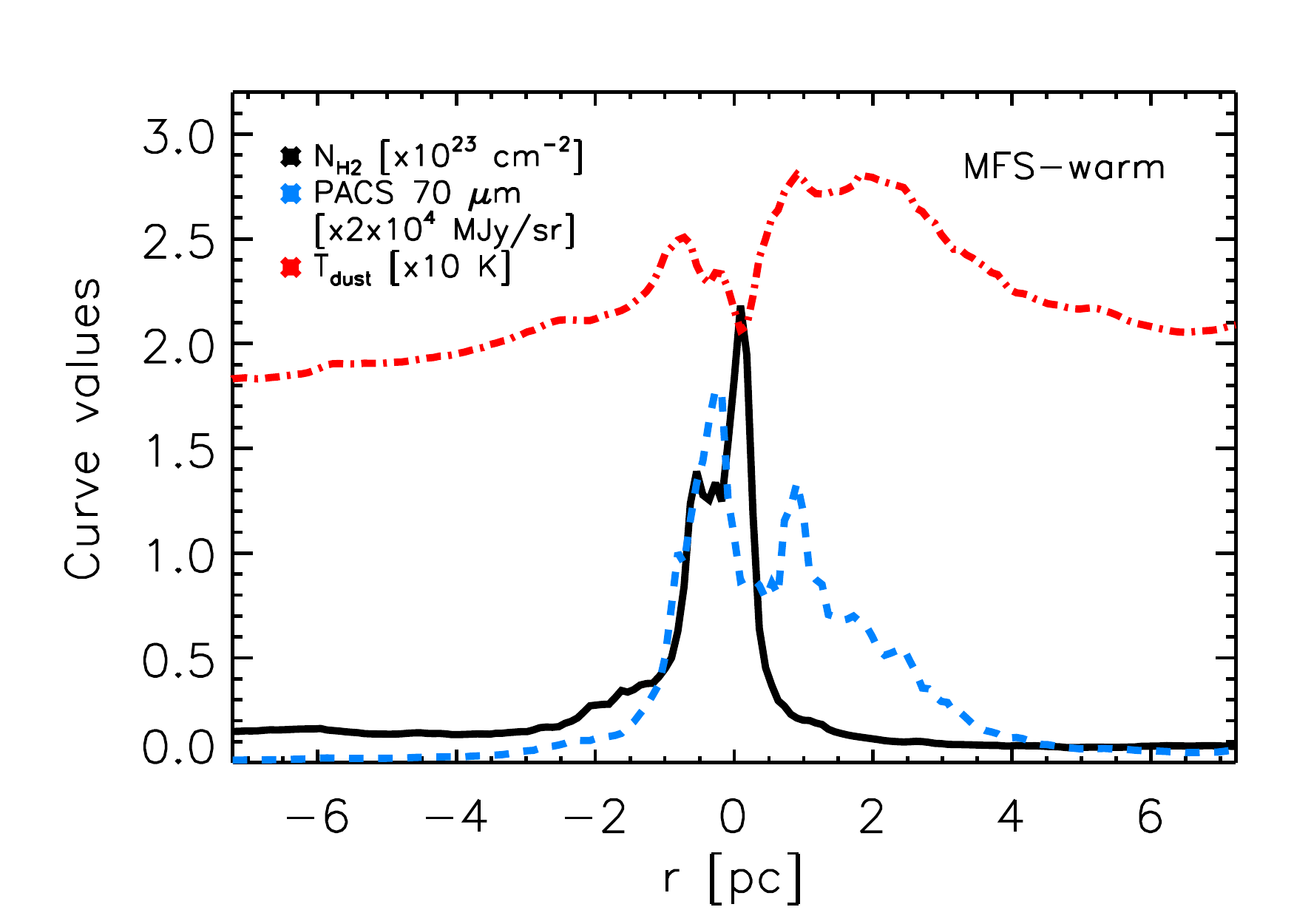}
       \includegraphics[angle=0]{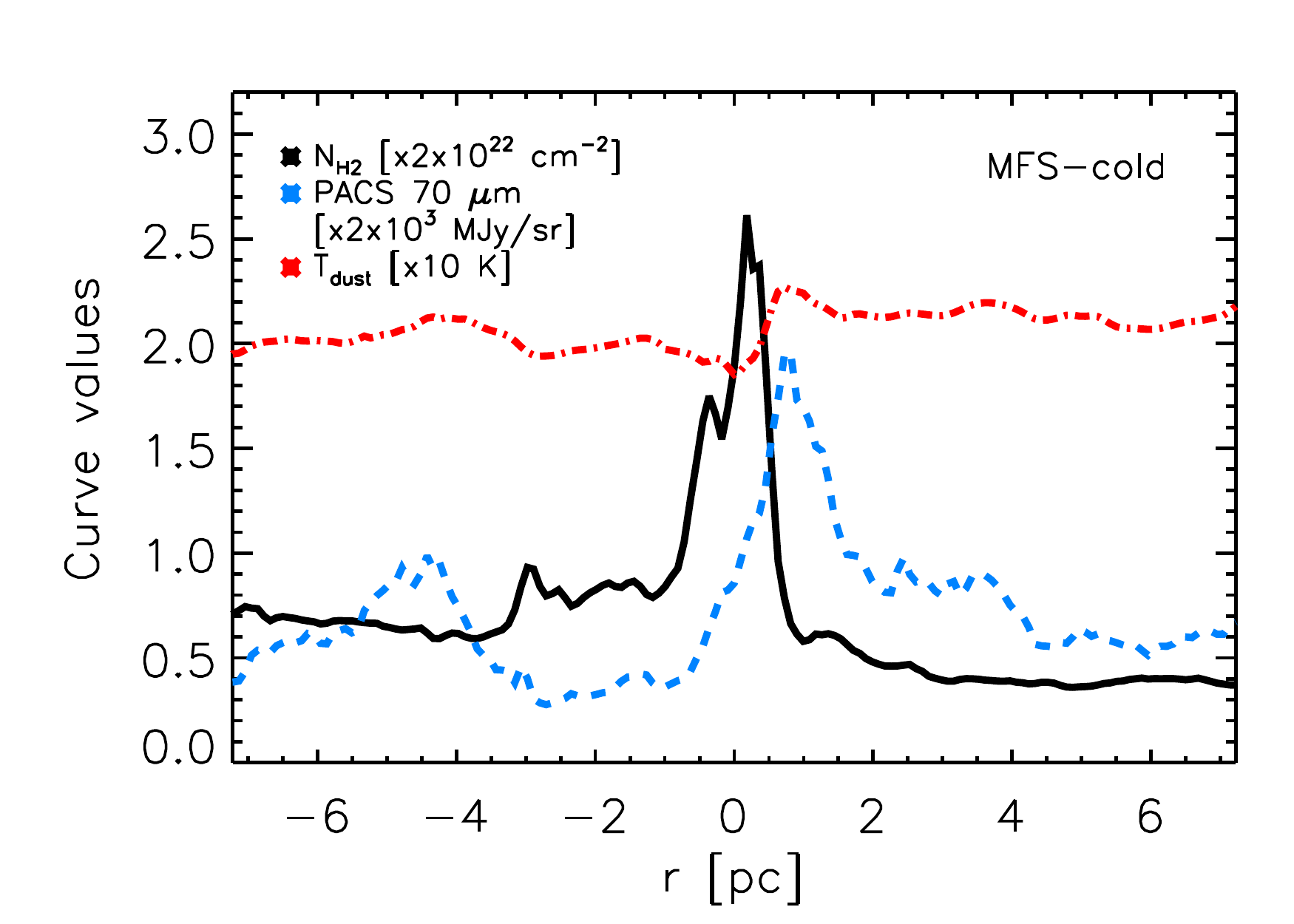}
     \includegraphics[angle=0]{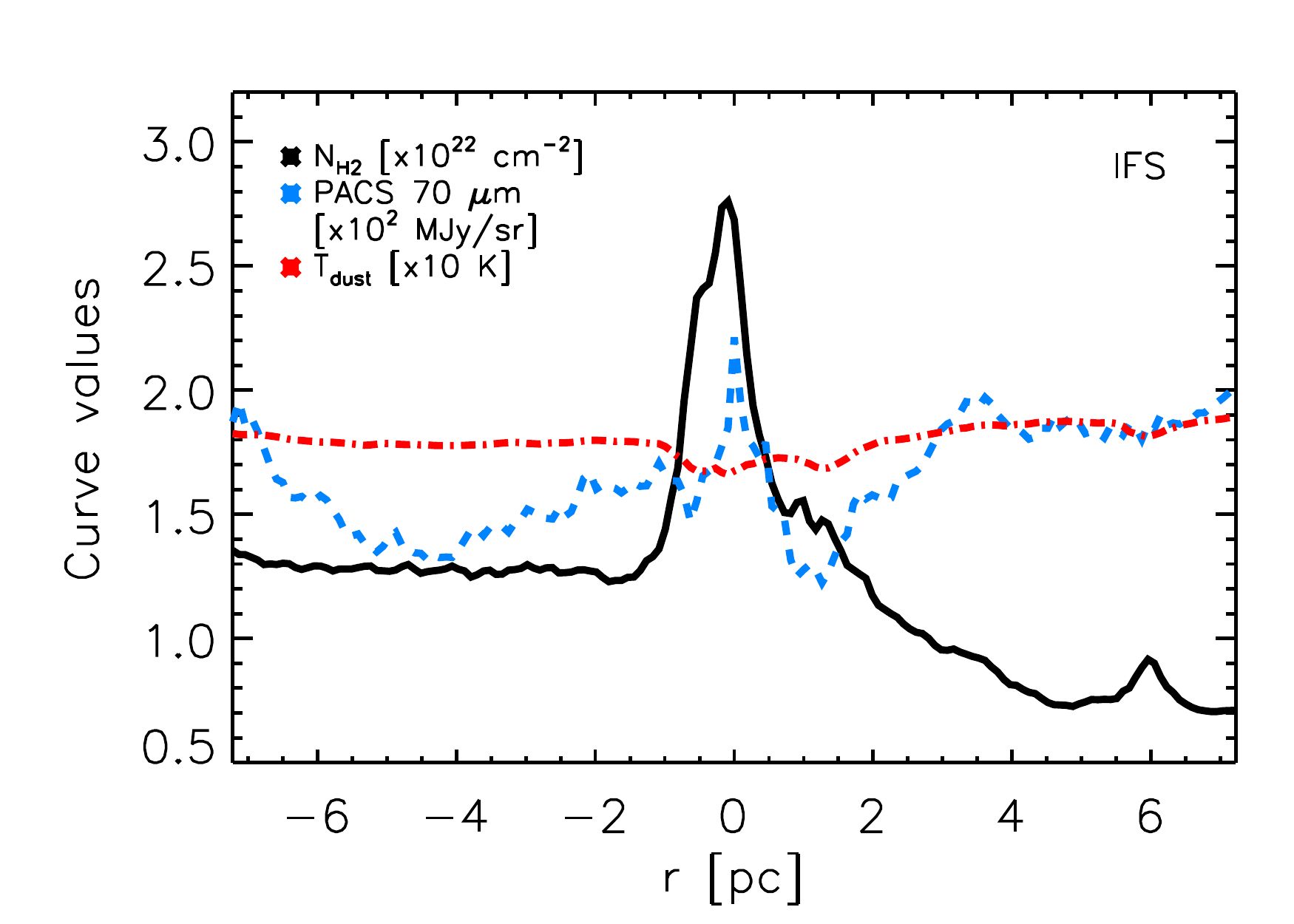}     
       } 
\vspace{-.5cm}
  \caption{ Radial profiles across the {\rev MFS-warm, MFS-cold, and IFS  regions,} from left to right (see boxes in Fig.\,\ref{MapSkel}).
  The column density, the PACS 70\,$\mu$m emission, and the dust temperature are shown in solid black, blue dashed, and red dotted dashed lines, respectively. The scaling of the different curves is indicated in the caption of each panel.  The  $r>0$ values correspond to the Galactic north.
}          
  \label{Radial_profs}
    \end{figure*}

  \begin{figure*}[!h]
   \centering
     \resizebox{18.5cm}{!}{
     \hspace{-3.5cm}
      \includegraphics[angle=0]{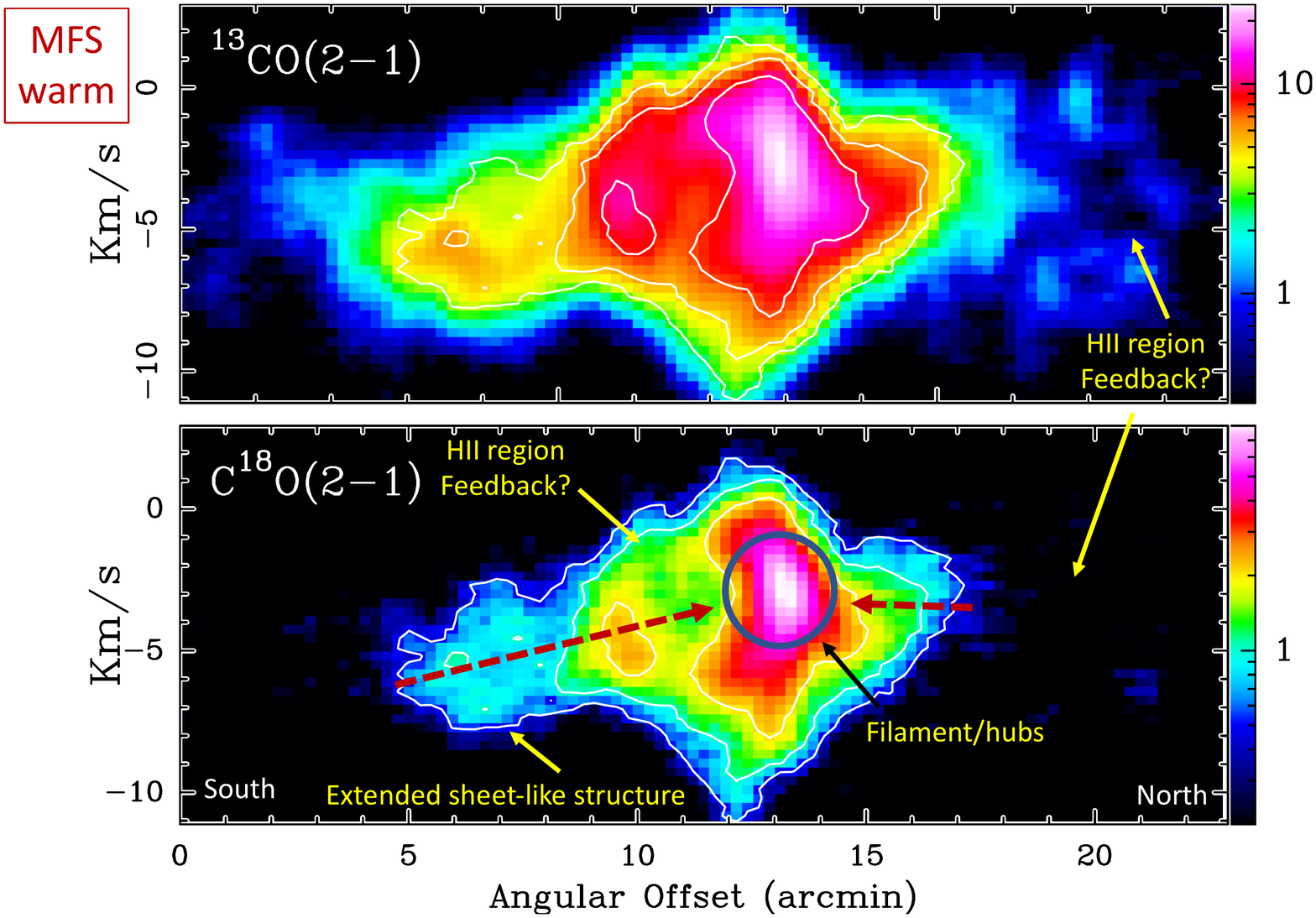}
              \hspace{-3.cm}
  \includegraphics[angle=0]{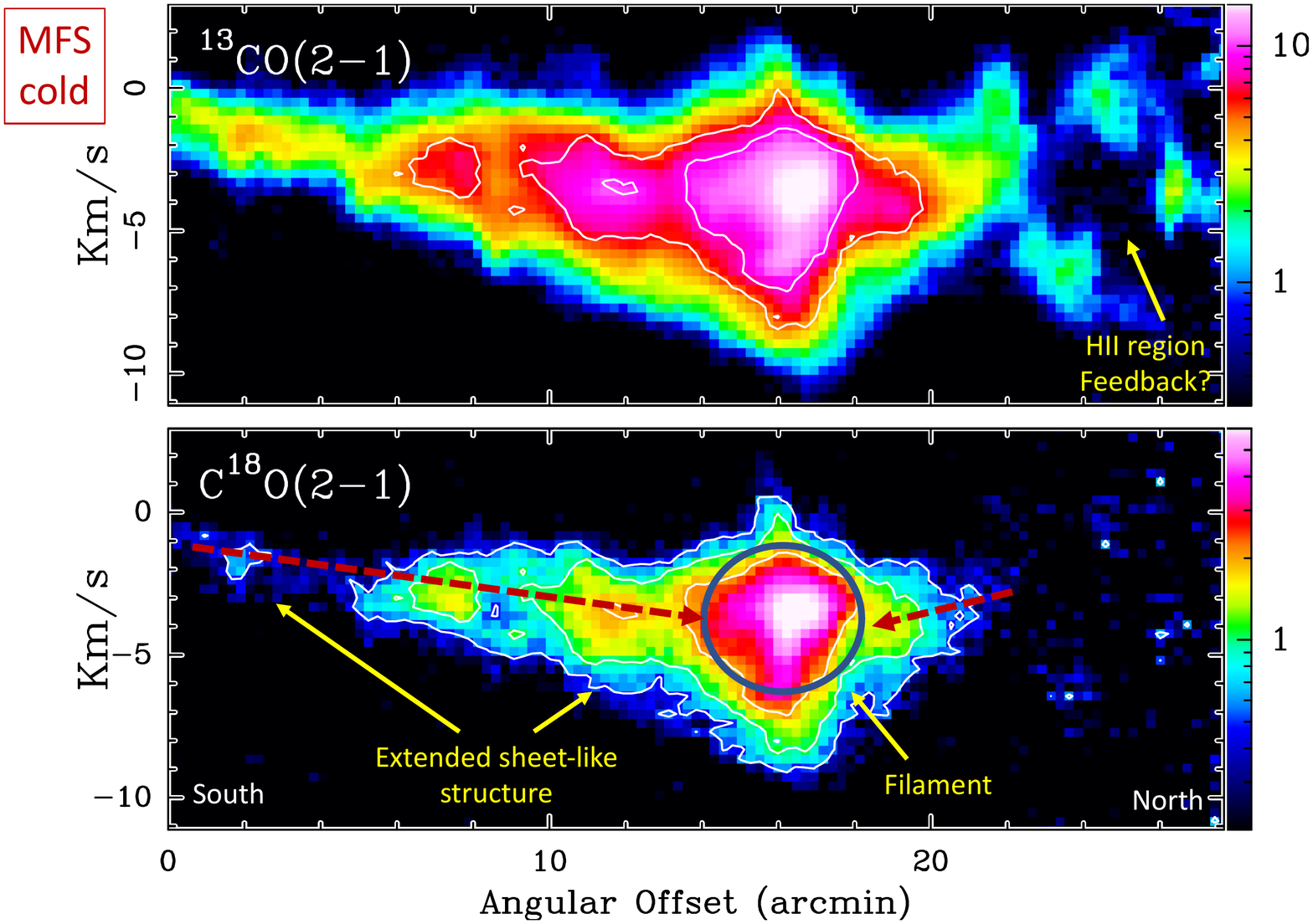}
     \hspace{-3.cm}
     \includegraphics[angle=0]{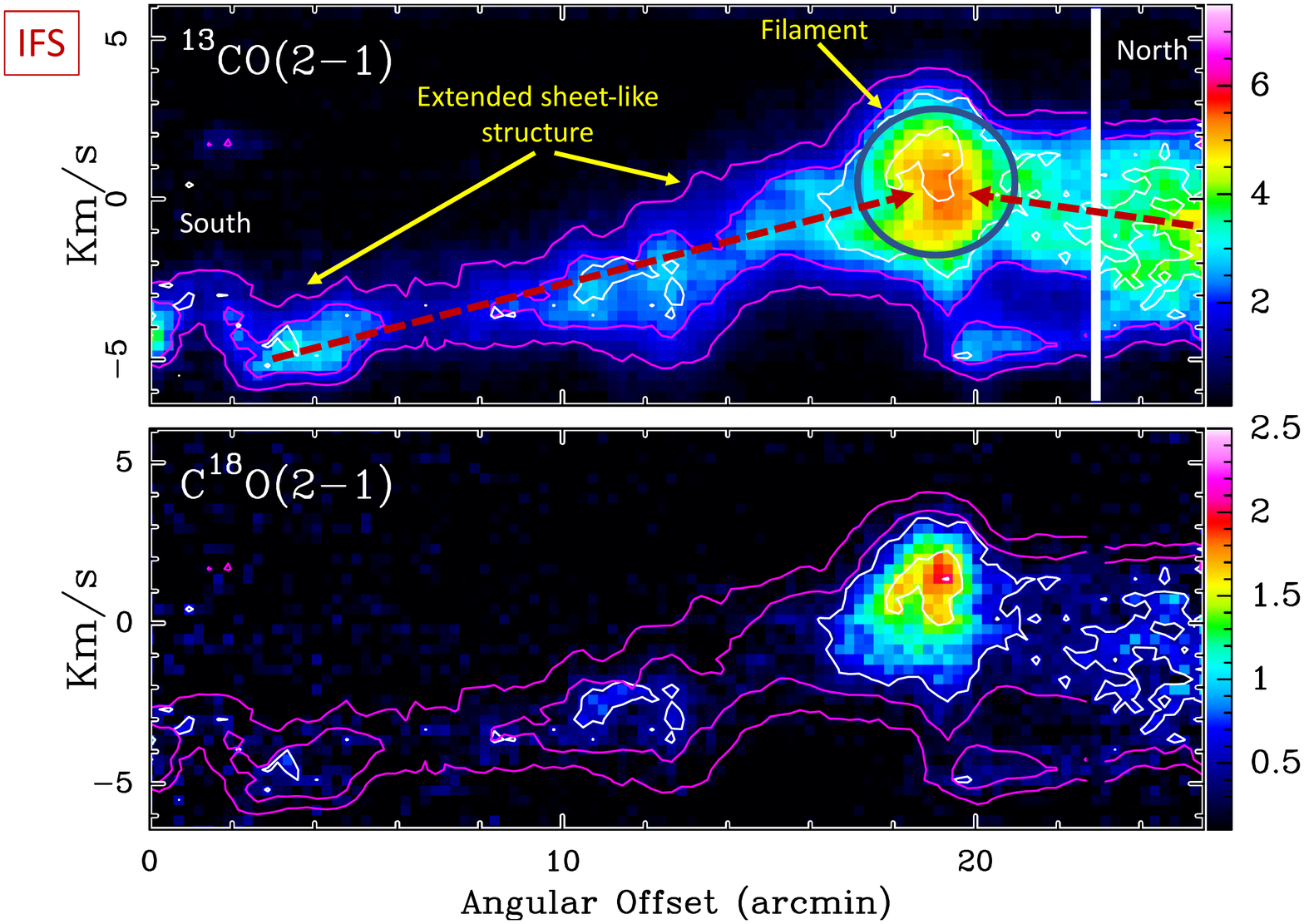}
   }
\vspace{-.6cm}
  \caption{  
Position$-$velocity (PV) diagrams perpendicular to and averaged along the main axes of the 
{\rev MFS-warm, MFS-cold, and IFS  regions,}
from left to right (see boxes in Fig.\,\ref{MapSkel}).
The PV maps are in units of K(T$_{\rm MB}$) and correspond to the $^{13}$CO($2-1$) and C$^{18}$O($2-1$) 
emission on the top and bottom, respectively.  
The white contours indicate the C$^{18}$O($2-1$) intensity and are the same for the two panels of the same filament system.
The magenta contours on the left panel correspond to the   $^{13}$CO($2-1$) intensity.
 The zero offset position (in the $x$-axes) corresponds to the Galactic South side of the filament systems.  At the distance of this cloud   10\arcmin\ corresponds to 3.8\,pc. The position of the filaments are indicated with a dark blue circle and the velocity gradients induced by the compression are shown as converging red dashed arrows towards the filament. 
 {\rev Longitudinal  PV diagrams along the crest of VCF~32 of the MFS-cold region   are shown in Fig.\,\ref{PValongCrest}.}
}          
  \label{PV_perpFil12}
    \end{figure*}

\begin{figure*}[!h]
   \centering
     \resizebox{18.cm}{!}{
\includegraphics[angle=0]{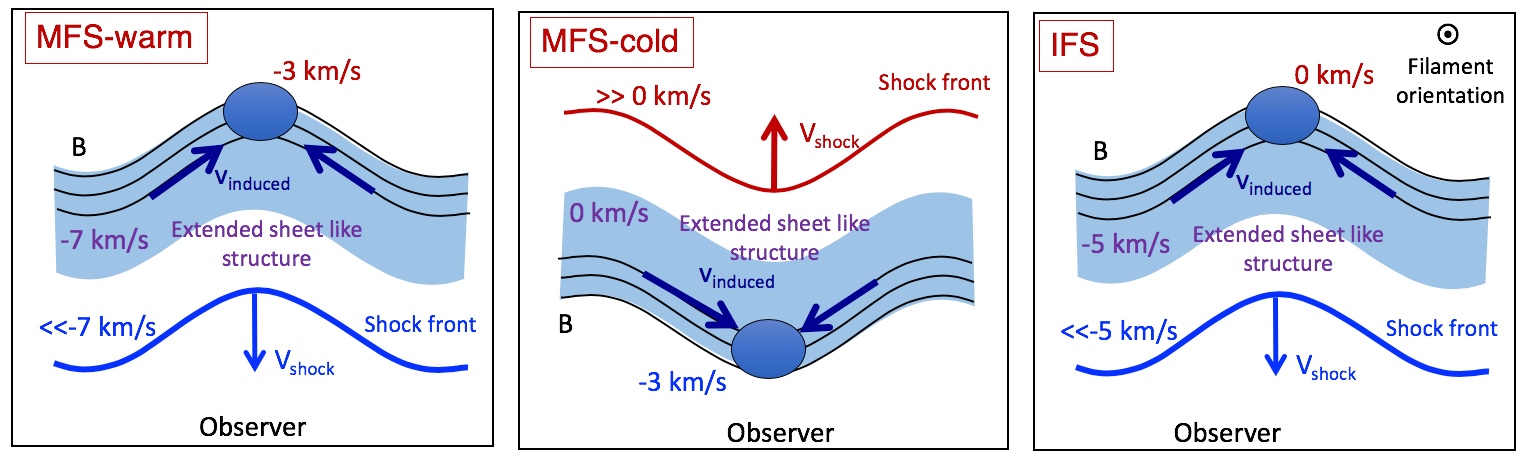}
}
  \caption{ 
  Schematic view of the suggested scenario where  propagating extended gas structures interact with the NGC 6334 complex, resulting in the formation  or altering the properties of the filament systems 
   {\rev MFS-warm, MFS-cold, and IFS}
   from left to right. 
These sketches are in the PP space perpendicular to the filament along axes. The velocities of the different structures are indicated in the LSR frame according to the PV diagrams (Fig.\,\ref{PV_perpFil12}). 
 {\rev The shock front is indicated in the post-shock configuration.}
   The $<<$ and $>>$ symboles refer to more blueshifted and more redshifted velocities, respectively, with respect to the velocity of the extended sheet like structures.  {\rev The velocity of the sheet like structure is connected to the filament velocity through a velocity gradient. }
      The ambient B-field (indicated by the black lines) is parallel to the shock front. 
}          
  \label{sketch} 
    \end{figure*}

\subsection{Description and properties of selected VCFs }\label{ana4a}
 
\hspace{0.5cm}{\bf VCF 1}
 
The VCF~1 corresponds to a section of the well studied star-forming  NGC~6334 main-filament 
  \citep[e.g.,][]{Russeil2013,Andre2016,Sadaghiani2020,Arzoumanian2021}. 
This VCF has a length of $\sim3.7$\,pc, a mean column density of  $\sim2\times10^{23}\NHUNIT$, a column density contrast of $\sim30$, and a mean {\rev LOS averaged} dust temperature of $\sim24$\,K. 
This VCF corresponds to the MFS-warm region in Fig.\,\ref{MapSkel}.
The PACS 70\,\mum\ emission along the crest shows values of $\sim50$\,MJy/sr over the southern $\sim1$\,pc tracing the heated dust from the surrounding \hii\ regions and also a more narrower  70\,\mum\ peak of $\sim100$\,MJy/sr towards its other end tracing the feedback from the young star cluster \citep[identified as source I, see, e.g.,][]{Persi2008}. 

The radial profiles across the MFS-warm region, perpendicular to the  VCF~1 crest, present the properties of its immediate environment  (Fig.\,\ref{Radial_profs}-right). 
The effect of compression is visible on the radial column density profile, which is sharper on the side ($r>0$) associated with the stronger and extended PACS 70\,\mum\ emission and larger dust temperatures associated with the feedback from the \hii\ regions. 

In our observations we detected a single VCF, while 2 or 3 velocity components have been fitted towards localised spectra in this region (see Fig.\,\ref{MapVelComp}). The other velocity components have not been, however, identified as elongated and continuous velocity components in {\rev our analysis of the C$^{18}$O($2-1$)  data given the conditions in terms of intensity threshold and minimum velocity separation (cf., Sect.\,\ref{ana2a})}.  ALMA high angular resolution  N$_2$H$^+$ observations, however, resolve at least two elongated  velocity-coherent structures \citep{Shimajiri2019}. 

A large scale velocity gradient of a magnitude of $\sim2$\,\kms\,pc$^{-1}$ is seen over its crest
 (Fig.\,\ref{plotsMeanParam}). 
{\revbis This large scale velocity} gradient has been identified by \citet{Zernickel2013} using HCO$^+$ APEX data and suggested to be tracing longitudinal infall motions along the filament.
{\revbis 
Another  strong velocity gradient of  $\sim8.5$\,\kms\,pc$^{-1}$ across 0.7\,pc towards the location of the source I
is also detected (see red line in Fig.\,\ref{plotsMeanParam}). 
This  strong velocity gradient }
{\rev is associated with  
the loop-like structure  in  PP space joining the straight section of VCF 1 \citep[e.g., the main-filament or the ridge in][]{Andre2016,Arzoumanian2021} to the star-cluster-forming infrared source I (see Fig.\,\ref{MapSkel}) and }
may be tracing the infall of matter onto the massive star-cluster. 
{\rev Such velocity gradients around star forming cores have been reported in a number of studies suggesting to be tracing  gravitational infall \citep[e.g.,][]{Hacar2011,Liu2019,Henshaw2020}.} 
{\revbis
The mass accretion rate of $\dot{M}\sim1.9\times10^{-3}\,$M$_\odot/$yr derived here towards source I is of the order 
as that reported earlier for high-mass protostellar cores
\citep[e.g.,][]{Schneider2010,Peretto2013}.
The  mass accretion rate  $\dot{M}=\rho\,v_{\rm inf}\pi r^2$ is estimated from  the density 
$\rho=\mu_{\rm H_2}m_{\rm H}\nhh/W_{\rm VCF}$ with $\nhh=2\times10^{23}\,$cm$^{-2}$, 
the infall velocity
$v_{\rm inf}= G_{v}L$ with the velocity gradient $G_{v}=8.5$\,\kms\,pc$^{-1}$ over the length $L=0.7\,$pc, and $r=W_{\rm VCF}/2=0.13/2=0.065\,$pc (the cross-section radius of the filament). }\\  

{\bf VCF 32}
 
The VCF~32 corresponds to the MFS-cold region in Fig.\,\ref{MapSkel}. It is located nearby  VCF~1, but with slightly lower column density ($\sim1\times10^{23}\NHUNIT$) and dust temperature ($\sim18$\,K) values. 
The PACS 70\,\mum\ emission along the crest shows values that increase towards the MFS-warm region and there is an excess of the emission (as in the case of MFS-warm) on its northern side (Fig.\,\ref{Radial_profs}-middle) that may result from dust heating from the surrounding \hii\ regions. This northern side also shows a sharper profile ($r>0$) compared to the southern side that may be tracing the effect  
of compression from neighbouring expanding \hii\ regions. 
This VCF also shows an increase of the column density and the velocity dispersion towards  the VCF~1 or the MFS-warm region (the zero offset in Fig.\,\ref{Radial_profs}-middle is the closest to VCF~1), 
where we also notice a change of the sign of the velocity gradient. This VCF has also been suggested  to be undergoing merging with a neighboring filament and onto VCF~1  \citep{Arzoumanian2021}. Higher angular and spectral resolution observations are needed to further study the different components towards VFC~32 and its possible connection/interaction with VCF~1.\\

{\bf VCF 47}

The VCF~47 is identified in the east of the field (indicated as IFS in Figs.\,\ref{ColdensTempMaps} and\,\ref{MapSkel}). VCF~47 and the associated filament system (IFS)  is {\rev located between the two bright high-mass star forming regions NGC~6334 and NGC 6357 and is not  well studied } \citep[identified as the inter-region filaments in][]{Russeil2010}. The mean column density $\sim5\times10^{22}\NHUNIT$,  column density contrast of $\sim3$, and mean dust temperature $\sim16\,$K makes it significantly less dense and colder than the {\rev MFS region. }
The VCF~47 is surrounded by a cold medium not affected by nearby  \hii\ regions (Figs.\,\ref{ColdensTempMaps} and\,\ref{Radial_profs}-left). 
VCF~47 is  star forming as can be traced with the PACS 70\,\mum\ localised peaks towards the young protostellar sources (Fig.\,\ref{plotsMeanParam}-bottom). Localised strong velocity gradients are also observed towards the position of these protostellar sources \citep[with spatial offset between the velocity and column density fluctuations, see][]{Hacar2011} suggesting filament fragmentation and matter infall from the filament onto the compact sources {\rev \citep[see also][showing localised velocity gradients, which do not appear to correspond to over-densities]{ChenMike2020}}. In contrast  to  VCF~1, there is no large scale velocity gradient along the $\sim3.5$\,pc crest of the VCF~47. This difference in the large scale velocity structure of these two VCFs may point to  the  absence of a single dominant potential well dragging the matter  along the filament and to a less evolved stage of the star formation activity and matter accumulation in VCF~47.

\subsection{Position velocity diagrams across selected VCFs }\label{ana4b}

Here we study the velocity structure in the surroundings of  the IFS and the MFS regions, to investigate the influence of the environment on the filament properties.  As presented above, these  two regions are located in different environments, yet part of the same molecular cloud complex. In addition we divided the  MFS region into a warm part and a cold part towards the VCF~1 and  the VCF~32, respectively, as indicated in Fig.\,\ref{MapSkel}.
For these three regions, we derived position-velocity (PV) diagrams perpendicular to the crests of the VCFs and averaged the emission along their crests. 
Figure\,\ref{PV_perpFil12} shows the PV diagrams of the C$^{18}$O($2-1$) and  $^{13}$CO($2-1$) 
emission, across the VCFs and up to a distance of $\sim6\,$pc on either side of their crests. 

The PV diagrams of the three regions and for both C$^{18}$O and  $^{13}$CO  lines show a $\Lambda$- or V-shaped velocity structure (Fig.\,\ref{PV_perpFil12}). 
{\rev This velocity structure can be described as an extended structure bent in velocity space with the filament at the tip of the $\Lambda$ or the V connected physically and in velocity, through a velocity gradient, to the extended sheet-like structure.}
This velocity structure has been first identified by \citet[][]{Arzoumanian2018} towards a low column density filament in the Taurus molecular cloud, and suggested to  trace the formation of a filament from the accumulation of matter flowing within a sheet like structure compressed by a propagating shock front. This velocity structure is  
 consistent with the filament formation scenario in compressed magnetized sheets proposed by \citet{Inoue2018} 
 \citep[see also][]{Inoue2013,Vaidya2013,Abe2021}. {\rev We however note that both the mass and the spatial scales of  the analysed  filaments in \citet[][]{Arzoumanian2018}  and this work are different. In the latter the filament has $M_{\rm line}\sim1\,$M$_{\odot}$\,pc$^{-1}$ and the PV diagram shows the velocity structure across $\sim0.2\,$pc, while the filaments analysed here have $M_{\rm line}>100\,$M$_{\odot}$\,pc$^{-1}$ and the scale of the PV diagram is $\sim10\,$pc. Hence, variations in the PV diagrams at these different scales are expected. 
}
 
  We also notice  the difference in the velocity shape of the PV diagrams of the different regions, 
  with a   $\Lambda$-shape for the IFS and the MFS-warm and a  V-shape for the MFS-cold (identified with the red arrows in Fig.\,\ref{PV_perpFil12}). 
  Since the IFS and the MFS regions are part of the same cloud, the opposite-sided compression scenario is
  {\revbis the only logical explanation.}
  For the IFS and the MFS-warm, the observed $\Lambda$-shape structure suggests that the original velocity of the compressing structure should be blueshifted (in the LSR) with respect to the velocity of the VCF crest {\rev with a velocity gradient connecting both structures}. For the MFS-cold, the observed V-shape structure suggests that the original velocity of the compressing structure should be redshifted (in the LSR) with respect to the velocity of the VCF crest {\rev with a velocity gradient connecting both structures {\revbis (see Fig.\,\ref{sketch} for a schematic representation).}}
  {\rev Longitudinal  PV diagrams along the crest of the MFS-cold region all show   V-shaped velocity structures although the structures vary in space from east to west (Fig.\,\ref{PValongCrest}).}
 
   An extended emission at about $-16$ and $-20$\,\kms\ is seen on the $^{12}$CO($2-1$)  PV diagram in Fig.\,\ref{PVmaps} (hereafter, we will refer to this cloud as the $-20$\,\kms-cloud). Velocity bridge-like  structures connecting the velocities of this cloud and the mean velocity of the MFS and IFS are also  seen.  
   These velocity bridges 
   {\rev have already been reported by \citet[][]{Fukui2018}  towards the  NGC~6334 complex and suggested to be tracing }
    a physical connection and interactions between these  clouds with different mean velocities   {\rev \citep[see also results from  numerical simulations, e.g.,][]{Haworth2015a,Haworth2015b}. }
   This blueshifted $-20$\,\kms-cloud may be the cloud interacting with the NGC~6334 complex at a mean LSR velocity of $-3$\,\kms, and at the origin of the observed $\Lambda$-shape velocity structure.  
     {\rev The causal connection  between the blueshifted cloud and the $\Lambda$-shape velocity structure towards  the analysed filament system is, however, speculative at this stage. 
     }   

   The V-shape velocity structure that we see now towards the MFS-cold  could be the result of  a different (redshifted) compression. 
   {\rev  There is atomic  \hi\ emission at velocities $>0$\,\kms  (Fig.\,\ref{IntSpecta}),}  
   it is, however, 
{\rev not possible to  }  
   identify the compressing redshifted cloud with these data alone.

  \section{Interpretation and discussion}\label{disc}
   
   The NGC~6334 complex is one of the most prominent Galactic high-mass star forming regions at $<2$\,kpc. It has an extent of $\sim50$\,pc parallel to the Galactic plane. The dense gas is observed to be  filamentary. While new generations of stars are forming along these dense gravitationally unstable filaments, the  surrounding warm and diffuse medium is shaped by expanding ionised (\hii) regions (at scales of $\sim1-10$\,pc) generated by the feedback from more evolved massive stars  \citep[][and see Fig.\,\ref{ColdensTempMaps}]{Russeil2016}.
   The asymmetry in the emission (an indication of the total present mass) on either sides of the VCFs crest of the MFS-cold and -warm regions is likely due to the compression of the matter by  expanding  \hii\ region(s) on the northern side of the MFS (Figs.\,\ref{Radial_profs} and \,\ref{PV_perpFil12}-middle and left). 
 
    The $^{12}$CO$(2-1)$  PV diagram  shows the presence of an extended structure at velocities of about  $-20$\,\kms\ (Fig.\,\ref{PVmaps}).   
   The presence of this blueshifted $-20$\,\kms-cloud with velocity bridges connecting the velocities of this latter cloud with the mean velocity of the NGC~6334 complex  {\rev (cf., Sect.\,\ref{ana1})} 
 {\revbis has been interpreted as evidence for}
   the impact of expanding  neutral \hi\ clouds/shells 
 on $\sim100$\,pc scales corresponding to the extent of the NGC~6334 complex and its neighbour NGC~6357 star forming region  \citep{Fukui2018}.
  The origin of this diffuse component around $-20$\,\kms\ is, however, not clear. It could probably originate from the Galactic plane as suggested by \citet[][]{Russeil2016}.   Such a large scale compression may be at the origin of the  coherence in velocity of the  $\sim50\,$pc long molecular NGC~6334 complex. 
  
 {\revbis We suggest that the observed}   large-scale {\rev smoothly varying velocity} from $\sim0$\,\kms\ to $\sim-5,-10$\,\kms\ from east to west of the complex (Fig.\,\ref{PVmaps})  {\revbis may be tracing} a spatial and temporal variation of the collision.  
  {\revbis  This would be expected, for example, when the  plane of the propagating shock front is inclined with respect to the  50\,pc extend of the NGC~6334 cloud} (or over the 100\,pc cloud if the NGC~6357 complex is also considered).
{\rev {\revbis The GM24FS region at $l\,\sim351^{\circ}$, with mean velocities closer to that of the diffuse blueshifted $-20$\,\kms-cloud (see Fig.\,\ref{PVmaps}),} may have experienced the interaction earlier, while the interaction from $l\,\sim351^{\circ}$ to $l\,\sim352\parcd5$  could be more recent and ongoing from west to east.  
  }
  
   We compared the PV diagrams  of the IFS and MFS regions, towards the middle and the eastern side of the field  (see Fig.\,\ref{ColdensTempMaps}), respectively, and noticed a difference in the {\rev bending} of the velocity structure with   V- and $\Lambda$-shaped   velocity structures {\rev in the PV space}. 
   As discussed in Sect.\,\ref{ana4b}, such velocity structures have been {\rev suggested} to be tracing filament formation due to compression. Moreover, here, the different velocity structures of the gas in the surroundings of the IFS and MFS regions suggest that the propagating shock fronts compressing these {\rev regions may be propagating}  in different (possibly opposite) directions. We thus suggest that  these regions have been formed (or at least affected) by compressions, as hinted by the observed velocity pattern, however, these compressions may not be coeval nor from the same origin (see Fig.\,\ref{sketch}). 
   The $^{12}$CO$(2-1)$  PV diagram
   shows velocity bridges between the cloud at $-20$\,\kms and the systemic velocity of the NGC~6334 complex
    (Fig.\,\ref{PVmaps}), suggesting that the compression is 
   currently undergoing and the $\Lambda$-shaped velocity structure observed towards the IFS and the MFS-warm may be  a result of this compression.   
    The    PV diagram of the MFS-cold shows V-shaped velocity structure suggesting that this part of the MFS has probably formed or been impacted by an other episode of compression from a propagating cloud with redshifted velocities (see Fig.\,\ref{sketch}). 
    
     The difference in   {\rev the velocity structure (V- or $\Lambda$-shaped)} towards these two regions of the same cloud points to different origins of the compressions (Fig.\,\ref{sketch}), which can be presently observed. {\rev These observational hints may indicate the importance} %
   of multiple compressions in the formation and evolution of molecular clouds as predicted by theory \citep[e.g.,][]{Inutsuka2015,Iwasaki2019}.
          
   The PV diagrams in Fig.\,\ref{PV_perpFil12} also show  stronger CO emission (a larger mass reservoir) in the surroundings of the MSF region suggesting more available matter to be accreted onto the star-forming filament  compatible with the intense star forming activity in this region. Matter accretion onto the VCF~1 and the hubs of the region are also suggested by the velocity gradients (of about $\sim1-3$\,\kms\,pc$^1$, see Table\,\ref{tab:Tparam}) identified along the VCFs 11, 18, 24, 28, and 40 converging towards the VCF~1. The increase of the velocity dispersion as a function of the column density of the VCFs may also  trace this matter flow along the filaments and onto hubs.
   The magnetic field structure as derived from polarization observations is perpendicular to the VCF~1 and along the {\rev filaments (i.e., VCFs 11, 18, 24, 28, and 40)} connected to the VCF~1 from the side \citep{Arzoumanian2021}, compatible with the filament formation scenario described in Sect.\,\ref{ana4b}.     
   The IFS, on the other hand, shows much less CO emission (smaller mass) in the surrounding of the VCF~47, which has probably accreted already a large fraction of the matter present in the compressed sheet like cloud. The VCF~47 is undergoing star formation traced by the dense fragments and the presence of protostellar sources, but it is less active than its MSF neighbour.  %
   
{ \rev We  analysed the velocity and column density power spectra along the crests of the VCFs, to compare with theoretical models.}
{ \rev While velocity structure function analysis along filaments have been previously analysed \citep{Hacar2016}, }
here, the slopes of the velocity power spectra along filaments are quantified for the first time observationally to directly be compared with theoretical models.  
\citet{Misugi2019} recently found  that  
cylindrical filaments characterized  by a 1D Kolmogorov velocity power spectrum slope of 
$-5/3$  will fragment into cores showing a distribution of  angular momentum compatible with the observations. They demonstrated that this 1D Kolmogorov  velocity fluctuations (with a slope of $-5/3$) along the cylindrical  filaments may be inherited from the 3D Kolmogorov  velocity fluctuations (with a slope of $-11/3$) of the surrounding cloud provided the mass in the filaments is a small fraction of the total mass of the cloud \citep[compatible with the results of][]{Arzoumanian2019}. 
\citet{Inutsuka2001}  has also suggested that filaments with (column) density fluctuations with a power law slope of $-1.5$ (close to $-5/3$) would fragment and form stars with a distribution of mass following the expected \citet{Salpeter1955} power-law slope at the high-mass end. 
{\rev This theoretically suggested slope for column density fluctuations is compatible with  the statistical results {\rev ($-1.6\pm0.3$)} found by \citet{Roy2015} from the (1D) analysis of the column density (mass per unit length) along the crest of a sample of 80 %
filaments observed by $Herschel$ in the Gould Belt.}

 { \rev The mean values of the  slopes $ \alpha$ of the $N_{\rm H_2}$ and the $v$ power spectra  measured here towards the 47 VCFs are about $-2$, with a mean difference between the  slopes of $N_{\rm H_2}$ and  $v$ of $0.15\pm0.99$ for the VCFs with a length $>10$\,pixels and  $0.07\pm0.81$ for the VCFs with a length $>20$\,pixels.} 
The similarity between the observed slopes of  $N_{\rm H_2}$ and  $v$ power spectra suggested by theoretical models and measured here observationally indicates the dynamical coupling of the  $N_{\rm H_2}$ and the $v$ fluctuations along a filament. These fluctuations along the filaments may be inherited from the  turbulent fluctuation of the surrounding ISM {\rev as a result of the filament formation process through compression. These density and velocity fluctuations may} play an important role in the fragmentation of the filaments into star-forming cores.  
Consequently, some of the properties of these cores (e.g., mass, angular momentum) would be inherited from the properties of their host filaments  { \rev \citep[as suggested by][]{Misugi2019}.}
 { \rev About $50\%$ of the measured slopes are close to $-5/3$ (e.g., for the VCF~1 in Fig.\,\ref{PSslopHisto}) and would be compatible with the  theoretical prediction of  \citet{Misugi2019} from the formation of cylindrical filaments out of a more massive cloud with 3D Kolmogorov velocity fluctuations. 
  { \revbis  We plan future  investigations,}  using  higher angular and spectral resolution data and dedicated MHD numerical simulations,    { \revbis to better  understand the origin of the observed distributions of the power spectrum slopes. } 
 } 

\section{Summary and conclusions}\label{Summary}

 In this paper, we presented the velocity structure of the $\sim50$\,pc long NGC~6334 high-mass star-forming region as traced by NANTEN2 $^{12}$CO($2-1$) and APEX $^{13}$CO($2-1$) and C$^{18}$O($2-1$) molecular line emission. 
Our  main results can be summarized as follows:

 \begin{itemize}
\item  This cloud has a coherent velocity structure over its $10\times50$\,pc extent with a smoothly varying velocity field from $\sim0$\,\kms\ to $\sim-5,-10$\,\kms\ from east to west (Fig.\,\ref{PVmaps}).  The C$^{18}$O($2-1$)  emission traces the elongated filamentary structures of the cloud  with column densities $\nhh>1.8\,\times10^{22}\NHUNIT$ as  derived from \herschel\ data. The $^{13}$CO and $^{12}$CO emission show more extended structures towards the cloud   (Fig.\,\ref{mom0maps}). 
\\

\item 
We traced velocity-coherent-filaments (VCF)  in the 3D PPV C$^{18}$O($2-1$) cubes. We identified a total of 75 VCFs in the full studied field. Out of these 75 VCFs, 47  have a length larger than 5 beams (0.9\,pc).  We  analysed the observed properties along these 47 VCFs (Table\,\ref{tab:Tparam}). 
\\

\item 
The mean length of the VCFs is $\sim2$\,pc with some of them as long as $\sim5$\,pc. They span a column density range (as derived from \herschel\ data) of an order of magnitude about a mean value of $\sim4\times10^{22}$\,cm$^{-2}$.
The LOS averaged dust temperature as derived from \herschel\ data indicates a mean value of $\sim20$\,K.
The mean velocity dispersion of the VCFs show an increasing trend as a function of their mean column density with a  linear relation of $\sigma_{v}\propto\,N_{\rm H_2}^{\,0.32}$.   Most of the identified VCFs show longitudinal velocity gradients {\rev along their crests. The filaments with larger velocity gradients tend to also show larger velocity dispersions and mean column densities (Fig.\,\ref{sigmaV_NH2}).}
\\

\item  We measured the column density ($N_{\rm H_2}$) and velocity ($v$)  power spectra of the VCFs. 
We show that the power spectra of  both $N_{\rm H_2}$ and $v$ can  be well represented by a power law function and no characteristic scales are detected  down to the resolution of the data. 
The slopes of both the $N_{\rm H_2}$ and $v$ power spectra for each VCFs are similar. The mean power spectra slope of the full sample is $\alpha\sim-2$. {\rev The mean and the standard deviation of the statistical difference between both quantities are $0.15\pm0.99$ }
 (see Table\,\ref{tab:Tparam}). \\

\item
We described in more details the properties of {\rev three  VCFs (1, 32, and 47)} and we compared their observed properties. 
VCF~1 corresponds to a section of the well studied NGC~6334 main-filament actively forming (high-mass) stars and affected by stellar feedback from young massive stars and stellar clusters. A large scale velocity gradient of a magnitude of $\sim2$\,\kms\,pc$^{-1}$ can be seen over its crest tracing matter flowing along the filament and  infalling  onto the massive star cluster  (Fig.\,\ref{plotsMeanParam}). Other identified VCFs connected to VCF~1 show velocity gradients indicating matter flow onto the VCF~1 and hubs formed along its crest.{\rev VCF~32  shows also signatures of compressions, but is colder on average.}
VCF~47 seems to be colder, younger, and less impacted by  surrounding stellar feedback. 
 \\
 
 \item
 We compared the PV diagrams  of {\rev the three filament systems associated to the VCFs 1, 32, and 47 (IFS, MFS-cold, and MFS-warm, respectively, see Fig.\,\ref{PV_perpFil12}).} The three PV diagrams show V- and $\Lambda$-shaped velocity structures   compatible with the filament formation by compression (Sect.\,\ref{disc} and Fig.\,\ref{sketch}).  
 Moreover, {\rev we suggest that } the presently observed difference in the curvature of the velocity structure {\rev  (i.e., V- and $\Lambda$-shaped)}
towards these filament systems {\rev  may result from  different } origins of the compressions. 
This {\rev may also indicate} the importance of multiple compressions in the formation and evolution of filamentary molecular clouds.

\end{itemize}

The observational results presented in this paper suggest the formation and evolution of filamentary molecular clouds induced by multiple compressions from expanding shells. The origin of these shells is, however, not well constrained.  We anticipate future studies 
to better understand the origin and the impact of these propagating shells on the surrounding medium and on the star formation process.  
{\rev
Quantitative comparisons with dedicated MHD numerical simulations will also be important to better understand the role of compressions from expanding shells in the formation and evolution of filamentary molecular clouds. 
High angular and spectral resolution observations towards this filamentary cloud would also be valuable to better constrain  {\revbis the origin and the variations of the} velocity and density structures along and across these filaments.}

   \begin{table*}[!h] 
    \hspace{-1.cm}  
\centering
{\small
 \caption{Properties of the 47 velocity-coherent-filaments  identified towards  NGC~6334  in the 3D PPV C$^{18}$O($2-1$) cube at the spatial and velocity resolutions of 30\parcs2  and 0.3\,\kms, respectively.  
}      \vspace{-.3cm}  
\begin{tabular}{ccccccccccc}   
\hline\hline   
Crest & $L$  &$N_{\rm H_2}$&$M_{\rm line}$& $ T_{\rm dust}$ & $T^{\rm peak}$ & $v_{\rm los}$ & $\sigma_{v}$ &  $G_{v_{\rm los}}$ & $\alpha_{N_{\rm H_2}}$ & $\alpha_{v_{\rm lsr}}$ \\ 
$\#$ & [pc] &[$10^{22}\,\NHUNIT$] &[M$_{\odot}$\,pc$^{-1}$]& [K] &[K]&[\kms]&[\kms]& [\kms\,pc$^{-1}$] && \\
  (1)&(2)& (3) & (4) &(5) &(6)&(7)&(8)&(9)&(10)& (11)\\
 \hline      
 1&3.7&18.6$\,\pm\,$15.0&507$\,\pm\,$447&24.3$\,\pm\,$3.5&12.7$\,\pm\,$3.4&-3.6$\,\pm\,$2.5&1.1$\,\pm\,$0.5&-2.1$\,\pm\,$0.2&-1.50$\,\pm\,$0.29&-1.66$\,\pm\,$0.15\\
2&2.9&2.7$\,\pm\,$1.5&47$\,\pm\,$24&21.5$\,\pm\,$3.0&6.3$\,\pm\,$2.9&-9.7$\,\pm\,$0.8&0.9$\,\pm\,$0.1&-0.1$\,\pm\,$0.2&-1.76$\,\pm\,$0.47&-2.05$\,\pm\,$1.18\\
3&1.3&2.1$\,\pm\,$0.6&32$\,\pm\,$16&19.3$\,\pm\,$1.0&4.5$\,\pm\,$1.4&-9.9$\,\pm\,$0.9&0.8$\,\pm\,$0.2&1.2$\,\pm\,$0.4&-2.58$\,\pm\,$0.85&-1.29$\,\pm\,$0.50\\
4&3.4&8.0$\,\pm\,$8.6&200$\,\pm\,$231&22.1$\,\pm\,$1.1&6.2$\,\pm\,$2.5&-5.8$\,\pm\,$1.2&1.0$\,\pm\,$0.4&1.0$\,\pm\,$0.1&-2.21$\,\pm\,$0.18&-2.15$\,\pm\,$0.39\\
5&3.9&2.9$\,\pm\,$1.9&54$\,\pm\,$34&18.0$\,\pm\,$0.9&4.4$\,\pm\,$1.4&-8.8$\,\pm\,$1.0&0.6$\,\pm\,$0.3&-0.8$\,\pm\,$0.1&-2.13$\,\pm\,$0.42&-2.33$\,\pm\,$0.24\\
6&1.7&1.4$\,\pm\,$0.3&11$\,\pm\,$19&21.7$\,\pm\,$1.3&4.4$\,\pm\,$1.9&-9.0$\,\pm\,$1.0&0.6$\,\pm\,$0.2&0.7$\,\pm\,$0.2&-3.44$\,\pm\,$0.53&-2.45$\,\pm\,$1.02\\
7&1.1&1.4$\,\pm\,$0.2&11$\,\pm\,$23&24.3$\,\pm\,$3.7&3.5$\,\pm\,$0.6&-9.3$\,\pm\,$0.9&0.9$\,\pm\,$0.4&-0.0$\,\pm\,$0.4&-1.15$\,\pm\,$0.21&0.72$\,\pm\,$0.23\\
8&1.8&1.9$\,\pm\,$0.3&26$\,\pm\,$23&20.4$\,\pm\,$1.5&4.6$\,\pm\,$1.3&-6.5$\,\pm\,$0.6&0.7$\,\pm\,$0.3&-1.0$\,\pm\,$0.1&-1.16$\,\pm\,$0.64&-1.44$\,\pm\,$0.07\\
9&1.3&7.5$\,\pm\,$2.6&185$\,\pm\,$85&22.3$\,\pm\,$1.0&8.4$\,\pm\,$1.8&-4.5$\,\pm\,$1.5&1.0$\,\pm\,$0.2&-1.5$\,\pm\,$1.2&-2.05$\,\pm\,$1.26&-2.08$\,\pm\,$0.08\\
10&2.5&2.0$\,\pm\,$0.4&27$\,\pm\,$20&19.2$\,\pm\,$0.6&4.7$\,\pm\,$1.7&-7.4$\,\pm\,$0.5&0.8$\,\pm\,$0.3&-0.0$\,\pm\,$0.1&-0.78$\,\pm\,$0.37&-1.20$\,\pm\,$0.73\\
11&1.7&8.1$\,\pm\,$3.7&204$\,\pm\,$110&18.7$\,\pm\,$0.5&5.1$\,\pm\,$1.7&-6.6$\,\pm\,$0.6&0.6$\,\pm\,$0.3&1.1$\,\pm\,$0.1&-2.70$\,\pm\,$0.82&-1.98$\,\pm\,$0.40\\
12&1.4&2.0$\,\pm\,$0.4&29$\,\pm\,$20&23.0$\,\pm\,$0.5&3.8$\,\pm\,$0.9&-6.7$\,\pm\,$0.6&0.8$\,\pm\,$0.2&-1.2$\,\pm\,$0.2&-2.16$\,\pm\,$0.59&-1.31$\,\pm\,$0.69\\
13&1.6&2.4$\,\pm\,$0.7&40$\,\pm\,$18&19.8$\,\pm\,$1.0&6.0$\,\pm\,$1.7&-6.5$\,\pm\,$0.2&0.6$\,\pm\,$0.2&-0.2$\,\pm\,$0.1&-1.47$\,\pm\,$0.73&-0.25$\,\pm\,$0.57\\
14&1.4&2.1$\,\pm\,$0.6&31$\,\pm\,$15&17.9$\,\pm\,$0.5&3.9$\,\pm\,$1.0&-5.4$\,\pm\,$0.8&0.5$\,\pm\,$0.1&0.8$\,\pm\,$0.2&-2.78$\,\pm\,$0.56&-3.50$\,\pm\,$1.73\\
15&1.1&1.5$\,\pm\,$0.1&15$\,\pm\,$26&21.0$\,\pm\,$0.9&3.1$\,\pm\,$0.7&-5.7$\,\pm\,$0.5&0.7$\,\pm\,$0.2&0.6$\,\pm\,$0.5&-1.32$\,\pm\,$1.39&-4.80$\,\pm\,$5.61\\
16&1.9&1.9$\,\pm\,$0.3&26$\,\pm\,$23&19.4$\,\pm\,$0.7&3.5$\,\pm\,$1.0&-5.5$\,\pm\,$0.2&0.5$\,\pm\,$0.1&-0.2$\,\pm\,$0.1&-1.68$\,\pm\,$0.29&-0.87$\,\pm\,$0.34\\
17&2.1&2.7$\,\pm\,$0.6&47$\,\pm\,$21&21.9$\,\pm\,$1.9&7.4$\,\pm\,$3.4&-3.2$\,\pm\,$0.9&0.5$\,\pm\,$0.1&-1.3$\,\pm\,$0.1&-1.49$\,\pm\,$0.19&-2.01$\,\pm\,$0.34\\
18&1.6&4.2$\,\pm\,$1.2&91$\,\pm\,$35&22.2$\,\pm\,$1.2&6.0$\,\pm\,$1.6&-5.2$\,\pm\,$0.7&1.3$\,\pm\,$0.4&-1.1$\,\pm\,$0.3&-0.95$\,\pm\,$0.34&-0.56$\,\pm\,$0.61\\
19&3.0&1.9$\,\pm\,$0.5&25$\,\pm\,$16&18.7$\,\pm\,$0.7&3.9$\,\pm\,$0.9&-5.1$\,\pm\,$0.8&0.6$\,\pm\,$0.2&-0.4$\,\pm\,$0.1&-1.13$\,\pm\,$1.14&-1.88$\,\pm\,$0.54\\
20&1.1&1.8$\,\pm\,$0.3&23$\,\pm\,$21&18.9$\,\pm\,$0.7&4.1$\,\pm\,$0.7&-3.7$\,\pm\,$1.0&0.7$\,\pm\,$0.3&-2.9$\,\pm\,$0.4&-2.15$\,\pm\,$0.54&-2.54$\,\pm\,$0.26\\
21&3.2&2.3$\,\pm\,$0.7&36$\,\pm\,$16&20.2$\,\pm\,$0.7&5.9$\,\pm\,$2.2&-4.5$\,\pm\,$0.4&0.5$\,\pm\,$0.2&0.0$\,\pm\,$0.1&-2.66$\,\pm\,$0.74&-1.56$\,\pm\,$0.33\\
22&1.1&2.3$\,\pm\,$0.2&37$\,\pm\,$26&23.5$\,\pm\,$0.3&4.0$\,\pm\,$0.9&-4.9$\,\pm\,$0.3&0.8$\,\pm\,$0.1&-0.1$\,\pm\,$0.3&-1.75$\,\pm\,$1.46&-2.73$\,\pm\,$0.08\\
23&1.0&2.6$\,\pm\,$1.0&46$\,\pm\,$18&21.5$\,\pm\,$0.5&5.3$\,\pm\,$1.4&-3.2$\,\pm\,$0.3&0.5$\,\pm\,$0.3&0.8$\,\pm\,$0.2&-1.18$\,\pm\,$0.71&-2.03$\,\pm\,$1.11\\
24&1.4&3.5$\,\pm\,$1.1&72$\,\pm\,$28&27.8$\,\pm\,$0.6&5.4$\,\pm\,$1.4&-3.6$\,\pm\,$1.5&1.0$\,\pm\,$0.3&-3.2$\,\pm\,$0.5&-2.63$\,\pm\,$0.66&-3.07$\,\pm\,$0.93\\
25&1.1&2.4$\,\pm\,$0.8&40$\,\pm\,$16&18.1$\,\pm\,$1.2&3.3$\,\pm\,$0.4&-4.4$\,\pm\,$0.1&0.6$\,\pm\,$0.1&0.2$\,\pm\,$0.1&-3.45$\,\pm\,$0.45&-3.75$\,\pm\,$1.39\\
26&3.9&3.4$\,\pm\,$1.6&68$\,\pm\,$31&18.9$\,\pm\,$1.3&4.6$\,\pm\,$1.2&-3.6$\,\pm\,$0.4&0.7$\,\pm\,$0.3&0.2$\,\pm\,$0.0&-3.43$\,\pm\,$0.43&-1.42$\,\pm\,$0.46\\
27&1.6&3.3$\,\pm\,$0.6&65$\,\pm\,$27&26.5$\,\pm\,$2.0&6.4$\,\pm\,$2.1&-2.8$\,\pm\,$1.4&1.2$\,\pm\,$0.3&2.6$\,\pm\,$0.3&-1.48$\,\pm\,$1.35&-1.53$\,\pm\,$0.45\\
28&1.5&6.2$\,\pm\,$5.1&150$\,\pm\,$130&19.9$\,\pm\,$0.7&4.2$\,\pm\,$0.6&-3.9$\,\pm\,$0.7&1.3$\,\pm\,$0.9&-1.1$\,\pm\,$0.3&-1.18$\,\pm\,$0.09&-2.57$\,\pm\,$1.35\\
29&3.6&2.6$\,\pm\,$1.2&46$\,\pm\,$19&19.1$\,\pm\,$1.2&4.0$\,\pm\,$1.0&-3.4$\,\pm\,$0.9&0.9$\,\pm\,$0.2&0.2$\,\pm\,$0.1&-1.14$\,\pm\,$0.55&-2.69$\,\pm\,$0.44\\
30&1.4&1.5$\,\pm\,$0.2&15$\,\pm\,$23&18.5$\,\pm\,$0.3&3.2$\,\pm\,$0.9&-3.4$\,\pm\,$0.5&0.3$\,\pm\,$0.4&-1.0$\,\pm\,$0.1&-2.23$\,\pm\,$1.48&-1.42$\,\pm\,$0.87\\
31&1.4&15.3$\,\pm\,$8.0&410$\,\pm\,$255&25.9$\,\pm\,$1.7&14.9$\,\pm\,$4.7&-3.5$\,\pm\,$1.3&1.1$\,\pm\,$0.2&-2.5$\,\pm\,$0.5&-0.81$\,\pm\,$1.13&-1.85$\,\pm\,$0.32\\
32&5.6&10.6$\,\pm\,$11.5&275$\,\pm\,$321&18.5$\,\pm\,$0.7&6.6$\,\pm\,$2.8&-3.3$\,\pm\,$0.6&0.9$\,\pm\,$0.6&0.0$\,\pm\,$0.0&-2.08$\,\pm\,$0.25&-3.16$\,\pm\,$0.41\\
33&1.4&2.0$\,\pm\,$0.3&30$\,\pm\,$23&18.5$\,\pm\,$0.3&3.7$\,\pm\,$0.6&-3.5$\,\pm\,$0.6&0.6$\,\pm\,$0.1&-1.5$\,\pm\,$0.2&-1.06$\,\pm\,$0.57&-3.09$\,\pm\,$0.81\\
34&1.7&3.4$\,\pm\,$0.8&68$\,\pm\,$27&16.4$\,\pm\,$0.4&3.7$\,\pm\,$0.9&-2.7$\,\pm\,$0.7&0.7$\,\pm\,$0.3&-1.2$\,\pm\,$0.1&-2.08$\,\pm\,$1.37&-2.01$\,\pm\,$0.63\\
35&3.7&2.1$\,\pm\,$0.4&32$\,\pm\,$20&18.0$\,\pm\,$0.4&4.1$\,\pm\,$0.9&-3.7$\,\pm\,$0.2&0.4$\,\pm\,$0.3&0.0$\,\pm\,$0.0&-2.15$\,\pm\,$0.40&-1.02$\,\pm\,$0.44\\
36&1.8&1.4$\,\pm\,$0.2&11$\,\pm\,$22&20.5$\,\pm\,$0.6&3.6$\,\pm\,$0.8&-2.9$\,\pm\,$0.5&0.5$\,\pm\,$0.2&-0.8$\,\pm\,$0.1&-1.78$\,\pm\,$0.48&-2.52$\,\pm\,$0.30\\
37&3.1&1.9$\,\pm\,$0.3&27$\,\pm\,$22&19.6$\,\pm\,$1.2&5.3$\,\pm\,$1.1&-3.1$\,\pm\,$0.4&0.4$\,\pm\,$0.2&0.4$\,\pm\,$0.0&-2.05$\,\pm\,$0.56&-1.83$\,\pm\,$0.32\\
38&4.1&3.4$\,\pm\,$1.7&69$\,\pm\,$32&17.5$\,\pm\,$0.8&5.2$\,\pm\,$3.3&-3.1$\,\pm\,$0.4&0.4$\,\pm\,$0.2&-0.3$\,\pm\,$0.0&-2.25$\,\pm\,$0.52&-1.91$\,\pm\,$0.29\\
39&3.2&2.4$\,\pm\,$1.0&40$\,\pm\,$15&19.6$\,\pm\,$1.1&4.8$\,\pm\,$1.4&-2.6$\,\pm\,$0.5&0.6$\,\pm\,$0.3&-0.5$\,\pm\,$0.0&-2.09$\,\pm\,$0.33&-1.84$\,\pm\,$0.33\\
40&1.0&2.3$\,\pm\,$0.5&36$\,\pm\,$20&24.9$\,\pm\,$0.5&5.0$\,\pm\,$1.6&-3.0$\,\pm\,$0.3&0.7$\,\pm\,$0.1&0.6$\,\pm\,$0.2&-0.77$\,\pm\,$1.20&-0.97$\,\pm\,$3.27\\
41&1.1&6.2$\,\pm\,$0.9&148$\,\pm\,$57&28.7$\,\pm\,$1.2&12.2$\,\pm\,$3.5&-2.1$\,\pm\,$0.5&0.9$\,\pm\,$0.2&1.6$\,\pm\,$0.2&-1.71$\,\pm\,$0.41&$-$\\ 
42&2.6&1.7$\,\pm\,$0.4&20$\,\pm\,$20&21.1$\,\pm\,$2.5&4.5$\,\pm\,$1.6&-1.9$\,\pm\,$0.4&0.6$\,\pm\,$0.2&-0.1$\,\pm\,$0.1&-1.77$\,\pm\,$0.47&-2.64$\,\pm\,$0.26\\
43&5.0&3.1$\,\pm\,$1.1&59$\,\pm\,$23&17.8$\,\pm\,$1.4&4.6$\,\pm\,$1.3&-1.5$\,\pm\,$0.9&0.7$\,\pm\,$0.5&0.4$\,\pm\,$0.1&-1.68$\,\pm\,$0.20&-2.08$\,\pm\,$0.27\\
44&1.0&2.3$\,\pm\,$0.2&36$\,\pm\,$27&18.1$\,\pm\,$0.4&3.3$\,\pm\,$0.6&-1.4$\,\pm\,$0.3&0.3$\,\pm\,$0.1&-1.1$\,\pm\,$0.1&-1.37$\,\pm\,$0.15&-1.55$\,\pm\,$0.69\\
45&2.3&2.7$\,\pm\,$0.5&49$\,\pm\,$23&19.2$\,\pm\,$0.3&6.2$\,\pm\,$2.0&-0.9$\,\pm\,$0.9&0.6$\,\pm\,$0.2&-0.7$\,\pm\,$0.2&-1.36$\,\pm\,$1.01&-1.74$\,\pm\,$0.69\\
46&1.3&2.9$\,\pm\,$0.6&55$\,\pm\,$24&18.3$\,\pm\,$0.2&4.0$\,\pm\,$1.2&-1.0$\,\pm\,$0.2&0.5$\,\pm\,$0.1&-0.3$\,\pm\,$0.2&-1.90$\,\pm\,$0.24&-2.54$\,\pm\,$0.49\\
47&3.5&5.3$\,\pm\,$2.0&124$\,\pm\,$56&15.6$\,\pm\,$0.8&3.8$\,\pm\,$0.7&1.2$\,\pm\,$0.7&1.0$\,\pm\,$0.3&-0.2$\,\pm\,$0.1&-2.58$\,\pm\,$0.64&-2.48$\,\pm\,$0.56\\
 \hline
Range &1.0$\,/\,$5.6&1.4$\,/\,$18.6&11$\,/\,$507&15.6$\,/\,$28.7&3.1$\,/\,$14.9&-9.9$\,/\,$1.2&0.3$\,/\,$1.3&-3.2$\,/\,$2.6&-3.4$\,/\,$-0.8&-4.8$\,/\,$0.7\\
Mean &2.2&3.8&79&20.6&5.3&-4.4&0.7&-0.3&-1.9&-2.0\\
Stdev &1.2&3.5&99&2.9&2.4&2.4&0.3&1.1&0.7&0.9\\
 \hline  \hline     
                  \end{tabular}
\vspace{-.3cm}\hspace{-.3cm}
\begin{list}{}{}
 \item[]{{\bf Notes:} 
Columns 3 to 8 give the mean and standard deviation of the values  
along the VCF crests (see Sect.\,\ref{ana2a}). Columns 9 to 11 give the slopes of  linear fits and the associated errors. 
 Columns: (1) Crest number. (2) Crest length. The VCFs of this sample are selected for $L>0.9$\,pc or $L>10$\,pixels. (3) Column density derived from \herschel\ observations. 
{\rev (4) Mass per unit length calculated using the relation: $M_{\rm line}=\mu_{\rm H_2}m_{\rm H}  \nhh^{\rm bs} \times W_{\rm fil}$, where  $W_{\rm fil}=0.13$\,pc is the filament width,  
$\mu_{\rm H_2}=2.8$  the mean molecular weight per hydrogen molecule,  $m_{\rm H}$ the mass of a hydrogen atom, and  $\nhh^{\rm bs}=\nhh-1\times10^{22}\,\NHUNIT$, with \nhh are the values given in Column (3) and $1\times10^{22}\,\NHUNIT$ is the mean value of the local background. 
}
 (5) LOS averaged dust temperature derived from \herschel\ observations. 
 ($6-8$) Mean and standard deviation of the C$^{18}$O($2-1$)  line peak temperature ($T^{\rm peak}$),  centroid velocity ($v_{\rm los}$), and  velocity dispersion ($\sigma_{v}$) {\rev derived from the multi-velocity Gaussian fit analysis}  
(see Sect.\,\ref{ana2a}). (9) Velocity gradient along the VCFs. (10) Slope of the column density power spectrum. 
  (11) Slope of the velocity power spectrum.
   The bottom three rows  give the observed range (minimum / 
maximum), the mean value, and the standard deviation (Stdev) of each property.
   }
 \end{list}      }
 \label{tab:Tparam}    
  \end{table*}

 \begin{acknowledgements}
 {\rev We thank the anonymous referee for their detailed report, which helped improving the presentation of our results. }
 DA thanks M. S. N. Kumar for motivating discussions on hub-filament systems. 
 We thank Y. Fukui for sharing the NANTEN2 data and for insightful discussions on the velocity structure of the region.  
 AZ thanks the support of the Institut Universitaire de France.
 We thank F. Wyrowski and the APEX staff for carrying out the observations and A. Zernickel for making the reduced data available.
 This research has made use of data from the Herschel HOBYS project
(http://hobys-herschel.cea.fr). HOBYS is a Herschel Key Project jointly
carried out by SPIRE Specialist Astronomy Group 3 (SAG3), scientists of the
LAM laboratory in Marseille, and scientists of the Herschel Science Center
(HSC).
 The present study has also made use of NANTEN2 data. NANTEN2 is an international collaboration of ten universities: Nagoya University, Osaka Prefecture University, University of Cologne, University of Bonn, Seoul National University, University of Chile, University of New South Wales, Macquarie University, University of Sydney, and Zurich Technical University. 
 This research has made use of the SIMBAD database, operated at CDS, Strasbourg, France.
 \end{acknowledgements}

\bibliographystyle{aa}
\bibliography{NGC6334_accepted_arxiv} 


\begin{appendix}
\section{Ratio maps and optical depth }\label{App1}
   
\begin{figure*}[!h]
   \centering
     \resizebox{19.cm}{!}{
\includegraphics[angle=0]{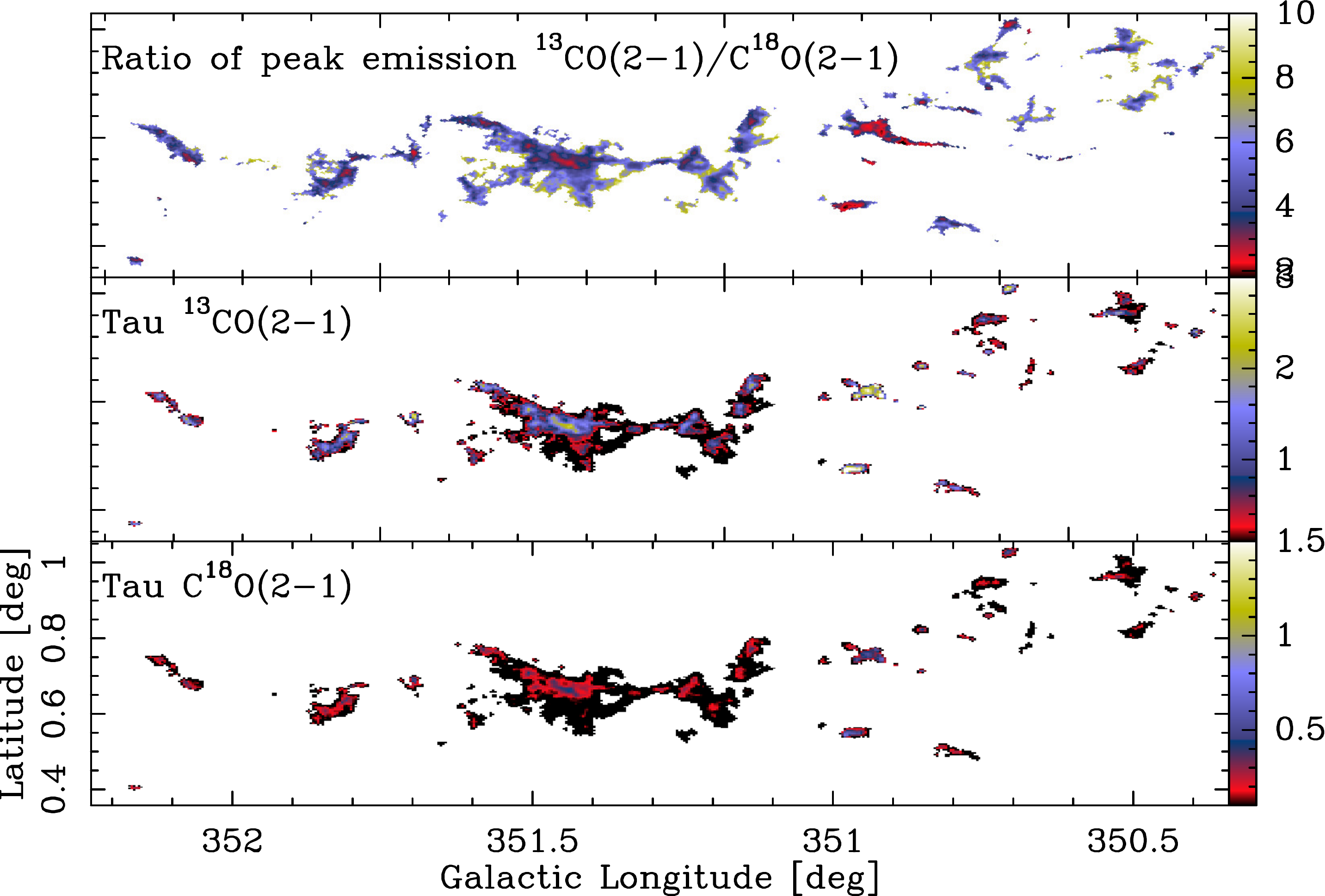}
\includegraphics[angle=0]{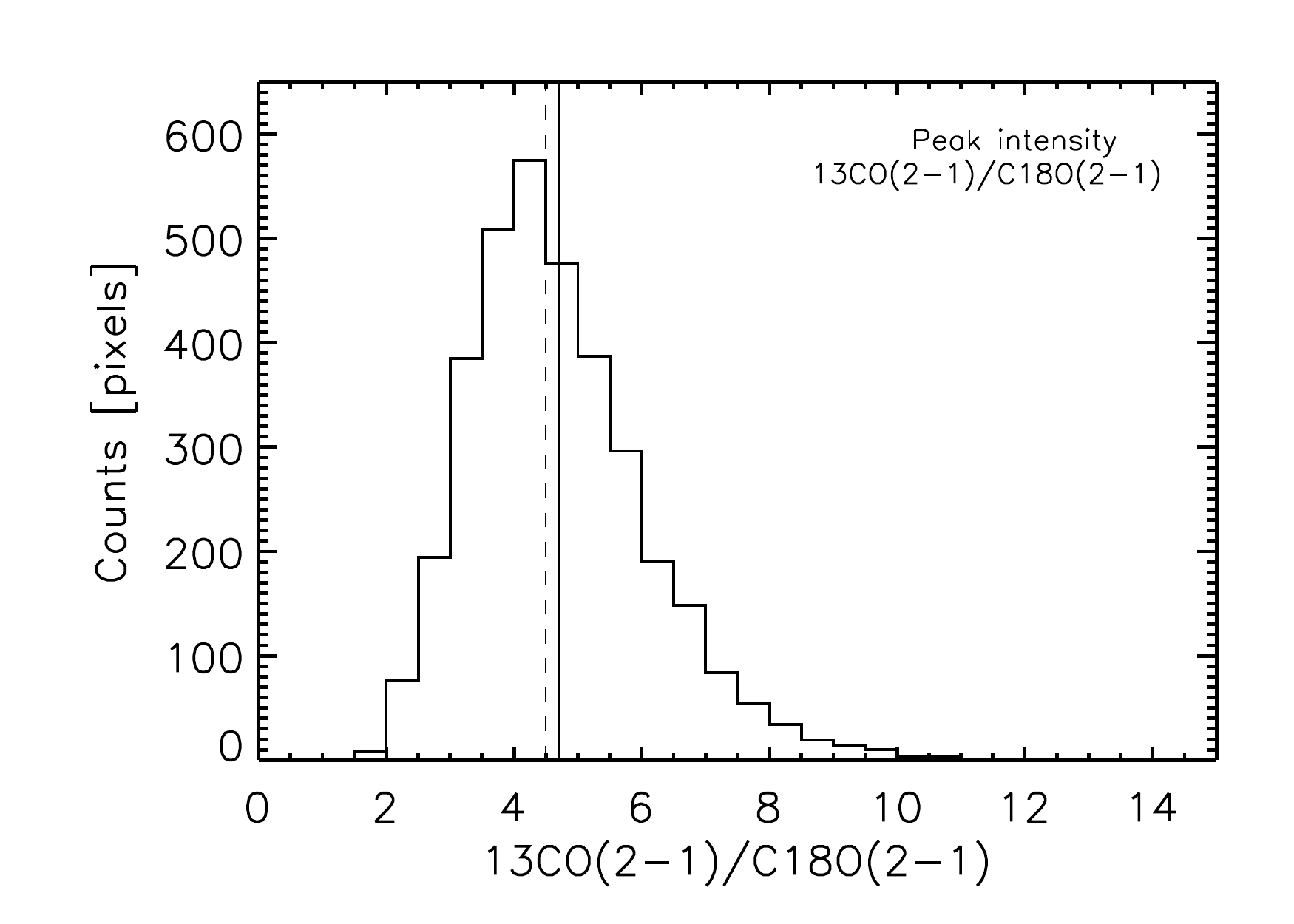}
\includegraphics[angle=0]{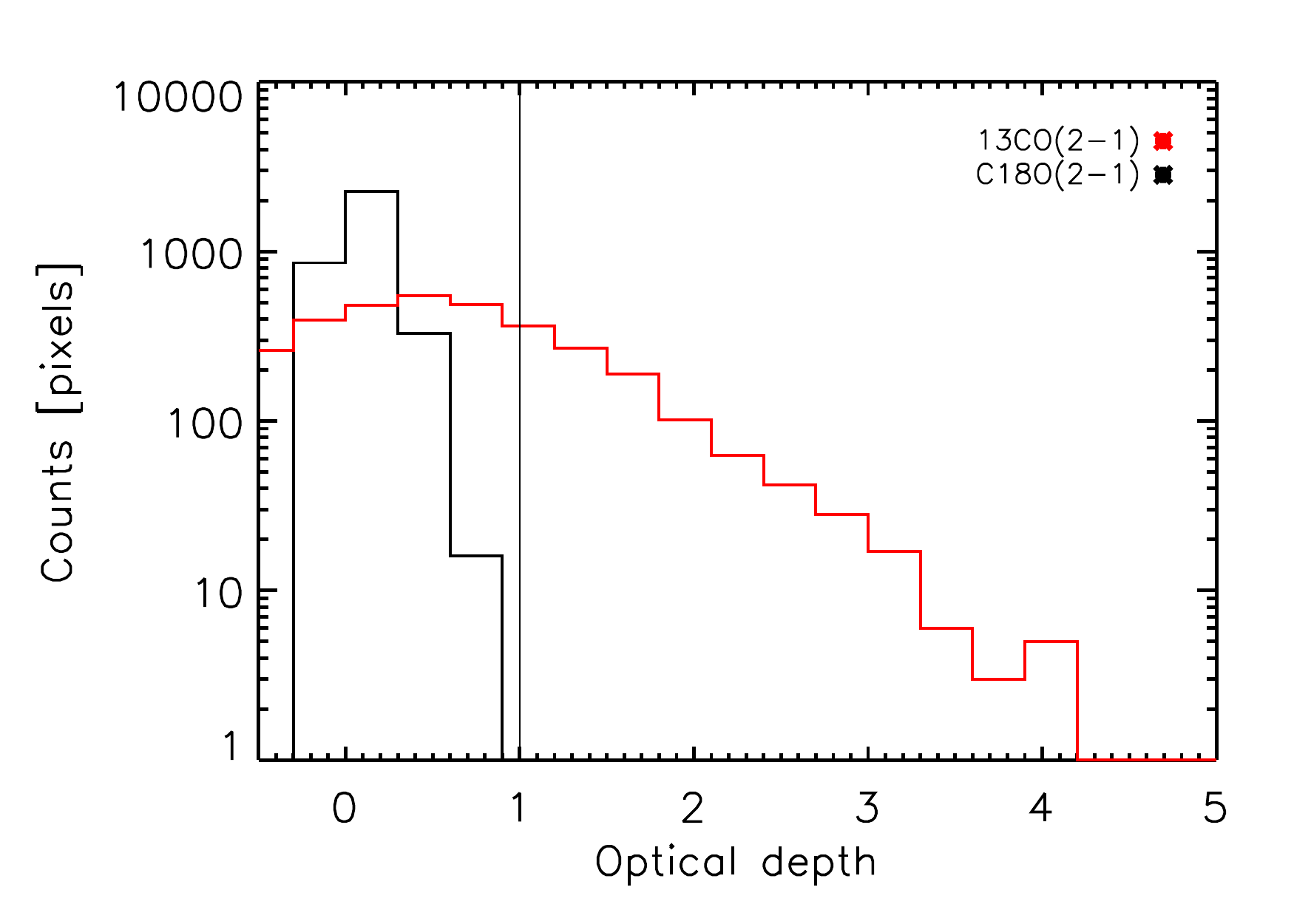}
}
\vspace{-0.3cm}
  \caption{{ \it Left:} Ratio map $R_{13/18}=T^{^{13}{\rm CO}}/T^{{\rm C}^{18}{\rm O}}$, $\tau^{13}$, and $\tau^{18}$, from top to bottom.
  The ratio map is derived from the peak emission at the spatial resolution of 30\arcsec. Only pixels 
with $T^{18}>3\sigma$  have been considered, where $\sigma=0.4$\,K. 
{ \it Middle:} Distribution of the values of the ratio map. 
{ \it Right:} Distribution of the optical depth of  $^{13}$CO$(2-1)$ and C$^{18}$O$(2-1)$ emission shown in red and black, respectively. The C$^{18}$O$(2-1)$ emission is optically thin over the entire studied region.
}          
  \label{RatioOpacity}
    \end{figure*}

 Figure\,\ref{RatioOpacity}-left,  {\rev shows the $R_{13/18}=T^{^{13}{\rm CO}}/T^{{\rm C}^{18}{\rm O}}$ ratio map of the  peak intensity observed over the LSR velocity range $-12$ to 4\,kms$^{-1}$,  and the estimated  mean optical depth values $\tau^{13}$ and $\tau^{18}$ of the $^{13}$CO and the C$^{18}$O lines, respectively. 
The mean optical depth values of the  lines are derived  from the  $R_{13/18}$ ratio  by solving the following relation 
 \begin{equation}
R_{13/18} =   \frac{ T_{\rm ex}^{13}[1-\exp(-\tau^{13})]}{ T_{\rm ex}^{18}[1-\exp(-\tau^{18})]}\,\,,
\label{tau}
\end{equation}
on a pixel by pixel basis \citep[see also, e.g.,][]{Arzoumanian2013,Zernickel2015}. In Eq.\,\ref{tau},
  $\tau^{13} = X  \tau^{18}$ , with $X = [^{13}\rm CO]/[\rm C^{18}\rm O] = 5.5$  \citep[the mean value of the abundance ratio  in the local ISM, e.g.][]{Wilson1994} and we assume a uniform excitation temperature ($T_{\rm ex}^{13}=T_{\rm ex}^{18}$) along the line of sight for both isotopomers.

}

As shown in Fig.\,\ref{RatioOpacity}, the C$^{18}$O$(2-1)$ emission is optically thin all over the observed field. The $^{13}$CO$(2-1)$ emission is mostly optically thin but shows optical depth values up to $\sim4$ in the densest regions. 
The $^{12}$CO$(2-1)$ emission is optically thick. 
We thus use the C$^{18}$O$(2-1)$ PPV cube to identify the velocity-coherent-filaments.

   \section{Multi-velocity component fitting results}\label{App2}
 
\begin{figure*}[!h]
   \centering
     \resizebox{19.cm}{!}{
             \hspace{-2.cm}
\includegraphics[angle=0]{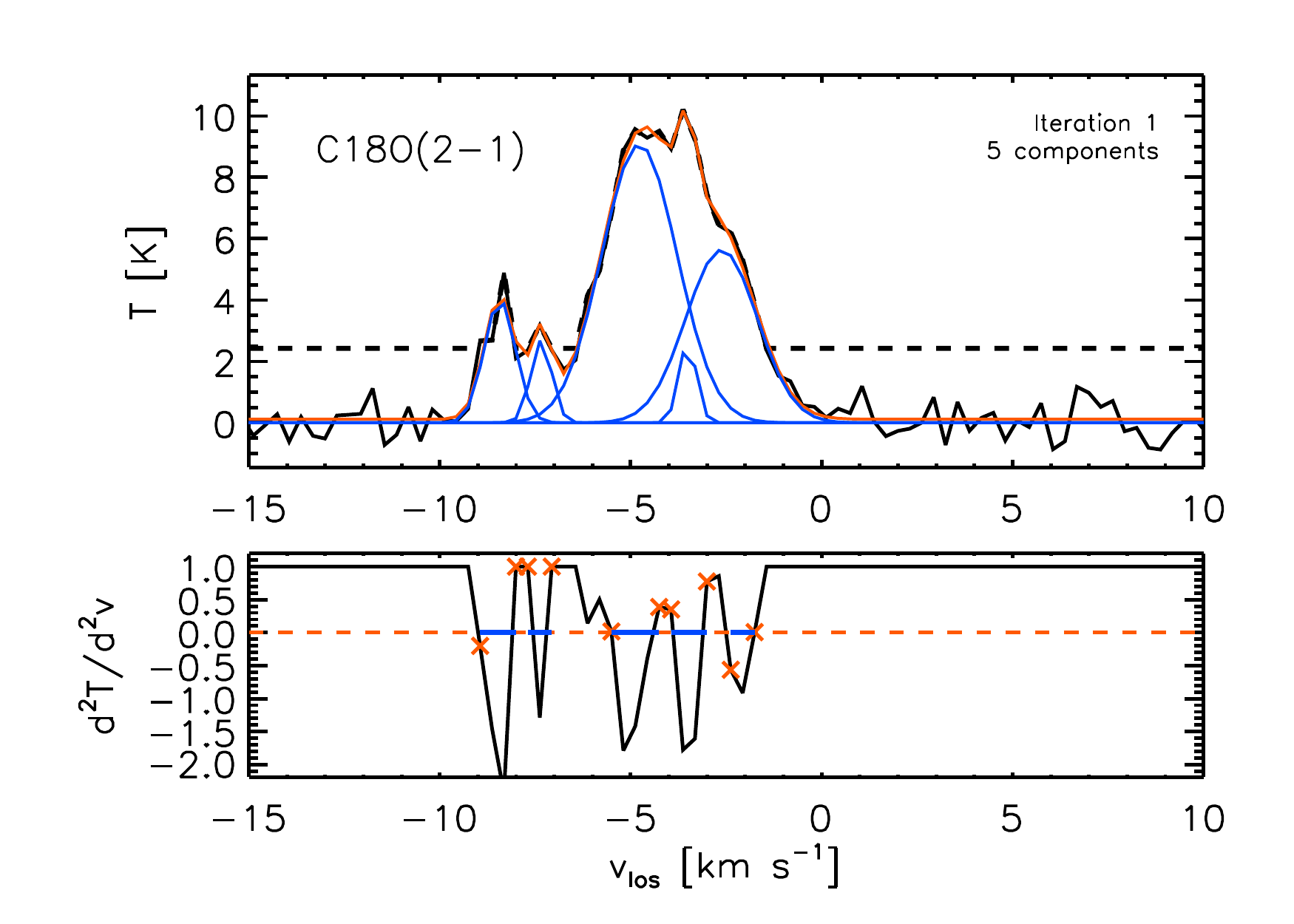}
\hspace{-2.cm}
\includegraphics[angle=0]{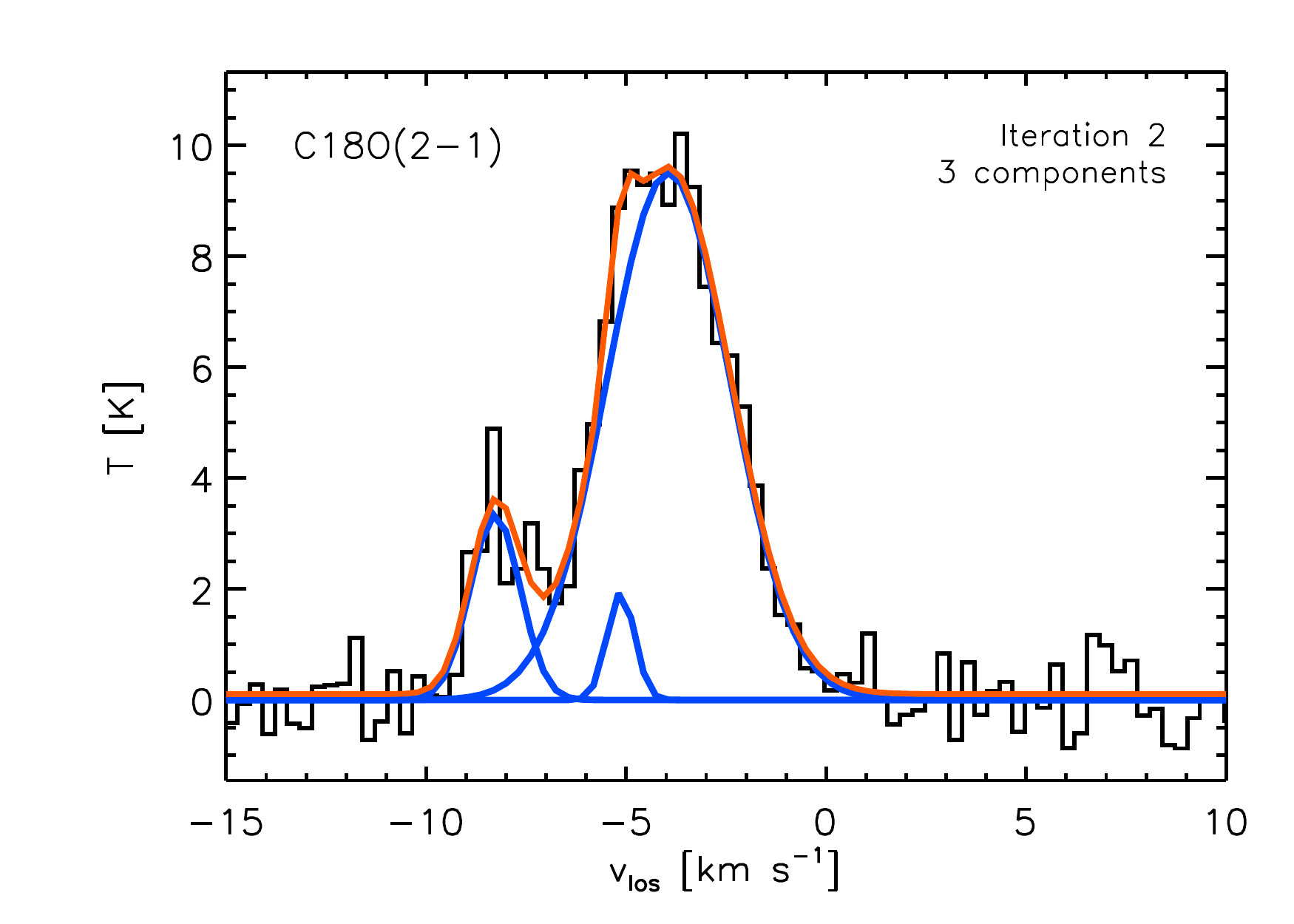}
\hspace{-2.cm}
\includegraphics[angle=0]{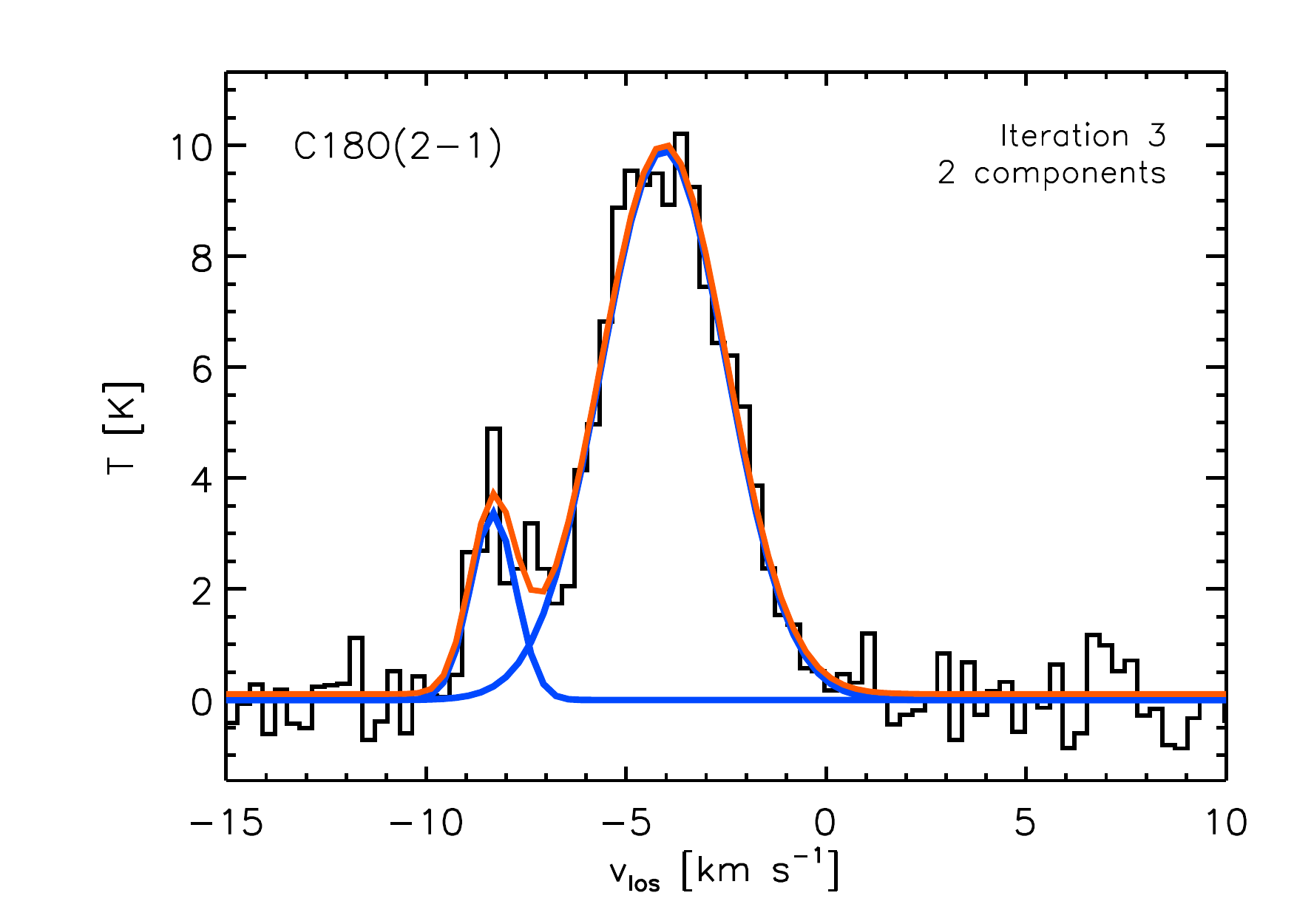}
}
\vspace{-0.3cm}
  \caption{{\rev
  { \it Left-top:} Observed C$^{18}$O$(2-1)$ spectra towards a position in the centre of the map. The blue and red curves show the five individual Gaussian functions and the sum of all the fitted Gaussians, respectively. The dotted line shows the intensity threshold corresponding to $S/N=4\sigma$. Here $\sigma=0.6$\,K.
   { \it Left-bottom:} The second derivative of the C$^{18}$O$(2-1)$ spectra shown on the top panel. The blue horizontal lines show the region where the first guess for each of the Gaussian fits is identified indicated by the red symbols defining the points surrounding a peak to be fitted.  
     { \it Middle:} The second iteration of the Gaussian fits where the Gaussian component not following the given criterium on the velocity resolution ($5\times0.3$\,\kms) are discarded. A new multi-component fit is done on the remaining positions. The results is a combined Gaussian with three components. 
         { \it Right:} The third iteration of the Gaussian fits following the same procedure of the second iteration. 
  The results is a combined Gaussian with two components. Both components follow the defined criteria and correspond to the final result of the $n$-Gaussian fit.  
}          
}
  \label{Spectra_Gfit}
    \end{figure*}

 {\rev We use  an automated procedure to identify multiple velocity peaks along each spectra to be fitted with a multi-Gaussian function as described in Sect.\,\ref{ana2a}.
We use the first and second derivatives to find the surroundings of the peaks to be fitted, where the absolute value of the first derivative reaches a maximum and the second derivative changes sign.  
Figure\,\ref{Spectra_Gfit} shows an example of a spectra fitted with two velocity components. }

\begin{figure*}[!h]
   \centering
     \resizebox{19.cm}{!}{
     \hspace{-0.3cm}
\includegraphics[angle=0]{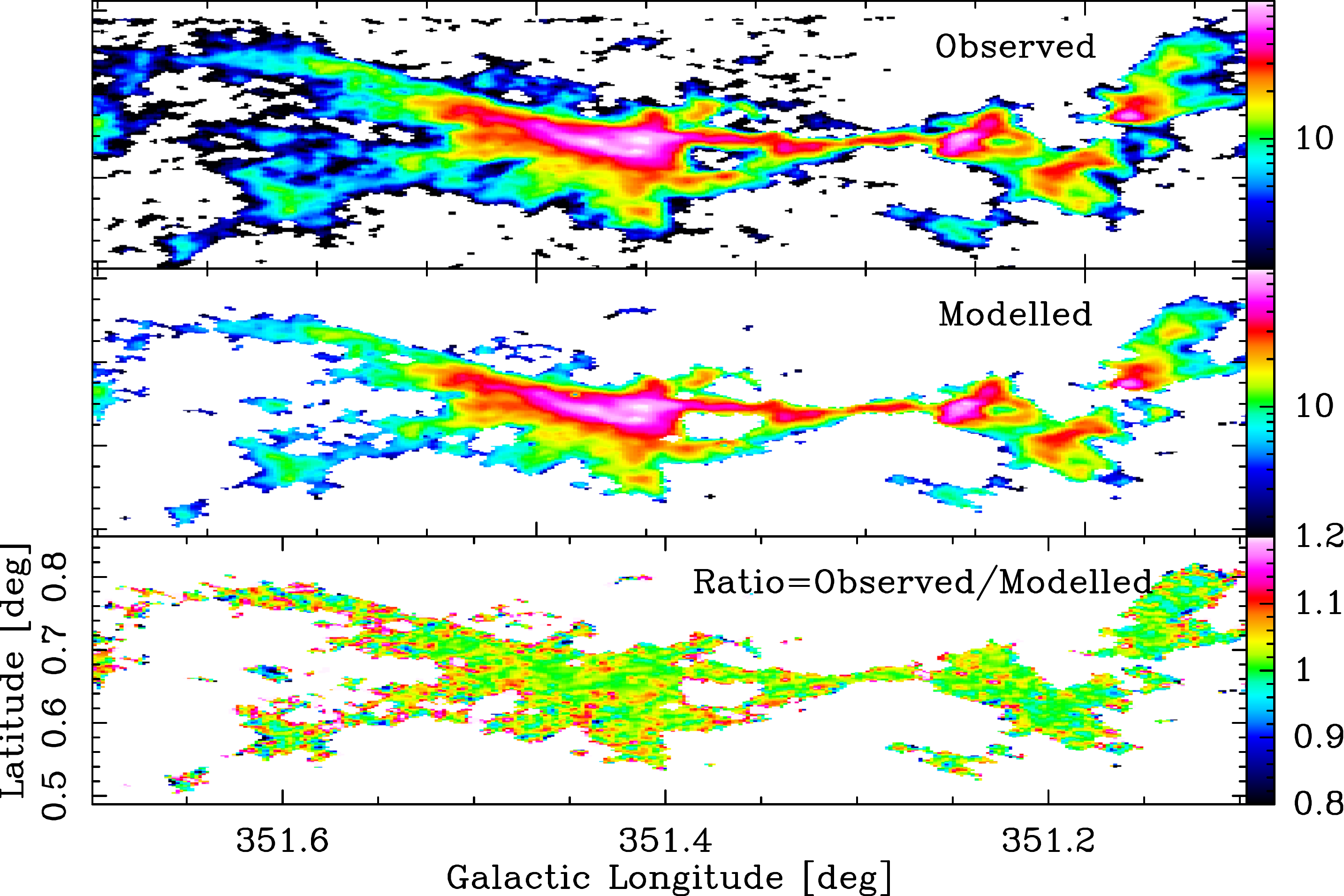}
\hspace{0.4cm}
\includegraphics[angle=0]{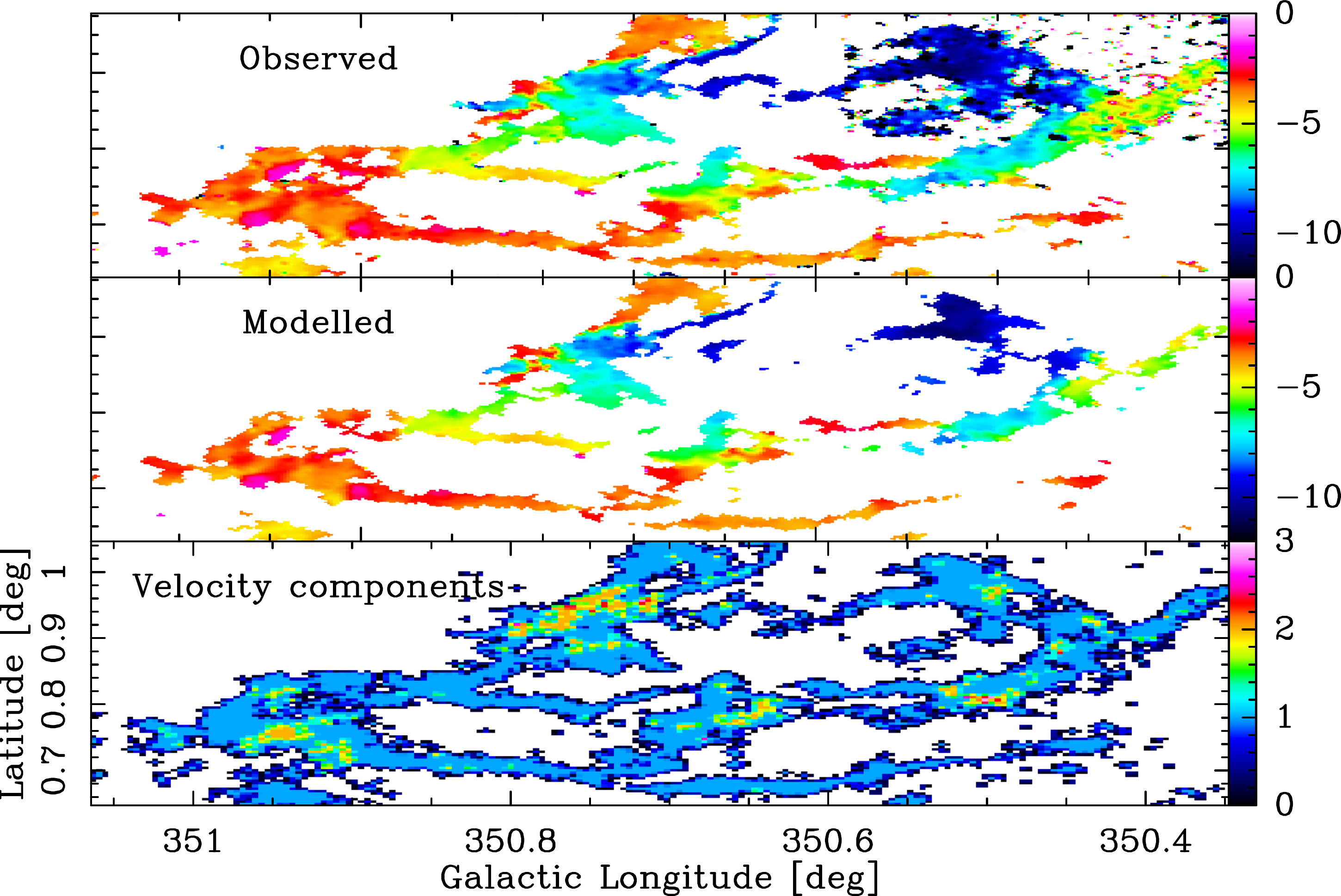}
}
  \caption{{ \it Left:} Observed and modelled moment zero map of the C$^{18}$O$(2-1)$ emission towards the MFS, on the top and bottom, respectively. The intensity is shown in K\,\kms. The bottom map shows the ratio of the above two maps. The median, mean, and standard deviation  values  of this latter ratio map are $1.02$, $1.10$, and $0.46$, resepctively.
{ \it Right:} Observed and modelled moment one map of the C$^{18}$O$(2-1)$ emission towards the GM24FS, on the top and bottom, respectively. The unit of the maps is \kms. The bottom map shows the number of velocity components fitted for each pixel (towards each line-of-sight).
}          
  \label{ObsModelmom0mom1}
    \end{figure*}

Figure\,\ref{ObsModelmom0mom1} shows the velocity integrated intensity maps  derived from the 
observed ($T_{\rm MB}(x,y,v)$) and modelled ($T_{\rm MB}^{\rm model}(x,y,v)$)  C$^{18}$O$(2-1)$ cubes towards the middle and west-side regions of the field, i.e., the MFS and GM24FS regions. The median, mean, and standard deviation  values  of the ratio map $R(x,y,v)=T_{\rm MB}(x,y,v)/T_{\rm MB}^{\rm model}(x,y,v)$ are $1.02$, $1.10$, and $0.46$, respectively, {\revbis for an intensity threshold of $S/N=4$}. 
{\revbis We also derived modelled cubes for $S/N=3$ and 5. The mean and standard deviation  values  of the ratio maps are $1.16$ and $0.75$  for the $S/N=3$ run, and $1.08$ and $0.37$ for the $S/N=5$ run, respectively. The standard deviation of the ratio map for the $S/N=3$ run is larger than the runs with $S/N=4$ and 5.  The run with $S/N=5$ has $\sim30\,\%$ less fitted spectra compared to the run with $S/N=4$. For the analysis presented in this paper, we thus selected the modelled cube derived with $S/N=4$.} 

Figure\,\ref{ObsModelmom0mom1} also shows a map of the  number of velocity components fitted for each pixel (towards each line-of-sight). More than $88\%$ of the spectra are fitted with a single velocity component and only $\sim12\%$ of the spectra required two or three velocity components to describe the observations.

    \begin{figure*}[!h]
   \centering
     \resizebox{19cm}{!}{
     \hspace{-1.cm}
    \includegraphics[angle=0]{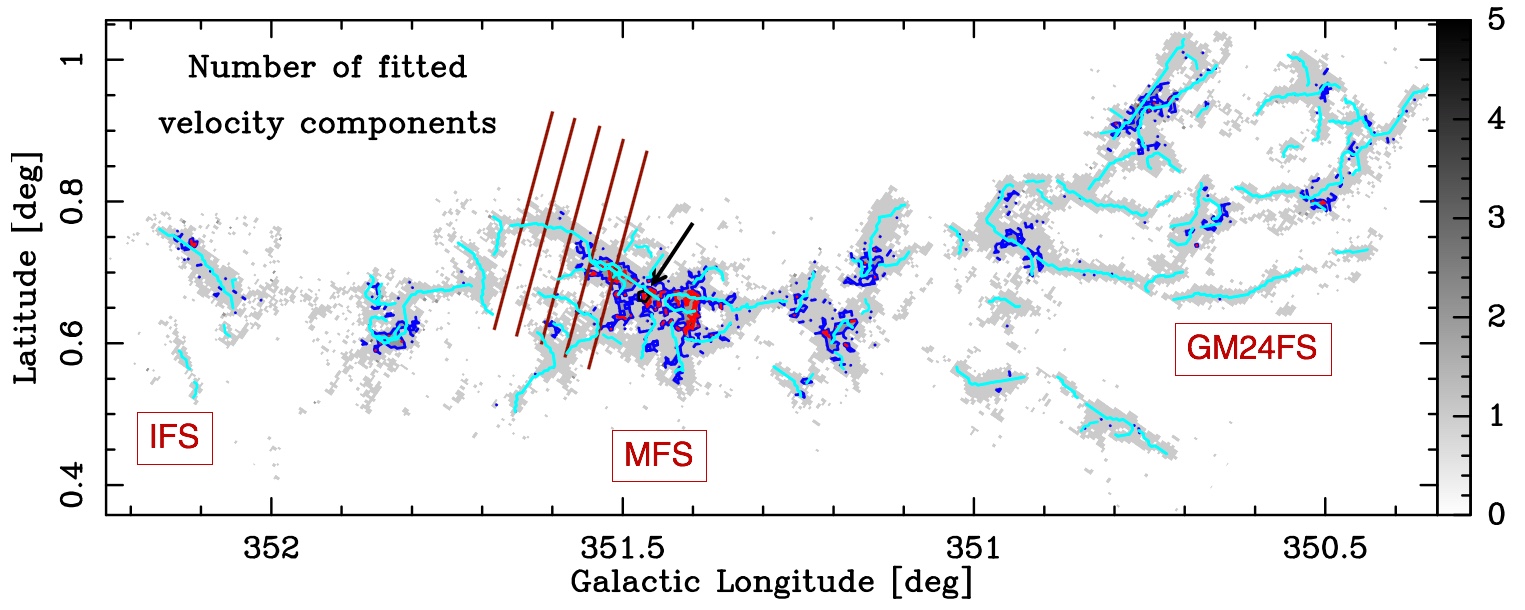}}
  \caption{Map showing the number of fitted velocity components for each C$^{18}$O($2-1$) spectra. 
  { \rev The blue and red contours indicate velocity components $\ge 2$ and $\ge 3$ respectively. }
  More than $88\%$ of the spectra are fitted with a single velocity component and only $\sim12\%$ of the spectra required two or three velocity components to fit the observed spectra.  The  { \rev cyan} curves trace the crests of the the identified velocity coherent filaments (same as in Fig.\,\ref{MapSkel}).  
  { \rev  The black arrow towards the MFS indicates the location of the spectrum shown in Fig.\,\ref{Spectra_Gfit}.
  The red lines indicate the central positions of the slices used to derive the PV diagrams shown in Fig.\,\ref{PValongCrest}.
  }
}          
  \label{MapVelComp}
    \end{figure*}

\section{PV diagrams along the VCF crest}\label{App3}
{\rev
We here show longitudinal  position$-$velocity (PV) diagrams along the crest of VCF~32 for the $^{13}$CO($2-1$) and C$^{18}$O($2-1$) 
emission from east to west. 
The PV diagrams are  perpendicular to the crest and averaged over 4~pixels  (2$\times$beams). 
}
 
\begin{figure*}[!h]
   \centering
    \resizebox{19cm}{!}{   
                                              \includegraphics[angle=0]{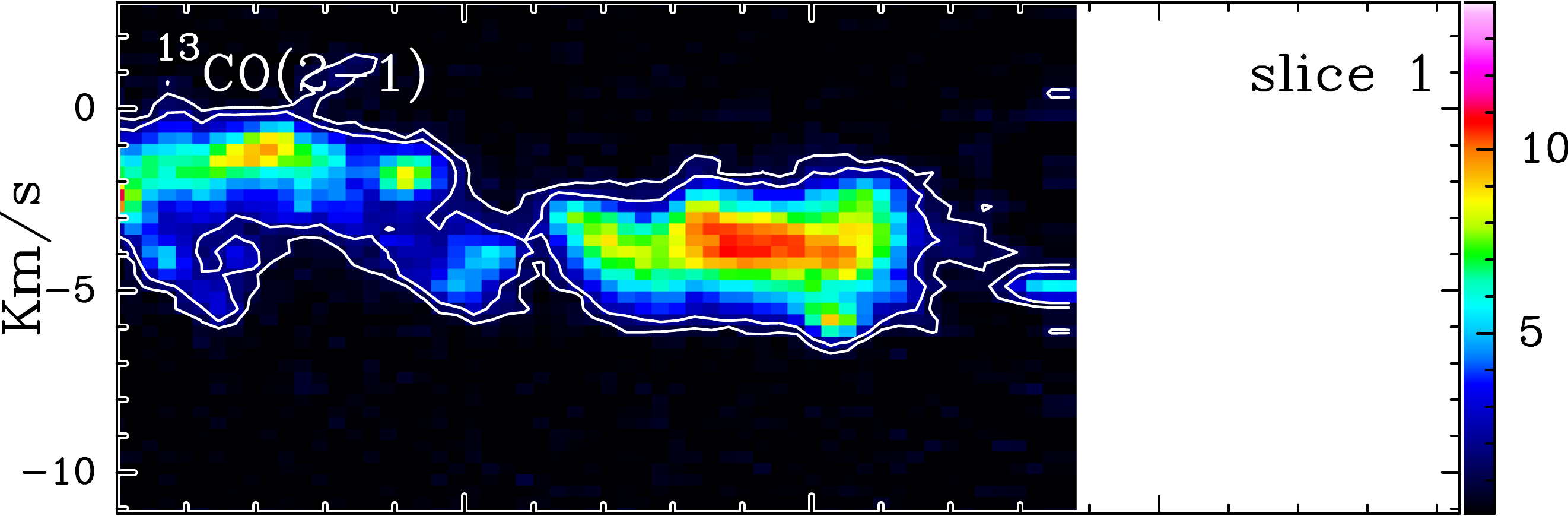}
                                                  \includegraphics[angle=0]{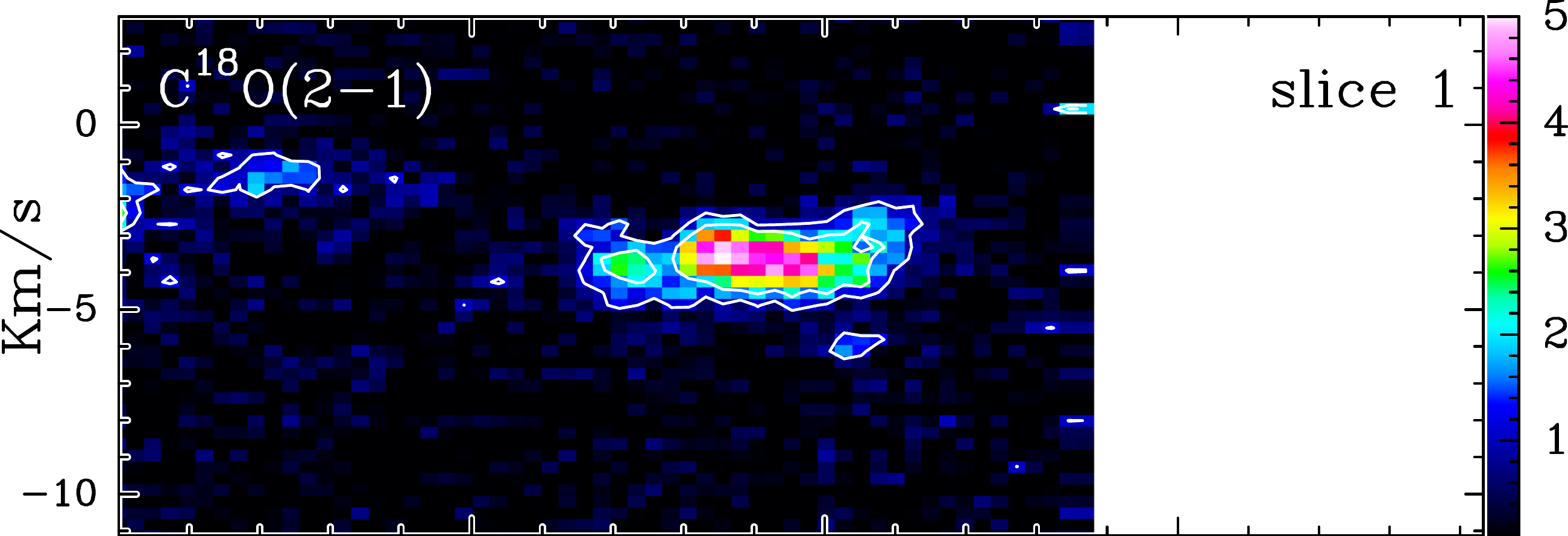}
                            }
                      \resizebox{19cm}{!}{            
                                              \includegraphics[angle=0]{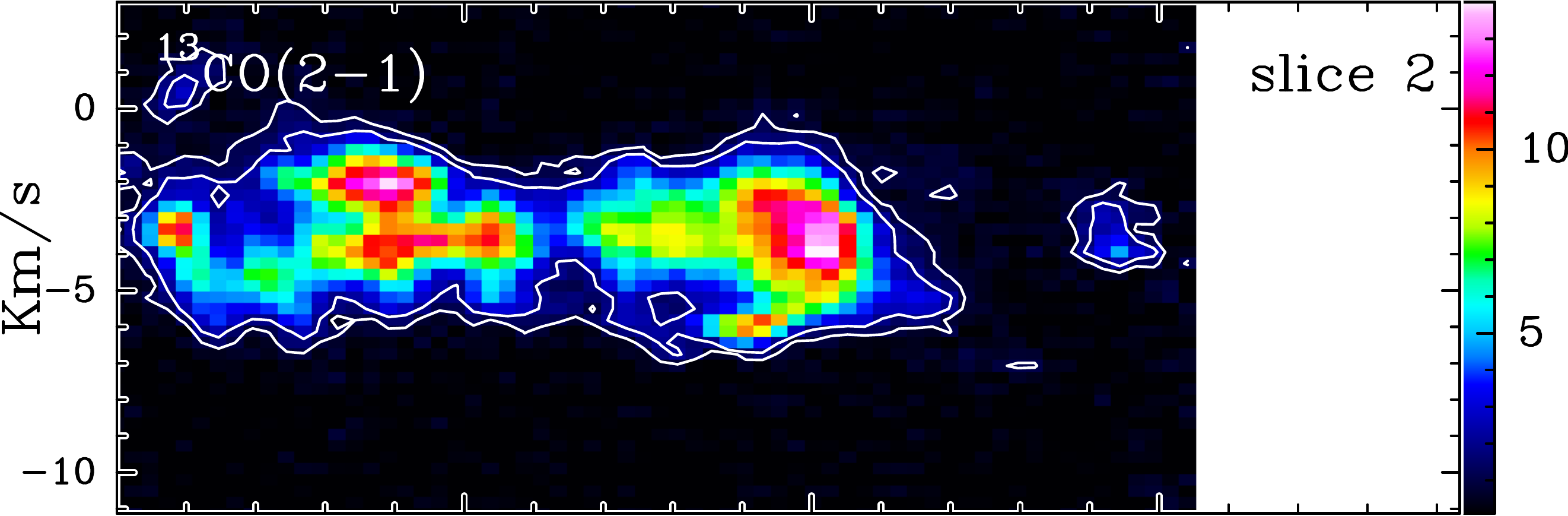}
                                               \includegraphics[angle=0]{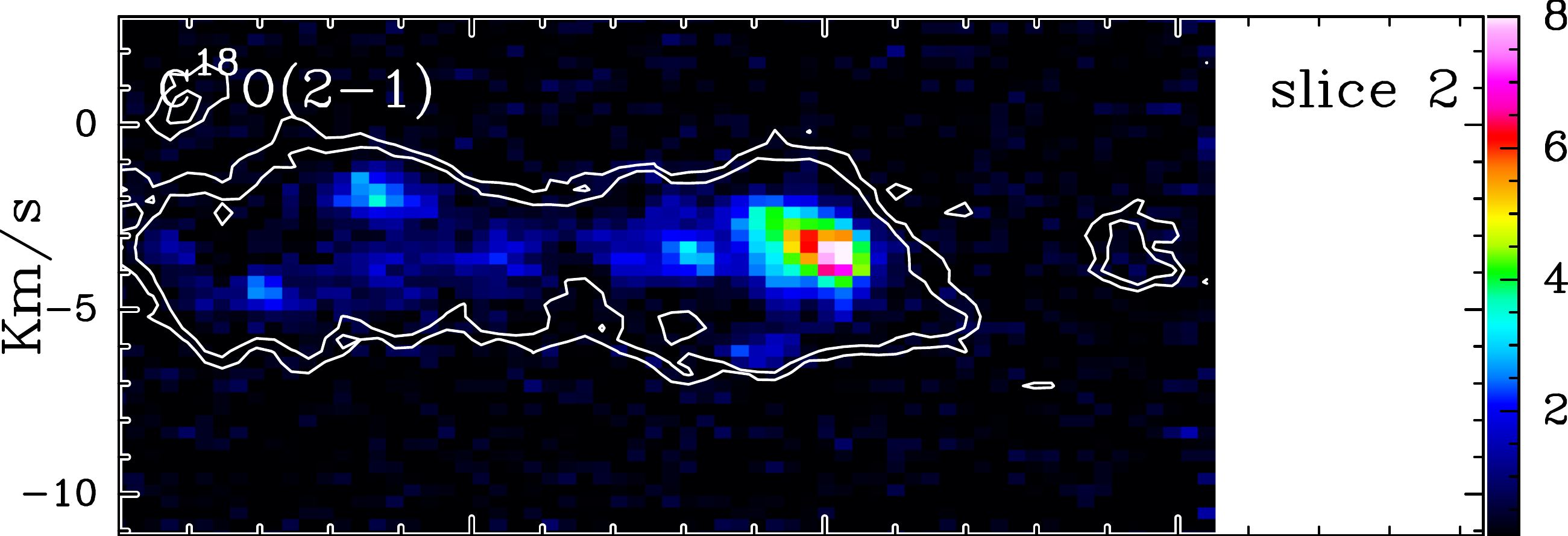}
                             }
     \resizebox{19cm}{!}{ 
                    \includegraphics[angle=0]{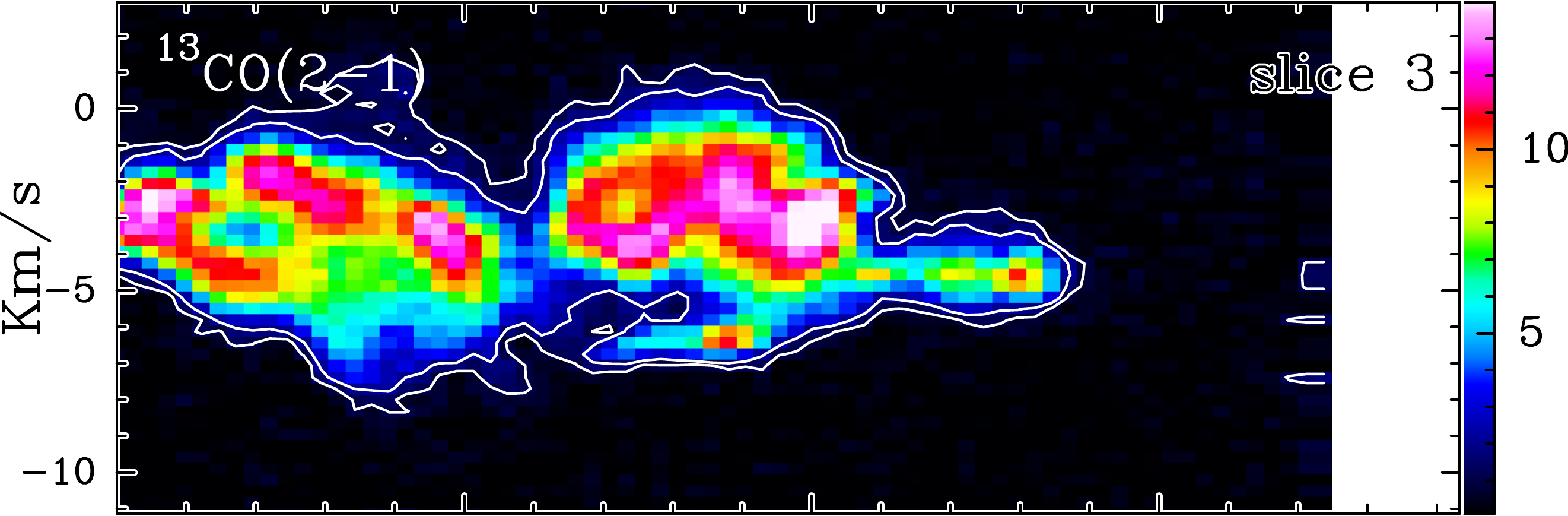}
                    \includegraphics[angle=0]{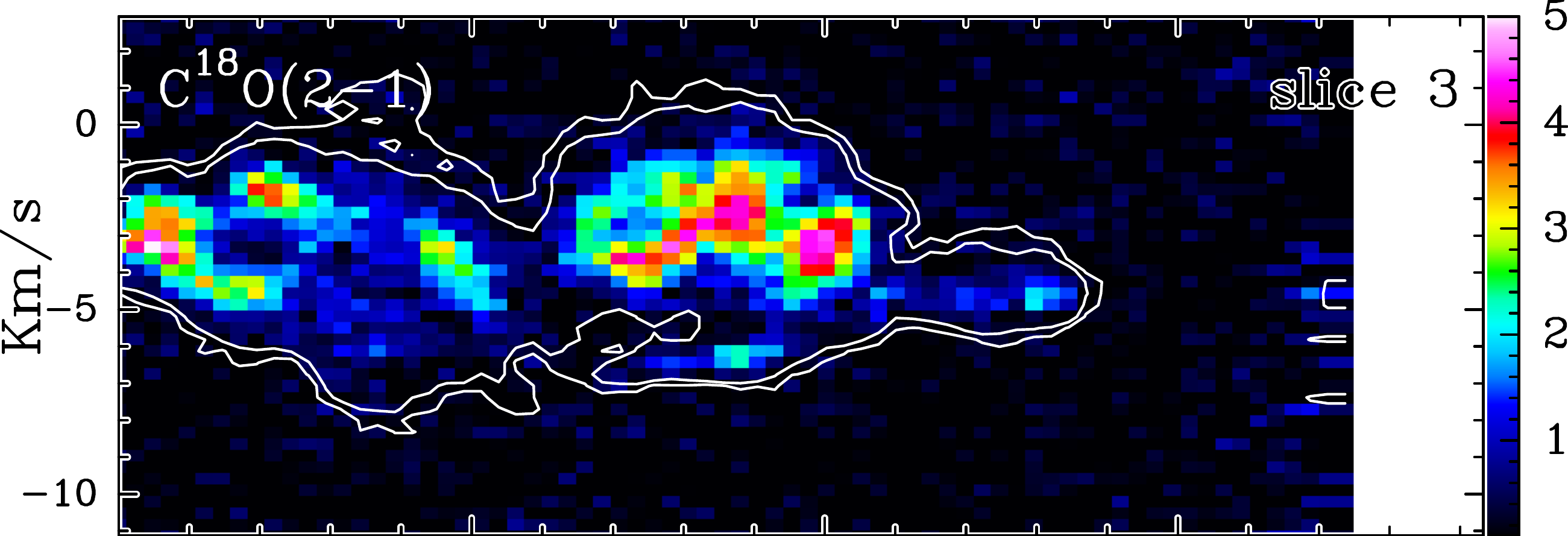}   }
                  \resizebox{19cm}{!}{   
                                              \includegraphics[angle=0]{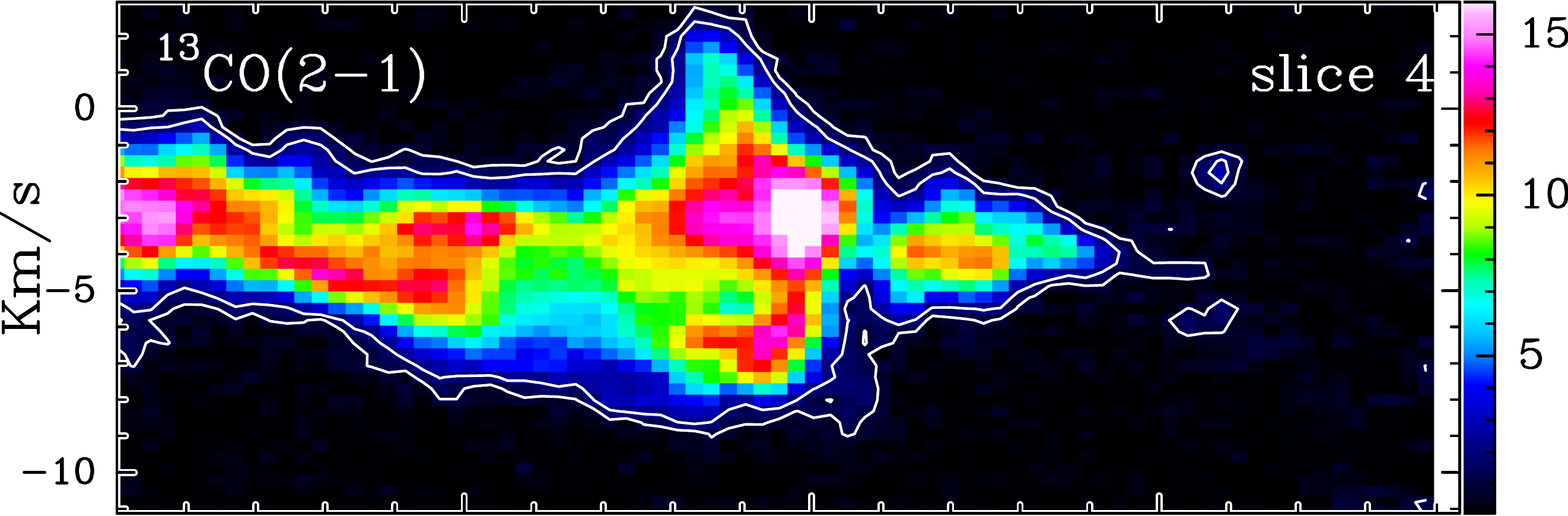}
                     \includegraphics[angle=0]{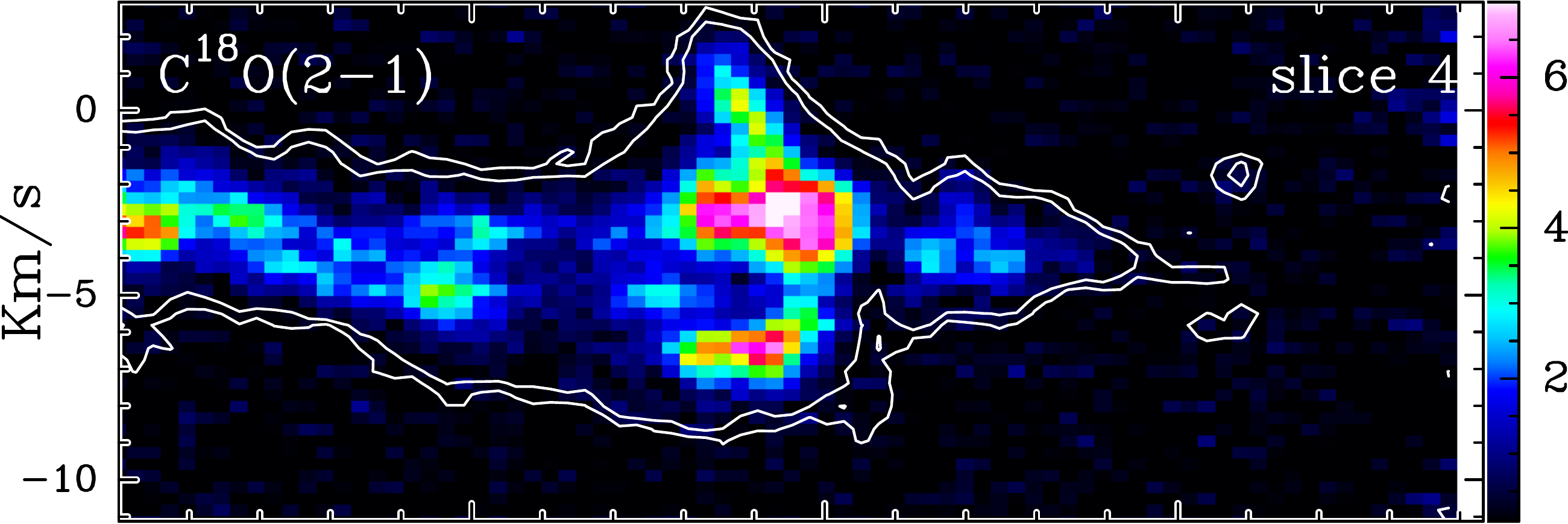}         }
                      \resizebox{19cm}{!}{            
                                              \includegraphics[angle=0]{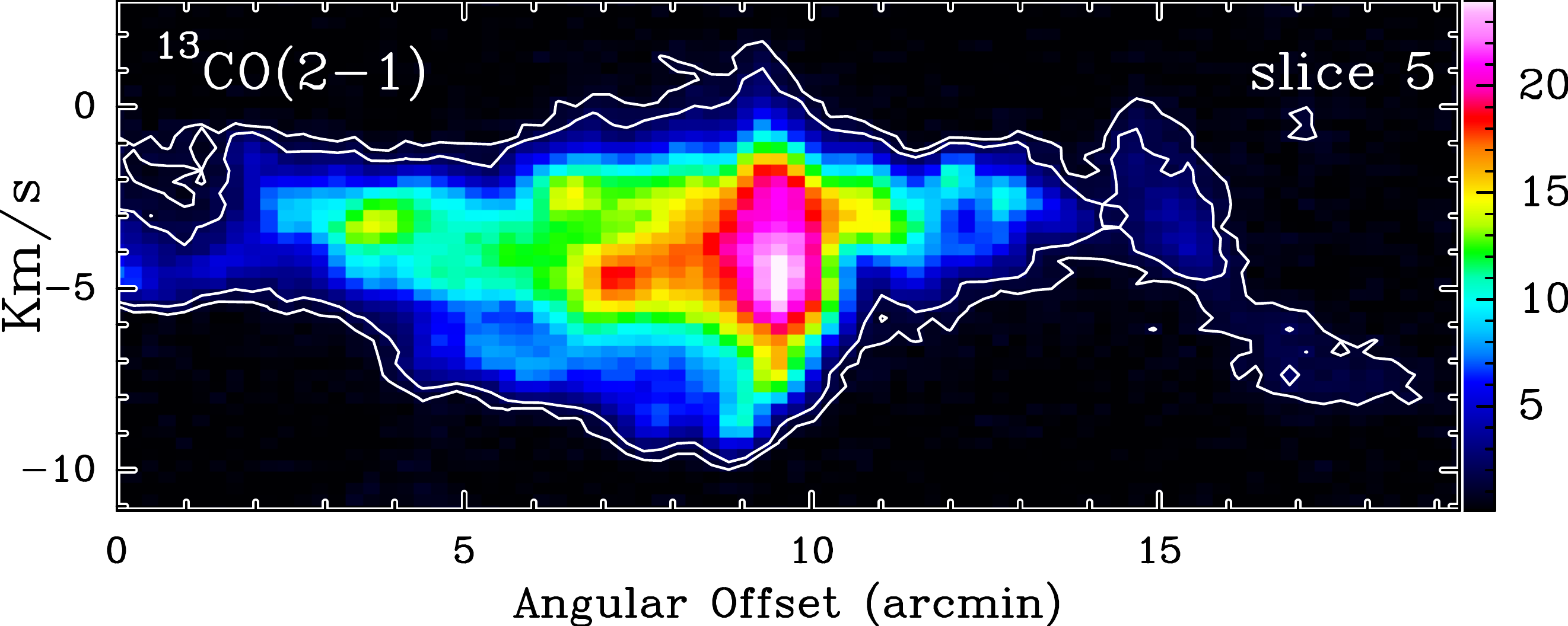}
                          \includegraphics[angle=0]{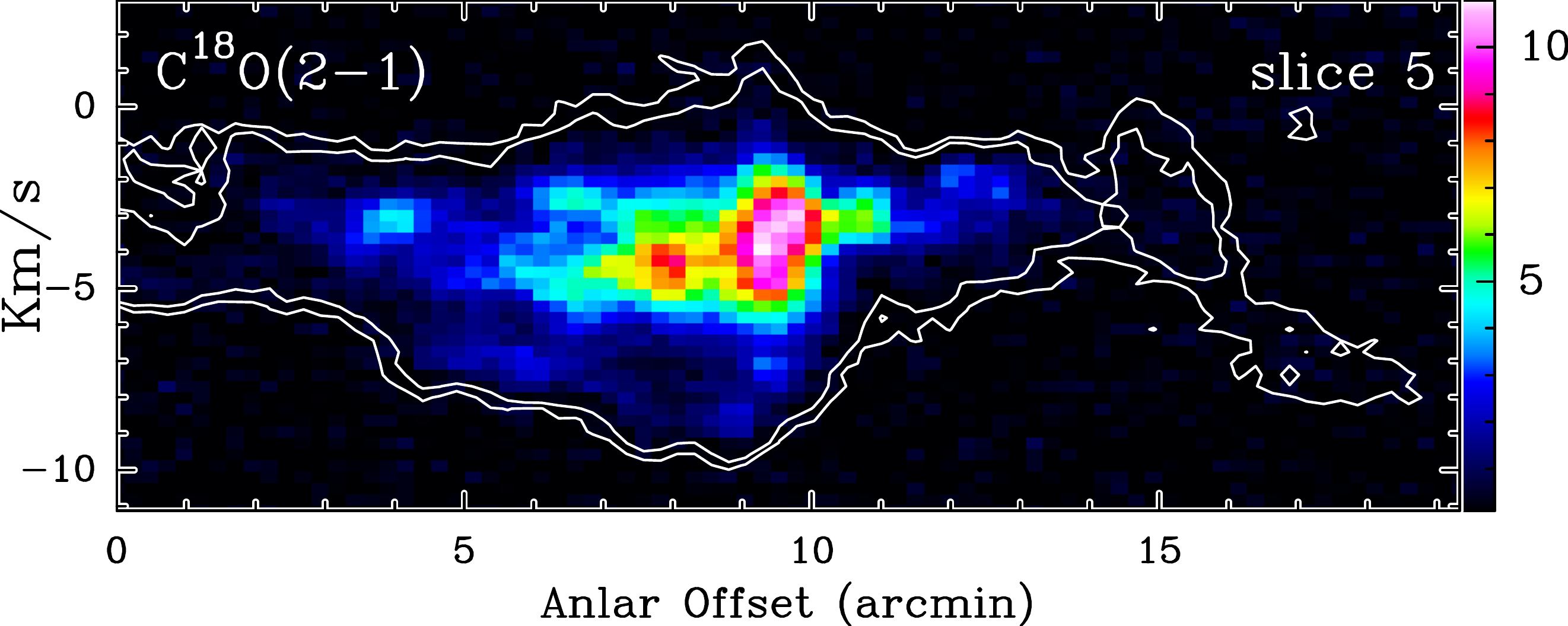}               }
  \caption{
  {\rev
Longitudinal  position$-$velocity (PV) diagrams along the crest of VCF~32 for the $^{13}$CO($2-1$) and C$^{18}$O($2-1$) 
emission on the left and right, respectively. 
  The PV maps are in units of K(T$_{\rm MB}$).
Each PV diagram is  perpendicular to the crest and averaged over 4~pixels  (2$\times$beams). 
The slices from 1 to 5 correspond to positions along the crest from east to west. The central positions of the slices are indicated on Fig.\,\ref{MapVelComp}. 
The white contours indicate the $^{13}$CO($2-1$)  intensity of 1~K and 2~K and are the same for the two panels of the same slice.
 The zero offset position (in the $x$-axes) corresponds to the Galactic South side of the filament systems.  At the distance of this cloud   10\arcmin\ corresponds to 3.8\,pc. The position of the filaments is between 9\arcmin\ and 10\arcmin. 
  }          }
  \label{PValongCrest}
    \end{figure*}

\end{appendix}

\end{document}